\documentclass[seceq]{ptptex}

\usepackage{graphicx}
\usepackage{epsf}
\usepackage{mathrsfs}

\makeatletter
\@ifundefined{lesssim}{\def\lesssim{\mathrel{\mathpalette\vereq<}}}{}
\@ifundefined{gtrsim}{\def\gtrsim{\mathrel{\mathpalette\vereq>}}}{}
\def\vereq#1#2{\lower3pt\vbox{\baselineskip1.5pt \lineskip1.5pt
\ialign{$\m@th#1\hfill##\hfil$\crcr#2\crcr\sim\crcr}}}
\makeatother
\newcommand{\beq}{\begin{equation}}
\newcommand{\eeq}{\end{equation}}
\newcommand{\beqn}{\begin{eqnarray}}
\newcommand{\eeqn}{\end{eqnarray}}

\title{
Current Status of Numerical-Relativity Simulations in Kyoto}

\author{
Yuichiro \textsc{Sekiguchi}$^{1}$, %
Kenta \textsc{Kiuchi}$^{1}$,
Koutarou \textsc{Kyutoku}$^{1,2}$,
Masaru \textsc{Shibata}$^{1}$
}

\inst{
$^{1}$Yukawa Institute for Theoretical Physics, Kyoto University, \\
Kyoto 606-8502, Japan,\\
$^{2}$Theory Center, Institute of Particle and Nuclear Studies, 
KEK, Tsukuba, Ibaraki 305-0801, Japan}

\abst{
We describe the current status of our numerical simulations for the collapse of a massive 
stellar core to a black hole (BH) and the merger of binary neutron stars (BNS), 
performed in the framework of full general relativity incorporating 
finite-temperature equations of state (EOS) and neutrino cooling.
 
For the stellar core collapse simulation, we present the latest numerical results. 
We employed a purely nucleonic EOS derived by Shen et al. (Nucl. Phys. A {\bf 637} (1998), 435).
As an initial condition, we adopted a 100 $M_{\odot}$ presupernova model calculated by 
Umeda and Nomoto (Astrophys. J. {\bf 637} (2008), 1014), which has a massive core 
($M \approx 3M_{\odot}$) with a high value of entropy per baryon ($s \approx 4 k_{B}$).
Changing the degree of rotation for the initial condition, we clarify
the strong dependence of the outcome of the collapse on this. 
When the rotation is rapid enough, 
the shock wave formed at the core bounce
is deformed to be a torus-like shape. Then, the infalling matter is 
accumulated in the central region due to the oblique shock at the torus surface, 
hitting the proto-neutron star
and dissipating the kinetic energy there. As a result, outflows can be launched.
The proto-neutron eventually collapses to a BH and an accretion torus is formed around it. 
We also found that the evolution of the BH and torus depends strongly on the rotation
initially given.

In the BNS merger simulations, we employ an EOS incorporating a degree of freedom for hyperons 
derived by Shen et al. (Astrophys. J. Suppl. {\bf 197} (2011), 20), 
in addition to the purely nucleonic EOS. 
The numerical simulations show that for the purely nucleonic EOS,
a hypermassive neutron star (HMNS) with a long lifetime ($\gg 10$ ms)
is the outcome for the total mass $M \lesssim 3.0M_{\odot}$. By contrast,
the formed HMNS collapses to a BH in a shorter time scale with
the hyperonic EOS for $M \gtrsim 2.7M_{\odot}$.  It is shown that the
typical total neutrino luminosity of the HMNS is $\sim 3$--$10\times
10^{53}$~ergs/s and the effective amplitude of gravitational waves
from the HMNS is 2--$6 \times 10^{-22}$ at $f\approx 2$--2.5~kHz for a
source distance of 100~Mpc.  
}

\begin{document}

\maketitle

\section{Introduction}

Along with the development of formulations and numerical techniques, as well
as progress in computational resources, numerical relativity (NR) is now the most viable
approach for exploring phenomena accompanying strong gravitational fields, such as
gravitational collapse of massive stellar core to a black hole (BH) or 
a neutron star (NS) and coalescence of compact-star binaries.
These phenomena show a wide variety of observable signatures, 
including  electromagnetic radiation, neutrinos, and gravitational radiation, 
and observations of neutrinos and gravitational radiation will provide us 
unique information of strong gravity and properties of dense nuclear matter 
otherwise cannot be obtained. 
Next-generation kilo-meter-size gravitational-wave detectors such as 
LIGO~\cite{LIGO}, VIRGO~\cite{VIRGO}, and KAGRA~\cite{LCGT} will report
the first detection of gravitational waves in the next $\sim 5$ years.
In addition, the above phenomena are promising candidates of the central 
engine of long gamma-ray bursts (LGRB) and short gamma-ray bursts (SGRB)~\cite{GRB}.

All four known forces of nature are involved and play important
roles in the stellar core collapse and the merger of binary compact objects: 
General relativistic gravity plays a crucial role 
in the formation of a BH and a neutron star.
Neutrinos produced by weak-interaction processes govern the energy and 
chemical evolution of the system. The electromagnetic and strong interactions 
determine the thermodynamical properties, in particular equation of state (EOS) 
of the dense nuclear matter. Strong magnetic field, if it is present, can modify 
the dynamics of the matter motion. To study the dynamical phenomena in general 
relativity, therefore, a multi-dimensional simulation incorporating a wide 
variety of physics is necessary.

We performed a simulation of stellar core collapse to a neutron star~\cite{Sekiguchi2010}
and a black hole~\cite{Sekiguchi2011},
incorporating a finite-temperature, a self-consistent treatment of the electron capture,
and neutrino cooling by a detailed leakage scheme, for the first time.
Such multi-dimensional simulations had not been done in full general 
relativity until quite recently~\footnote{There are a number of 
simulations of stellar core collapse in spherical symmetry~\cite{SpheGR} 
in which Boltzmann's equation is solved and detailed microphysical processes 
are implemented.}.
Ott et al.~\cite{Ott2007} (see also Dimmelmeier et al.~\cite{Dimm2007}) 
performed fully general relativistic simulations of stellar core collapse,
employing a finite-temperature EOS derived by Shen et al.~\cite{Shen1998} (Shen-EOS)
for the first time. 
In their calculation, however, the electron capture rate was not calculated in 
a self-consistent manner and neutrino cooling is not taken into account. 
Instead, they adopted a simple parameterized prescription proposed 
by Ref.~\citen{Lieb2005}: The electron fraction is assumed to be a function of density 
which is presumed based on a result of a single core collapse simulation with a specific 
initial condition. 
Recently, M\"uller et al.~\cite{Muller2012}
performed simulations of stellar core collapse with detailed microphysics and
neutrino transfer. However it is done in the framework of an approximate general 
relativistic gravity~\cite{xCFC}.
Kuroda et al.~\cite{Kuroda2012} have made a fully general relativistic code
with an approximate treatment of neutrino transfer, applying the schemes 
developed by the authors~\cite{Sekiguchi2010,SKSS2011}.
Ott et al.~\cite{Ott2011} performed simulations of rotating stellar core collapse,
employing the parameterized prescription~\cite{Lieb2005} in the collapse
phase, and a ray-by-ray neutrino leakage scheme after bounce.

As for the compact-star binary mergers, there have been only a few studies in
general relativistic frameworks~\footnote{There are several studies 
of binary neutron star mergers in Newtonian frameworks in which finite-temperature
EOS and weak interactions are taken into account together with neutrino cooling~\cite{Ruffert,Rosswog}.}.
In the framework of an approximate general relativistic gravity 
(the conformal flatness approximation~\cite{CFC}),
Oechslin and Janka~\cite{OJM} performed simulations of binary neutron star (BNS) 
mergers adopting the Shen-EOS and a finite-temperature EOS by Lattimer and 
Swesty~\cite{LS}, but they did not take account of weak interaction processes.
Duez et al.~\cite{DuezBHNS} studied effects of EOS on the dynamics of black 
hole-neutron star mergers (BHNS) adopting the Shen-EOS in full general relativity.
Recently, Bauswein et al.~\cite{Bauswein} performed BNS simulations adopting a wide variety of EOS 
in the conformal flatness approximation 
and investigated the dependence of gravitational wave spectra on EOS.
In these works, however, the weak interaction processes are not included.

In this paper, we describe our latest results of numerical-relativity
simulations for the stellar core collapse to a BH~\cite{Sekiguchi2012} 
and the BNS merger~\cite{SKKS1,SKKS2}, which are performed incorporating both a finite-temperature
EOS~\cite{Shen1998,Shen2011} and neutrino cooling~\cite{Sekiguchi2010}.
For reviews on other topics, namely, simulations of stellar core collapse to a neutron star, 
of BHNS merger, and of BH-BH binary merger, the reader may refer to 
Refs.~\citen{Ott,Kotake,LLR}, Refs.~\citen{ShibataTaniguchi,Duez} and 
Refs.~\citen{Pretorius,Centrella}, respectively.

The paper is organized as follows. In \S~\ref{Sec_BasicEq}, we first briefly summarize
basic equations, input microphysics, and numerical setup. 
The results of simulations of the stellar core collapse and BNS merger are described in
\S~ \ref{Sec_ResultSCC} and \S~\ref{Sec_ResultBNS}, respectively. 
Section~\ref{Sec_Summary} is devoted to a summary.
Throughout this paper, $\hbar$, $k_{B}$, $c$, and $G$ denote 
the Planck's constant, the Boltzmann's constant, the speed of light, and
the gravitational constant, respectively.
In appendices, details of the microphysics adopted in our latest implementation
are summarized for the purpose of completeness.
We adopt the geometrical unit $c=G=1$ in \S~\ref{Sec_3+1} and 
\S~\ref{Sec_Leakage}.

\section{Basic Equations and Numerical Method}\label{Sec_BasicEq}

\subsection{Einstein's equations and gauge conditions} \label{Sec_3+1}

The standard variables in the 3+1 decomposition of Einstein's equations are 
the three-dimensional metric $\gamma_{ij}$ and 
the extrinsic curvature $K_{ij}$
on a three-dimensional hypersurface defined by~\cite{York79}
\beqn
\gamma_{\mu\nu} &\equiv& g_{\mu\nu} + n_{\mu}n_{\nu}, \\
K_{\mu\nu} &\equiv& - \frac{1}{2} \mathscr{L} _{n} \gamma_{\mu\nu}, 
\eeqn
where $g_{\mu\nu}$ is the spacetime metric,
 $n_{\mu}$ is the unit normal to the three-dimensional
hypersurface, and $\mathscr{L}_{n}$ is the Lie derivative with
respect to the unit normal $n^{\mu}$. 
Then the line element is written in the form
\beq
ds^{2} = - \alpha^{2} dt^{2} + \gamma_{ij}(dx^{i}+\beta^{i}dt)
(dx^{j}+\beta^{j}dt),
\eeq
where $\alpha$ and $\beta^{i}$ are the lapse function
and the shift vector, which describe the gauge degree of freedom.

Numerical simulations are performed in the so-called BSSN-puncture 
formulation~\cite{Shibata95,Baumgarte99,Campanelli06},
in which the spatial metric, $\gamma_{ij}$, is conformally decomposed as
$\gamma _{ij} = W^{-2}\tilde{\gamma}_{ij}$
where the condition, $\det (\tilde{\gamma}_{ij}) = 1$, is imposed for the
conformal spatial metric $\tilde{\gamma}_{ij}$. 
From this condition, the conformal factor is written as
$W^{-6} = \det(\gamma_{ij})$. 
The extrinsic curvature, $K_{ij}$, is decomposed into the trace part, 
$K$, and the traceless part, $A_{ij}$, as
$K_{ij} = A_{ij} + (1/3)\gamma_{ij}K$. 
The traceless part is conformally
decomposed as $A_{ij} = W^{-2}\tilde{A}_{ij}$.
To summarize, the fundamental quantities for the evolution equation are now
split into 
$W$, $\tilde{\gamma}_{ij}$, $K$, and $\tilde{A}_{ij}$.
Furthermore, the auxiliary variable 
$F_{i} \equiv \delta^{jk}\partial_{k}\tilde{\gamma}_{ij} $ 
is introduced in the original version of the BSSN formulation~\cite{Shibata95}.
Merits of using $W$ as a conformal factor  are that (i) the equation 
for the Ricci tensor is slightly simplified, 
(ii) no singular term appears in the evolution equations even for 
$W \rightarrow 0$, and (iii) the determinant of $\gamma_{ij}$ is
always positive~\cite{Marronetti08,Yamamoto08}.

The basic equations to be solved are
\begin{eqnarray}
&& \left(\partial_{t} - \beta^{k}\partial_{k} \right) W =
 \frac{1}{3}\left(\alpha K - \partial_{k} \beta^{k} \right)W ,
\label{phidevelopP}\\
&& \left(\partial_{t} - \beta^{k}\partial_{k} \right)\tilde{\gamma}_{ij} = 
 -2 \alpha \tilde{A}_{ij} + \tilde{\gamma}_{ik}\partial_{j}\beta^{k} +
 \tilde{\gamma}_{jk}\partial_{i}\beta^{k} -
 \frac{2}{3}\tilde{\gamma}_{ij} \partial_{k}\beta^{k}, 
\label{tilGamdevelopP}\\
&& \left(\partial_{t} - \beta^{k}\partial_{k} \right) K = 
 - D^{k}D_{k} \alpha  + \alpha \left[ \tilde{A}_{ij}\tilde{A}^{ij} +
  \frac{1}{3}K^{2} \right]
 + 4 \pi \alpha \left( e^{\rm Total}_{h} + S^{\rm Total} \right) , \label{trKdevelopP}\\
&& \left(\partial_{t} - \beta^{k}\partial_{k} \right)\tilde{A}_{ij} =
  \alpha W^{2} \left(R_{ij} - \frac{1}{3}\tilde{\gamma}_{ij} R \right) 
- \left(W^{2}D_{i}D_{j} \alpha -\frac{1}{3}\tilde{\gamma}_{ij} D^{k}D_{k}\alpha  \right) \nonumber \\
&& \ \ \ \ \ \ \ \ \ \ \ \ \ \ \ \ \ \ \ \ \ \ \ \  
 + \alpha \left( K \tilde{A}_{ij} - 2 \tilde{A}_{ik} \tilde{A}^{k}_{\ j} \right)
 + \tilde{A}_{ik}\partial_{j}\beta^{k} + \tilde{A}_{jk}\partial_{i}\beta^{k}
 - \frac{2}{3} \tilde{A}_{ij} \partial_{k}\beta^{k} \nonumber \\
&& \ \ \ \ \ \ \ \ \ \ \ \ \ \ \ \ \ \ \ \ \ \ \ \ 
 - 8 \pi \alpha \left( W^{2} S^{\rm Total}_{ij} 
 - \frac{1}{3}\tilde{\gamma}_{ij}S^{\rm Total} \right), \label{AdevelopP} \\
&&
\left( \partial_{t} - \beta^{k}\partial_{k} \right)F_{i} = 
  -16\pi \alpha j^{\rm Total}_{i} \nonumber \\
&& \ \ \ \ \ \ \ \ \ \ + 2\alpha 
   \left\{
   f^{kj}\partial_{j}\tilde{A}_{ik} + \tilde{A}_{ik} \partial_{j}f^{kj} 
   - \frac{1}{2}\tilde{A}^{jl}\partial_{i}h_{jl}
   - 3 \tilde{A}^{k}_{\ i}\partial_{k}\ln W 
   - \frac{2}{3}\partial_{i}K
  \right\} \nonumber \\
&& \ \ \ \ \ \ \ \ \ \   
  + \delta^{jk}
  \left\{
   -2 \tilde{A}_{ij} \partial_{k} \alpha 
   + \left(\partial_{k}\beta^{l}\right)\partial_{l}h_{ij}
   \right. \nonumber \\
&& \ \ \ \ \ \ \ \ \ \  \ \ \ \ \ \ \ \ \ 
   \left. + \partial _{k} \left(
	\tilde{\gamma}_{il}\partial_{j}\beta^{l} 
        + \tilde{\gamma}_{jl}\partial_{i}\beta^{l}
        -\frac{2}{3}\tilde{\gamma}_{ij} \partial_{l}\beta^{l}
     \right) 
\right\},
\label{FdevelopP}
\end{eqnarray}
where $f^{ij}\equiv \tilde{\gamma}^{ij} - \delta^{ij}$. 
$^{(3)}R$, $^{(3)}R_{ij}$, and $D_{i}$ are the Ricci scalar, the
Ricci tensor, and the covariant derivative associated with
three-dimensional metric $\gamma_{ij}$, respectively. 
The matter source terms are the projections of the stress-energy tensor 
(see Eq. (\ref{Eq_T_comb})) with respect 
to $n^{\mu}$ and $\gamma_{\mu\nu}$, and $S^{\rm Total} \equiv \gamma^{ij}S^{\rm Total}_{ij}$:
\beqn
e^{\rm Total}_{h} &\equiv& \ \ (T^{\rm Total})^{\alpha \beta} n_{\alpha}n_{\beta}, \\
j^{\rm Total}_{i}    &\equiv& -(T^{\rm Total})^{\alpha \beta} \gamma_{i\alpha}n_{\beta}, \\ 
S^{\rm Total}_{ij}   &\equiv& \ \ (T^{\rm Total})^{\alpha \beta} \gamma_{i\alpha}\gamma_{j \beta},
\eeqn 
where $(T^{\rm Total})_{\alpha \beta}$ is the total energy-momentum
tensor (see Eq.~(\ref{Eq_Ttot}) for definition).

As a gauge condition for the lapse, 
we use a dynamical slicing~\cite{Bona,Alcubierre01a}
\footnote{In the axisymmetric stellar core collapse simulation described in 
\S~\ref{Sec_ResultSCC}, we do not include the advection term $\beta^{k}\partial_{k}\alpha$.}:
\beq
(\partial_{t}-\beta^{k}\partial_{k} )\alpha = -2K \alpha.
\eeq
It is known that this dynamical slicing enables to perform a long-term-evolution 
simulation of neutron stars and BH spacetime.
The shift vector is determined by solving a dynamical-shift 
equation~\cite{Shibata03b}
\begin{equation}
\partial_{t}\beta^{k} = \tilde{\gamma}^{kl} (F_{l} + \Delta t
 \partial_{t} F_{l}).
\label{Dynbeta}
\end{equation}
Here the second term in the right-hand side is necessary for
the numerical stability, and $\Delta t$ denotes the numerical timestep.

A fourth-order-accurate finite differencing in space and a fourth-order
Runge-Kutta time integration are used in solving Einstein's equations and 
the gauge conditions.

\subsection{Hydrodynamic equations and GR leakage scheme}\label{Sec_Leakage}

Numerical simulations were performed using a fully general relativistic 
hydrodynamic code~\cite{Sekiguchi2010} recently developed, in which
a nuclear-theory-based finite-temperature EOS,
a self-consistent treatment of electron and positron captures, and neutrino cooling by
a general relativistic leakage scheme, are implemented. 

Because the characteristic timescale of the weak-interaction processes 
($t_{\rm wp} \sim \vert Y_{e}/\dot{Y}_{e} \vert $) 
is much shorter than the dynamical timescale, $t_{\rm dyn}$, in hot 
dense matters, source terms in the hydrodynamic equations become 
too {\it stiff} for the equations to be solved explicitly in a 
straightforward manner~\cite{Bruenn85}:
A very short timestep ($\Delta t$ $<$ $t_{\rm wp} \ll t_{\rm dyn}$) will
be required to solve the equations explicitly.
The characteristic timescale, $t_{\rm leak}$, with which neutrinos leak out
from the system, by contrast, is much longer 
than $t_{\rm wp}$ in the hot dense matter region, as
 $t_{\rm leak} \sim L/c \sim t_{\rm dyn}$, where $L$ is the
characteristic length scale of the system. 
Using this fact, we developed a method of solving the hydrodynamic 
equations in which the source terms are characterized by 
the leakage timescale $t_{\rm leak}$.

Note that neutrino heating is not included in the current version of the leakage scheme. 
A conservative shock capturing scheme~\cite{KT} with third-order accuracy in space and 
fourth-order accuracy in time is employed for solving hydrodynamic equations.
In this section, we adopt the geometrical unit $c=G=1$.

\subsubsection{Energy-momentum conservation equation}

The basic equations of general relativistic hydrodynamics including the radiation transfer
for neutrinos are
\beq
\nabla_{\alpha}(T^{\rm Total})^{\alpha}_{\ \beta} 
= \nabla_{\alpha}\left[(T^{\rm F})^{\alpha}_{\ \beta} +
(T^{\nu})^{\alpha}_{\ \beta} \right] = 0, \label{Eq_Ttot}
\eeq
where $(T^{\rm Total})_{\alpha \beta}$ is the total energy-momentum
tensor, and $(T^{\rm F})_{\alpha \beta}$ and $(T^{\nu})_{\alpha \beta}$
are the energy-momentum tensor of fluids and neutrinos, respectively.
Equation (\ref{Eq_Ttot}) can be decomposed, by introducing the interaction source term,
 as
\beqn
\nabla_{\alpha}(T^{{\rm F}})^{\alpha}_{\beta} &=& -Q_{\beta}, \label{T_Eq1} \\
\nabla_{\alpha}(T^{\nu})^{\alpha}_{\beta} &=& Q_{\beta} \label{T_Eq2}.
\eeqn
Here the source term $Q_{\alpha}$ is characterized by $t_{\rm wp}$ and 
becomes too stiff in hot dense matter regions.
To overcome the situation, the following procedures are adopted.

\vspace{2mm}
\begin{enumerate}
\item The neutrino energy-momentum tensor is decomposed
into 'trapped-neutrino' ($(T^{\nu,{\rm T}})_{\alpha\beta}$) 
and 'streaming-neutrino' ($(T^{\nu,{\rm S}})_{\alpha\beta}$) parts as
\beq
(T^{\nu})_{\alpha\beta} = (T^{\nu,{\rm T}})_{\alpha\beta} +
                        (T^{\nu,{\rm S}})_{\alpha\beta}. 
\label{Eq_Tnu_div}
\eeq
Here, the trapped-neutrino part phenomenologically represents neutrinos which 
interact sufficiently frequently with matter, and the streaming-neutrino 
part describes a phenomenological flow of neutrinos which freely stream out of the system\footnote{
We note that Liebend\"orfer et al.~\cite{Lieb09} developed a more sophisticate method 
in terms of the distribution functions of trapped and streaming neutrinos 
in the Newtonian framework.}

\item A part of neutrinos produced by Eq. (\ref{T_Eq2}) is assumed to {\it leak out} 
to be the streaming-neutrinos with a leakage rate $Q^{\rm leak}_{\alpha}$:
\beq
\nabla_{\alpha}(T^{\nu,{\rm S}})^{\alpha}_{\ \beta} =  Q^{\rm leak}_{\beta}.
\label{Eq_T_nuS}
\eeq
On the other hand, it is assumed that the remaining neutrinos constitute the trapped-neutrino part:
\beq
\nabla_{\beta}(T^{\nu,{\rm T}})^{\beta}_{\alpha} = 
Q_{\alpha} - Q^{\rm leak}_{\alpha}.
\label{T_Eq_nuT}
\eeq

\item The trapped-neutrinos is combined with the fluid part as
\beq
T_{\alpha\beta} \equiv (T^{\rm F})_{\alpha\beta} 
+ (T^{\nu,{\rm T}})_{\alpha\beta}. \label{Eq_T_comb}
\eeq
Then the equation for $T_{\alpha \beta}$ is
\beq
\nabla_{\alpha}T^{\alpha}_{\ \beta} = -Q^{\rm leak}_{\beta} \label{Eq_T}.
\eeq
We solve Eqs. (\ref{Eq_T_nuS}) and (\ref{Eq_T}). 
Note that the new equations only include the source term,
$Q^{\rm leak}_{\alpha}$, which is characterized by the leakage 
timescale $t_{\rm leak}$.
Definition of $Q^{\rm leak}_{\alpha}$ is given in \S~\ref{Sec_leakagerate}.
\end{enumerate}
\vspace{2mm}

The energy-momentum tensor of the fluid and trapped-neutrino parts
($T_{\alpha \beta}$) is treated as that of a perfect fluid,
\beq
T_{\alpha\beta} = (\rho + \rho \varepsilon + P)
 u_{\alpha}u_{\beta} + P g_{\alpha\beta}, \label{T_fluid}
\eeq
where $\rho$ and $u^{\alpha}$ are the rest mass density and the 4-velocity of the fluid.
The specific internal energy ($\varepsilon$) and the pressure ($P$) 
are the sum of the contributions from the baryons 
(free protons, free neutrons, $\alpha$-particles, and heavy nuclei), 
leptons (electrons, positrons, and {\it trapped-neutrinos}), and photons as,
\beqn
P &=& P_{B} + P_{e} + P_{ph} + P_{(\nu)}, \\
\varepsilon &=&  
\varepsilon_{B} + \varepsilon_{e} + \varepsilon_{ph} + \varepsilon_{(\nu)},
\eeqn
where subscripts '$B$', '$e$', '$ph$', and '$(\nu)$' denote the components
of baryons, electrons and positrons, photons, and trapped-neutrinos 
(for $\nu_{e}$, $\bar{\nu}_{e}$, and $\nu_{x}$, see \S~\ref{Sec_BL}), respectively. 
Our treatment of the EOS is summarized in \S~\ref{Sec_EOS}.

The Euler equation ($\gamma_{i}^{\alpha} \nabla_{\beta}
T^{\beta}_{\ \alpha} = - \gamma_{i}^{\alpha} Q^{\rm leak}_{\alpha}$), 
and the energy equation 
($n^{\alpha}\nabla_{\beta}T^{\alpha}_{\beta}
=-n^{\alpha}Q^{\rm leak}_{\alpha})$ can be written explicitly, in terms of 
$e_{h} \equiv T^{\alpha \beta} n_{\alpha}n_{\beta}$ and 
$j_{i}\equiv -T^{\alpha \beta} \gamma_{i\alpha}n_{\beta}$, as,
\beqn
&&\partial_{t}(\sqrt{\gamma} j_{i}) 
+ \partial_{k}\left[ \sqrt{\gamma} (j_{i}v^{k}+\alpha\delta_{i}^{k}) \right] \nonumber \\
&& \ \ \ \ \ \ \ \ \ \ \ \ \ \ \ \ \ \ \ \ = 
\sqrt{\gamma}\left[ -e_{h}\partial_{i}\alpha 
+ j_{k}\partial_{i}\beta^{k} + \frac{\alpha}{2}S^{jk}\partial_{i}\gamma_{jk}
-\alpha Q^{\rm leak}_{i} \right], \label{Eq_euler}\\
&&\partial_{t}(\sqrt{\gamma}e_{h}) 
+ \partial_{k}\left[\sqrt{\gamma}(e_{h}v^{k}+P(v^{k}+\beta^{k}))\right] \nonumber \\
&& \ \ \ \ \ \ \ \ \ \ \ \ \ \ \ \ \ \ \ \ = 
\alpha \sqrt{\gamma}\left(
S^{ij}K_{ij}-\gamma^{ik}j_{i}\partial_{k}\ln \alpha + n^{\mu}Q_{\mu}^{\rm leak} \right),
\label{Eq_energy}
\eeqn
where $w\equiv \alpha u^{t}$ and $v^{i}\equiv u^{i}/u^{t}$.

The streaming-neutrino part, on the other hand, is written in the general form of
\beq
(T^{\nu,{\rm S}})_{\alpha\beta}= 
E n_{\alpha}n_{\beta} + F_{\alpha}n_{\beta} + F_{\beta}n_{\alpha} + P_{\alpha\beta},
\label{T_neutrino}
\eeq
where $F_{\alpha}n^{\alpha}=P_{\alpha \beta}n^{\alpha}=0$. 
Then the evolution equations of streaming-neutrinos 
($E$ and $F_{i}$) are explicitly written as
\beqn
&&\!\!\!\!\!\!\!\!
     \partial_{t}(\sqrt{\gamma}E) 
  + \partial_{k}\left[\sqrt{\gamma}(\alpha F^{k} - \beta^{k}E)\right]  
  = \sqrt{\gamma}\left(
      \alpha P^{kl}K_{kl} - F^{k}\partial_{k}\alpha - \alpha Q^{\rm leak}_{a}n^{a} 
    \right), \label{Eq_Erad}\\
&&\!\!\!\!\!\!\!\!
    \partial_{t}(\sqrt{\gamma}F_{i}) 
  + \partial_{k}\left[\sqrt{\gamma}(\alpha P^{k}_{i}-\beta^{k}F_{i})\right]
  \nonumber \\
&&\!\!  \ \ \ \ \ \ \ \ \ \ \ \ \ \ \ \ \ \ \ \ \ \ \ \ \ \ \ 
  = \sqrt{\gamma}\left(
    -E\partial_{i}\alpha + F_{k}\partial_{i}\beta^{k} 
    + \frac{\alpha}{2}P^{kl}\partial_{i}\gamma_{kl} + \alpha Q^{\rm leak}_{i}
    \right). \label{Eq_Frad}
\eeqn
In order to close the system,
we need an explicit expression of $P_{\alpha \beta}$ (closure relation).
In the current implementation, we adopt a simple form 
$P_{\alpha \beta}=\chi E \gamma_{\alpha \beta}$ with $\chi = 1/3$.
We solve Eq. (\ref{Eq_T_nuS}) in a high resolution shock capturing 
scheme~\cite{Sekiguchi2010}.

The closure relation employed in the current implementation is not very physical.
Moreover we do not consider the so-called neutrino heating. 
To take into account the neutrino heating as well as the propagation of the streaming
neutrinos accurately, 
a more sophisticated implementation of handling the neutrino transfer together with a better 
closure relation is required.
However, such a study is beyond the scope of this 
paper. A more sophisticated implementation based on the moment method~\cite{Thorne1981,SKSS2011},
will be presented in the near future.

\subsubsection{Baryon and Lepton-number conservation equations}\label{Sec_BL}

The continuity equation for the baryon is
\beq
\nabla_{\alpha}(\rho u^{\alpha}) = 0 \label{conti},
\eeq
which can be written explicitly as
\beq
\partial_{t}(\sqrt{\gamma}\rho w) + \partial_{k}(\sqrt{\gamma}\rho w v^{i}) = 0.
\label{Eq_conti}
\eeq

The conservation equations of the lepton fractions are written 
schematically as
\beqn
&&\!\! \frac{d Y_{e}}{dt} = \gamma_{e} , \label{dYe} \\
&&\!\! \frac{d Y_{\nu_{e}}}{dt} = \gamma_{\nu_{e}},  \label{dYnu} \\ 
&&\!\! \frac{d Y_{\bar{\nu}_{e}}}{dt} = \gamma_{\bar{\nu}_{e}},  \label{dYna} \\ 
&&\!\! \frac{d Y_{\nu_{x}}}{dt} = \gamma_{\nu_{x}},  \label{dYno}  
\eeqn
where $Y_{e}$, $Y_{\nu_{e}}$, $Y_{\bar{\nu}_{e}}$, and $Y_{\nu_{x}}$ denote
the fractions per baryon number for electrons, electron neutrinos ($\nu_{e}$), 
electron anti-neutrinos ($\bar{\nu}_{e}$), and {\it total} of 
$\mu$ and $\tau$ neutrinos and anti-neutrinos ($\nu_{x}$),
respectively. Note that only the trapped-neutrinos are responsible for these
neutrino fractions.

Using the continuity equation for the baryon,
we rewrite the conservation equations of the lepton fractions in the following form as
\beq
\partial_{t}(\sqrt{\gamma}\rho w Y_{(L)}) 
+ \partial_{k}(\sqrt{\gamma}\rho w Y_{(L)}v^{k} ) = \sqrt{\gamma} \alpha \rho \gamma_{(L)},
\label{e-Y} 
\eeq
where $Y_{(L)}$ and $\gamma_{(L)}$ are abbreviated expressions of the
lepton fractions and the source terms.

\begin{figure}[t]
\begin{center}
 \includegraphics[scale=0.5]{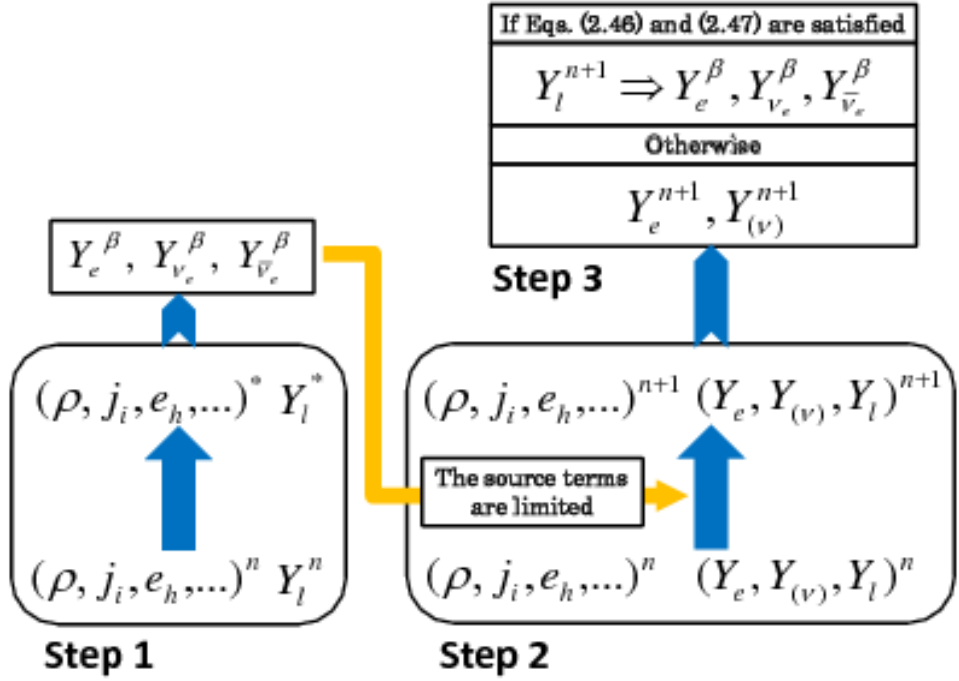}
\end{center}
  \caption{ A schematic picture of the leakage scheme. See text for details.
\label{fig_evolv}}
\end{figure}

The source terms are given by 
\beqn
-\gamma_{e} &=& \gamma_{\nu_{e}}^{\rm local} - \gamma_{\bar{\nu}_{e}}^{\rm local}, \\
\gamma_{\nu_{e}} &=& \gamma_{\nu_{e}}^{\rm local} - \gamma_{\nu_{e}}^{\rm leak}, \\
\gamma_{\bar{\nu}_{e}} &=& \gamma_{\bar{\nu}_{e}}^{\rm local} 
                        - \gamma_{\bar{\nu}_{e}}^{\rm leak}, \\
\gamma_{\nu_{x}} &=& \gamma_{\nu_{x}}^{\rm local} - \gamma_{\nu_{x}}^{\rm leak}, 
\eeqn
where $\gamma^{\rm local}$'s and $\gamma^{\rm leak}$'s are rates of 
the local production and leakage for each species of neutrinos, respectively.
As local reactions, we consider the electron capture, the positron capture, 
electron-positron pair annihilation, plasmon decay, 
and the Bremsstrahlung radiation of pair neutrinos
(see \S~\ref{Sec_leakagerate} for definitions and details).
Because $\gamma^{\rm local}$'s are characterized 
by the timescale of weak-interaction processes 
$t_{\rm wp}$, we follow the procedure proposed in Ref.~\citen{Sekiguchi2010} 
to stably solve the equations with a usual timestep ($\Delta t \approx 0.4\Delta x$) 
in an explicit manner (see Fig. \ref{fig_evolv}). 

\vspace{2mm}
\begin{enumerate}
\item At each timestep $n$, we first solve the conservation equation of 
the {\it total} lepton fraction ($Y_{l}=Y_{e}+Y_{\nu_{e}}-Y_{\bar{\nu}_{e}}$), 
\beqn
&&\!\! \frac{d Y_{l}}{dt} = \gamma_{l} 
= -(\gamma^{\rm leak}_{{\nu}_{e}}-\gamma^{\rm leak}_{\bar{\nu}_{e}})
,  \label{dYl}
\eeqn
instead of solving Eqs. (\ref{dYe})--(\ref{dYna}),
together with Eqs. (\ref{Eq_conti}), (\ref{Eq_euler})--(\ref{Eq_Frad}), and
(\ref{dYno}).
Note that the source term in Eq. (\ref{dYl}) is characterized by the leakage timescale
and can be solved explicitly.
In this step, we assume that the $\beta$-equilibrium condition is achieved, and 
the source terms are modified according to this assumption.
After the time integration, the lepton fractions in the 
'hypothetical' $\beta$-equilibrium ($Y_{e}^{\beta}$, $Y_{\nu_{e}}^{\beta}$, 
and $Y_{\bar{\nu}_{e}}^{\beta}$) are calculated from the evolved $Y_{l}$. 

\item We next solve the whole set of the equations 
(Eqs. (\ref{Eq_conti}), (\ref{Eq_euler})--(\ref{Eq_Frad}), and
(\ref{dYe})--(\ref{dYno})). In this step, we first calculate the maximum 
allowed values of the source terms, $\gamma_{\nu_{e}, {\rm max}}^{\rm local}$ and 
$\gamma_{\bar{\nu}_{e}, {\rm max}}^{\rm local}$, regarding that 
$Y_{\nu_{e}}^{\beta}$ and $Y_{\bar{\nu}_{e}}^{\beta}$ as the maximum allowed values 
of the neutrino fractions at the next timestep $n+1$.
Then the source terms are limited according to
\beqn
\gamma_{\nu_{e}}^{\rm local} &=& {\rm min} \left[ 
\gamma_{\nu_{e}}^{\rm local}, \ \gamma_{\nu_{e}, {\rm max}}^{\rm local} \right], \\
\gamma_{\bar{\nu}_{e}}^{\rm local} &=& {\rm min} \left[ 
\gamma_{\bar{\nu}_{e}}^{\rm local}, \ \gamma_{\bar{\nu}_{e}, {\rm max}}^{\rm local} \right], \\
Q_{\nu_{e}}^{\rm local} &=& {\rm min} \left[
Q_{\nu_{e}}^{\rm local}, \ 
Q_{\nu_{e}}^{\rm local} (\gamma_{\nu_{e}, {\rm max}}^{\rm local}/\gamma_{\nu_{e}}^{\rm local})
\right], \\
Q_{\bar{\nu}_{e}}^{\rm local} &=& {\rm min} \left[
Q_{\bar{\nu}_{e}}^{\rm local}, \ 
Q_{\bar{\nu}_{e}}^{\rm local} (\gamma_{\bar{\nu}_{e}, {\rm max}}^{\rm local}/\gamma_{\bar{\nu}_{e}}^{\rm local})
\right].
\eeqn
These limiter procedures enable us to solve the equations in an explicit manner.

\item After the evolution, following conditions are checked,
\beqn
\mu_{p}+\mu_{e} < \mu_{n}+\mu_{\nu_{e}} , \\
\mu_{n}-\mu_{e} < \mu_{p}+\mu_{\bar{\nu}_{e}},
\eeqn
where $\mu_{p}$, $\mu_{n}$, $\mu_{e}$, $\mu_{\nu_{e}}$, and 
$\mu_{\bar{\nu}_e}$ are the chemical potentials of protons, neutrons, electrons, 
electron neutrinos, and electron anti-neutrinos, respectively. 
If both conditions are satisfied, the values of
the lepton fractions at the timestep $n+1$ are reset to be those in 
the $\beta$-equilibrium value;  
$Y_{e}^{\beta}$, $Y_{\nu_{e}}^{\beta}$, and $Y_{\bar{\nu}_{e}}^{\beta}$.

\end{enumerate}

\subsection{Microphysics} \label{Sec_MicroPhys}

\subsubsection{Equation of state}\label{Sec_EOS}

We employ two versions of Shen's EOS. One is a purely 
nucleonic EOS~\cite{Shen1998}, which is adopted both in stellar core 
collapse simulations (see \S~\ref{Sec_ResultSCC}) and in BNS merger
simulations (see \S~\ref{Sec_ResultBNS}).
In the BNS merger simulation, we also adopt
an EOS in which effects of $\Lambda$ hyperons are taken into account~\cite{Shen2011} 
(hereafter, referred to as Hyp-EOS). 
These EOS are tabulated in terms of the
rest-mass density ($\rho$), temperature ($T$), and $Y_e$ or $Y_{l}$.

Shen-EOS, derived from a relativistic mean-field theory~\cite{Walecka},
is a stiff one among many other EOS, giving a large maximum gravitational
mass of zero-temperature spherical neutron stars $M_{\rm max} \approx
2.2M_{\odot}$. By contrast, $M_{\rm max} \approx 1.8M_{\odot}$ for
Hyp-EOS because the appearance of hyperons softens the EOS. The
latest discovery of a high-mass neutron star with mass $1.97 \pm
0.04M_{\odot}$~\cite{twosolar} suggests that stiff EOS are favored,
and Shen-EOS satisfies this requirement whereas Hyp-EOS does
not. However, we consider that Hyp-EOS is a viable candidate for the
neutron-star EOS except for a very high density that 
an only high-mass neutron star of $M\lesssim M_{\rm max}$ has. 
We note that the neutron star in the BNS and a hypermassive neutron star (HMNS)
studied here do not have the extremely 
high density (except for the HMNS just before the collapse to a BH).

The thermodynamical quantities of a dense matter
at various sets of $(\rho, Y_{p}, T)$ are
calculated to construct the numerical data table for simulations.
Here $Y_{p}$ is the total proton fraction per baryon number. 
The original table covers a range of density
$10^{5.1}$--$10^{15.4}$ g/cm$^{3}$, proton fraction $0.0$--$0.56$,
and temperature $0$--$100$ MeV, which are sufficient parameter ranges 
for supernova simulations. 
The original table has been extended to a higher density 
($10^{5.1}$--$10^{17}$ g/cm$^{3}$ )~\cite{Sumiyoshi07,Shen2011} and 
a higher temperature ($0$--$400$ MeV)~\cite{Nakazato08,Shen2011}, 
for following the BH formation.

It should be noted that the causality is guaranteed to be satisfied in
this framework, whereas the sound velocity
sometimes exceeds the speed of the light in the non-relativistic
framework, e.g., in the EOS by Lattimer and Swesty~\cite{LS}. 
This is one of the benefits of the relativistic EOS.

To consistently calculate the pressure and the internal energy of 
electrons and positrons, the charge neutrality condition $Y_{p} = Y_{e}$ has to 
be solved to determine the electron chemical potential
$\mu_{e}$ for a given set of $\rho$ and $T$ in the EOS table.
Namely, it is required to solve the equation
\beq 
n_{e}(\mu_{e},T) \equiv n_{-} - n_{+} = \frac{\rho Y_{e}}{m_{u}}
\label{n_to_mu}
\eeq
in terms of $\mu_{e}$ for given values of $\rho$, 
$T$, and $Y_{e}\ (= Y_{p})$.
Here, $m_{u} = 931.49432$ MeV is the atomic mass unit, 
and $n_{-}$ and $n_{+}$ are the total number densities
(i.e., including electron-positron pairs) of
electrons and positrons, respectively. 
Then, assuming that electrons and positrons obey the Fermi-Dirac distribution, 
the number density, the pressure, and the internal
energy density of electrons and positrons are calculated~\cite{Cox68}.

The pressure and the specific internal energy density of photons are
given by
\beqn
P_{ph} = \frac{a_{r}T^{4}}{3},\ \ 
\varepsilon_{ph} = \frac{a_{r}T^{4}}{\rho},
\eeqn
where $a_{r}=(\pi^{2}k_{B}^{4})/(15c^{3}\hbar^{3})$ is the radiation constant.

In our leakage scheme, the trapped-neutrinos are assumed to interact sufficiently 
frequently with the matter that be thermalized, and hence, they are described as ideal Fermi gases 
with the matter temperature. From the numerically evolved trapped-neutrino 
fractions $Y_{(\nu)}$ 
($\nu_{e}$, $\bar{\nu}_{e}$, and $\nu_{x}$ are abbreviated by $(\nu)$), 
the chemical potentials of the trapped-neutrinos ($\mu_{(\nu)}$) are calculated by solving
\beq
Y_{(\nu)} = \frac{m_{u}}{\rho}n_{(\nu)}(\mu_{(\nu)}, T),
\eeq
where $n_{(\nu)}$ is the number density of the trapped-neutrinos.
Then the pressure and the internal energy of the trapped-neutrinos are 
calculated in the same manner as for electrons, 
using $\mu_{(\nu)}$ and the matter temperature.

In the high-resolution shock-capturing scheme for hydrodynamics,
we in general need to evaluate the sound velocity $c_{s}$,
\beq
c_{s}^{\,2} = \frac{1}{h}\left[ 
\left.\frac{\partial P}{\partial \rho}\right|_{\epsilon}
+\frac{P}{\rho}
\left.\frac{\partial P}{\partial \epsilon}\right|_{\rho}
\right]. \label{defcs}
\eeq
Here, the derivatives of the pressure are calculated by
\beqn
\left.\frac{\partial P}{\partial \rho}\right|_{\epsilon}
&=&
\sum_{i}
\left[
  \left.\frac{\partial P_{i}}{\partial \rho}\right|_{T}
  -\left.\frac{\partial P_{i}}{\partial T   }\right|_{\rho}
  \left(
  \sum_{j}
  \left.\frac{\partial \epsilon_{j}}{\partial \rho}\right|_{T}
  \right)
  \left( \sum_{k}
  \left.\frac{\partial \epsilon_{k}}{\partial T}\right|_{\rho}
  \right)^{-1}  
\right], \label{Prho} \\
\left.\frac{\partial P}{\partial \epsilon}\right|_{\rho}
&=&
 \left(\sum_{i}
 \left.\frac{\partial P_{i}}{\partial T}\right|_{\rho}
 \right)
 \left(\sum_{j}
 \left.\frac{\partial \epsilon_{j}}{\partial T}\right|_{\rho}
 \right)^{-1} , \label{Peps}
\eeqn
where the sum is taken over $B,e,ph$ and $(\nu)$.

\subsubsection{Weak-interaction and leakage rates}\label{Sec_leakagerate}

The leakage rates are phenomenologically defined by~\cite{Sekiguchi2010}
\beqn
&&\!\! Q^{\rm leak}_{\alpha} \equiv \sum_{(\nu)}Q_{(\nu)}^{\rm leak}u_{\alpha} = 
\sum_{(\nu)}\left[
(1-e^{-b\tau_{(\nu)}}) Q_{(\nu)}^{\rm diff} + e^{-b\tau_{(\nu)}} Q_{(\nu)}^{\rm local}
\right]u_{\alpha}, \label{Q_leak} \\
&&\!\! \gamma_{(\nu)}^{\rm leak}= (1-e^{-b\tau_{(\nu)}}) \gamma_{(\nu)}^{\rm diff} 
+ e^{-b\tau_{(\nu)}} \gamma_{(\nu)}^{\rm local}, \label{g_leak}
\eeqn
where $\tau_{(\nu)}$ is the optical depth of neutrinos and $b$ is a parameter
which is typically set to be $b^{-1}=2/3$\footnote{
We again used an abbreviation $(\nu)$ for $\nu_{e}$, $\bar{\nu}_{e}$, and $\nu_{x}$}.
Note that $Q^{\rm leak}_{(\nu)}$ should be regarded as the emissivity of
neutrinos measured in the {\it fluid rest frame} so that we set
$Q^{\rm leak}_{\alpha}=Q_{(\nu)}^{\rm leak}u_{\alpha}$~\cite{Shibata_etal07,Sekiguchi2010}.

The optical depth is calculated by~\cite{Ruffert,Rosswog}
\beqn
\tau_{(\nu)} &=& {\rm min}\left[ \tau^{x}_{(\nu)}, \tau^{z}_{(\nu)}, \tau^{qr}_{(\nu)} \right],\\
\tau_{(\nu)} &=& {\rm min}\left[ \tau^{x}_{(\nu)}, \tau^{y}_{(\nu)}, \tau^{z }_{(\nu)} \right],
\eeqn
for axisymmetric (see \S~\ref{Sec_ResultSCC}) and 
three-dimensional (see \S~\ref{Sec_ResultBNS}) simulations, respectively.
Here $\tau_{(\nu)}^{x}$, $\tau_{(\nu)}^{y}$, $\tau_{(\nu)}^{z}$, and $\tau_{(\nu)}^{qr}$ 
are the optical depths along $x$, $y$, $z$, and a 'quasi-radial' directions 
from each grid point, respectively. We calculate, for example, $\tau_{(\nu)}^{z}$ 
by\footnote{$\tau_{(\nu)}^{\varpi}$ and $\tau_{(\nu)}^{r}$ are calculated in a similar manner.}.
\beqn
\tau_{(\nu)}^{z}(\varpi,z) &=& 
E_{(\nu)}(\varpi,z)^{2} \tilde{\tau}^{z}(\varpi,z), \\
\tilde{\tau}^{z}(\varpi,z) &=& \int_{z}^{z_{\rm out}}\tilde{\kappa}(\varpi, z')dz',
\eeqn 
where $z_{\rm out}$ denotes the outer boundary in the $z$-direction.
$\tilde{\kappa}\ (= \kappa_{(\nu)}/E_{(\nu)}^{2})$ is an 'opacity' in which 
the neutrino-energy dependence is factored out (see Appendix C).

The neutrino energy is determined by 
\beq
E_{(\nu)} = (1-e^{-\tau_{(\nu)}/c})E^{\rm diff}_{(\nu)} + e^{-\tau_{(\nu)}/c}E^{\rm local}_{(\nu)},
\eeq
where we set the parameter as $c=5$, which implies that it takes about three collisions
to thermalize a neutrinos~\cite{Cooperstein88}. Note that $\tau_{(\nu)}$ depends on the
neutrino energy $E_{(\nu)}$ and we solve this equation by the Newton-Raphson method.
$E^{\rm diff}_{(\nu)}$ and $E^{\rm local}_{(\nu)}$ are 
the average (thermalized) diffusion and local-production energy,
which are given respectively by,
\beqn
E^{\rm diff}_{(\nu)} &=& k_{B}T\frac{F_{3}(\mu_{(\nu)}/k_{B}T)}{F_{3}(\mu_{(\nu)}/k_{B}T)},\\
E^{\rm local}_{(\nu)} &=& \frac{m_{u}}{\rho} \frac{Q^{\rm local}_{(\nu)}}{\gamma^{\rm local}_{(\nu)}},
\eeqn
where $F_{k}(x)$ is the Fermi-Dirac integral.

As the local production reactions of neutrinos, we consider
the electron and positron captures ($\gamma_{\nu_{e}}^{\rm ec}$ and
$\gamma_{\bar{\nu}_{e}}^{\rm pc}$)~\cite{Fuller85},
the electron-positron pair annihilation
($\gamma_{\nu_{e} \bar{\nu}_{e}}^{\rm pair}$ for electron-type neutrinos
and $\gamma_{\nu_{x} \bar{\nu}_{x}}^{\rm pair}$ for other types)~\cite{Cooperstein86},
the plasmon decays
($\gamma_{\nu_{e} \bar{\nu}_{e}}^{\rm plas}$ and
$\gamma_{\nu_{x} \bar{\nu}_{x}}^{\rm plas}$)~\cite{Ruffert},
and the Bremsstrahlung processes
($\gamma_{\nu_{e} \bar{\nu}_{e}}^{\rm Brems}$ and
$\gamma_{\nu_{x} \bar{\nu}_{x}}^{\rm Brems}$)~\cite{Burrows06}.
Then, the local reaction rates for the neutrino fractions are
\beqn
&& 
\gamma_{\nu_{e}}^{\rm local} = \gamma_{\nu_{e}}^{\rm ec} + 
\gamma_{\nu_{e} \bar{\nu}_{e}}^{\rm pair} + 
\gamma_{\nu_{e} \bar{\nu}_{e}}^{\rm plas} +
\gamma_{\nu_{e} \bar{\nu}_{e}}^{\rm Brems}, \label{gnlocal}\\
&& 
\gamma_{\bar{\nu}_{e}}^{\rm local} = \gamma_{\bar{\nu}_{e}}^{\rm pc} + 
\gamma_{\nu_{e} \bar{\nu}_{e}}^{\rm pair} + 
\gamma_{\nu_{e} \bar{\nu}_{e}}^{\rm plas} +
\gamma_{\nu_{e} \bar{\nu}_{e}}^{\rm Brems}, \label{galocal}\\
&& 
\gamma_{\nu_{x}}^{\rm local} = 
4\,(\gamma_{\nu_{x} \bar{\nu}_{x}}^{\rm pair} + 
    \gamma_{\nu_{x} \bar{\nu}_{x}}^{\rm plas} +
    \gamma_{\nu_{x} \bar{\nu}_{x}}^{\rm Brems}) . \label{gxlocal}
\eeqn
Similarly, the local neutrino energy emission rate $Q_{(\nu)}^{\rm local}$ is given
by
\beqn
Q_{\nu_{e}}^{\rm local} &=& Q_{\nu_{e}}^{\rm ec} 
     + (Q_{\nu_{e} \bar{\nu}_{e}}^{\rm pair} + 
        Q_{\nu_{e} \bar{\nu}_{e}}^{\rm plas} +
        Q_{\nu_{e} \bar{\nu}_{e}}^{\rm Brems})\ , \label{Qnlocal} \\
Q_{\bar{\nu}_{e}}^{\rm local} &=& Q_{\bar{\nu}_{e}}^{\rm pc} 
     + (Q_{\nu_{e} \bar{\nu}_{e}}^{\rm pair} + 
        Q_{\nu_{e} \bar{\nu}_{e}}^{\rm plas} +
        Q_{\nu_{e} \bar{\nu}_{e}}^{\rm Brems})\ , \label{Qalocal} \\
Q_{\nu_{x}}^{\rm local} &=& 
    4\,(Q_{\nu_{x} \bar{\nu}_{x}}^{\rm pair} + 
        Q_{\nu_{x} \bar{\nu}_{x}}^{\rm plas} +
        Q_{\nu_{x} \bar{\nu}_{x}}^{\rm Brems})\ . \label{Qxlocal}
\eeqn
The explicit forms of the local rates in
Eqs. (\ref{gnlocal})--(\ref{Qxlocal}) are summarized in 
Appendices A and B (see also, Ref.~\citen{Sekiguchi2010}).

We follow the recent work by Rosswog and Liebend{\"o}rfer~\cite{Rosswog} for
the diffusive neutrino emission rates $\gamma_{(\nu)}^{\rm diff}$ and 
$Q_{(\nu)}^{\rm diff}$ in Eqs. (\ref{Q_leak}) and (\ref{g_leak}).
The explicit forms of $\gamma_{(\nu)}^{\rm diff}$ and $Q_{(\nu)}^{\rm diff}$
are described in Appendix C (see also, Ref.~\citen{Sekiguchi2010}).

\subsection{Recover of ($\rho$, $Y_{e}$/$Y_{l}$, $T$)} \label{Reconst}

The quantities numerically evolved in the hydrodynamic equations
are the conserved quantities: 
$\sqrt{\gamma}\rho w$, $\sqrt{\gamma}\rho w Y_{L}$, $\sqrt{\gamma} j_{i}$, and 
$\sqrt{\gamma}e_{h}$.
The argument variables,
($\rho$, ($Y_{e}$ or $Y_{l}$), $T$), of the EOS table, 
together with $w = \alpha u^{t} = \sqrt{1+\gamma^{ij}u_{i}u_{j}}$, 
should be calculated from the conserved
quantities at each timestep. 
Note that $\sqrt{\gamma}$ is readily given by numerical evolution of
Einstein's equations. Also, the lepton fractions ($Y_{L}$) 
can be calculated directly from the conserved quantities.

\subsubsection{Non-$\beta$-equilibrium case}
In the case that the $\beta$-equilibrium condition is not satisfied,
the argument quantities ($\rho$, $Y_{e}$, $T$) can be reconstructed
from the conserved quantities in the following straightforward manner.

\vspace{2mm}
\begin{enumerate}
\item Give a trial value of $w$, referred to as $\tilde{w}$. 
Then, one obtains a trial value of the rest mass density
$\tilde{\rho}$.

\item A trial value of the temperature, $\tilde{T}$, can be
obtained from the numerically evolved value of $e_{h}$,
by solving the following equation:
\beq
e_{h} - \sum_{(\nu)}e_{h,(\nu)}(\tilde{\rho},Y_{(\nu)} \tilde{T}) 
= e_{h,{\rm EOS}}(\tilde{\rho}, Y_{e}, \tilde{T}).
\eeq
Here, $e_{h,{\rm EOS}}(\tilde{\rho}, Y_{e}, \tilde{T})
\equiv \tilde{\rho}\tilde{w}^{2}h_{\rm EOS}(\tilde{\rho},Y_{e},\tilde{T})
- P(\tilde{\rho}, Y_{e}, \tilde{T})$ should be evaluated from the EOS table
which does not include the contributions of trapped-neutrinos.
In the left-hand-side, $e_{h,(\nu)}$ is the trapped-neutrino part.
Note that one dimensional search over the EOS table is
required to obtain $\tilde{T}$.

\item The next trial value of $w$ is given by
\beq
\tilde{w} =
\sqrt{1 + \gamma^{kl}
\left(\frac{j_{k}}{\tilde{\rho}\tilde{w}h_{\rm EOS}}\right)
\left(\frac{j_{l}}{\tilde{\rho}\tilde{w}h_{\rm EOS}}\right)}.
\eeq

\item Repeat the procedures (1)--(3) until a required degree of convergence
  is achieved. Convergent solutions of the temperature and $w$ are
  obtained typically within 10 iterations.
\end{enumerate}
\vspace{2mm}

\subsubsection{The $\beta$-equilibrium case}
In the case that the $\beta$-equilibrium condition is satisfied,
on the other hand,
we may reconstruct the argument quantities ($\rho, Y_{e}, T$) from 
the conserved quantities and $Y_{l}$, under the assumption of the 
$\beta$-equilibrium. 
In this case, two-dimensional recover
$(Y_{l}, e_{h}) \  \Longrightarrow \ (Y_{e}, T)$
would be required for a given value of $\tilde{w}$. 
In this case, there may be more than one combination 
of ($Y_{e}$, $T$) which gives the same values of $Y_{l}$ and $e_{h}$.
Therefore, we have to adopt a different method 
to recover ($\rho, Y_{e}, T$).
Under the assumption of the $\beta$-equilibrium, 
the electron fraction is related to the total lepton fraction: 
$Y_{e} = Y_{e}(\rho, Y_{l}, T)$. Using this relation, the EOS table
can be rewritten in terms of the argument variables of 
($\rho$, $Y_{l}$, $T$). 
Then, the similar strategy as in the non-$\beta$-equilibrium
case can be adopted. Namely,

\vspace{2mm}
\begin{enumerate}
\item Give a trial value $\tilde{w}$. 
Then one obtains a trial value of the rest mass density.
\item A trial value of the temperature can be
obtained by solving 
\beq
e_{h} = 
e^{\beta}_{\rm EOS}(\tilde{\rho}, Y_{l}, \tilde{T})
\eeq
with one dimensional search over the EOS table.
Here $e^{\beta}_{\rm EOS}$ should be evaluated from the $\beta$-equilibrium 
EOS table, which contains the trapped-neutrino contributions.

\item The next trial value of $w$ is given in the same way.

\item Repeat the procedures (1)--(3) until a required degree of convergence
  is achieved. The electron fraction is given as 
  $Y_{e} = Y_{e}(\rho, Y_{l}, T)$ in the $\beta$-equilibrium EOS table.
\end{enumerate}
\vspace{2mm}

In the case of a simplified or analytic EOS, 
the Newton-Raphson method may be applied to recover the primitive
variables. In the case of a tabulated EOS, by contrast,
the Newton-Raphson method may not be a good approach because it requires
derivatives of thermodynamical quantities which in general cannot be 
calculated precisely from a tabulated EOS by the finite differentiating 
method.

\section{Gravitational collapse of massive stellar core}\label{Sec_ResultSCC}

The observational associations (for a review, see Ref.~\citen{WoosleyB06}) 
between LGRBs and supernovae has provided the strong support to a scenario, 
so-called collapsar model, in which 
LGRBs are assumed to be driven in the collapse of a massive stellar 
core to a BH~\cite{Woosley93,MacFadyen99}.
In the collapsar model, a central core of a massive star
is required to be rotating rapidly enough that 
a massive accretion disk can be formed around a BH.

Because the observed supernovae associated with LGRBs are Type Ib/c and the 
relativistic jets have to reach the stellar surface~\cite{Zhang04},
the progenitors should have lost their hydrogen (and helium) envelopes before 
the onset of the stellar core collapse; 
otherwise a peculiar evolution path is required. 
Due to these reasons, the progenitors of LGRBs are now believed to be 
rotating massive Wolf-Rayet (WR) stars.
However, ordinary WR stars are known to be accompanied by strong stellar
winds driven by the radiation pressure which cause a rapid spin-down of
the stellar core.
Here, a serious problem concerning the collapsar model is that according to stellar
evolution calculations, it is very difficult to produce pre-collapse cores
which satisfy both the requirement of the collapsar model and the association of 
Type Ib/c supernova, if magnetic torques and standard mass-loss rates are 
taken into account~\cite{WoosleyH06}.

To resolve the above dilemma, several models have been proposed 
(see Ref.~\citen{Fryer07} for a review).
All of the proposed progenitor models of LGRBs are anomalous in the sense
that they are different from the progenitors of ordinary supernovae 
(see Ref.~\citen{Sekiguchi2011} for a discussion).
Qualitatively speaking, LGRB progenitor cores may be modeled by a rapidly rotating,
higher-entropy core, regardless of their formation processes.
Based on this assumption, we performed simulations of 
a massive stellar core with higher values of entropy collapsing to a BH.
In this section, we report our latest results of fully general 
relativistic simulations for the collapse of a rotating, higher-entropy cores, 
performed taking into account detailed microphysics.

\subsection{Initial models and grid setting}

As a representative model of a high entropy core, we adopt a presupernova core 
of $100M_{\odot}$ model calculated by Umeda and Nomoto~\cite{UN2008} 
(hereafter denoted by UN100).
The model has an iron core of a large mass $M_{\rm core} \approx 3.2M_{\odot}$ and radius 
$R_{\rm core} \approx 2500$ km with the central density and temperature of 
$\rho_{c} \approx 10^{9.5}$ g/cm$^{3}$ and $T_{c} \approx 10^{10}$ K. 
The central value of entropy per baryon is $s\approx 4k_{B}$, which is much larger than
that of an ordinary presupernova core for which $s\lesssim 1k_{B}$.

Because the model UN100 is non-rotating, we add rotational profiles according to~\cite{{O'Connor11}}
\beq
\Omega(\varpi) = \Omega_{0} \frac{R_{0}^{2}}{R_{0}^{2}+{\varpi^{2}}} {\cal F}_{\rm cut},
\eeq
where $\varpi = \sqrt{x^{2} + y^{2}}$,
$\Omega_{0}$, and $R_{0}$ are parameters which control 
the magnitude and degree of differential rotation. 
The cut-off factor ${\cal F}_{\rm cut}$ is
introduced by a practical reason for the numerical simulation:
If the specific angular momentum in the outer region of the core is too large,
the matter escapes from the computational domain.
To avoid this, the rotational velocity has to be suppressed in the outer region.

We fix the central angular velocity as $\Omega_{0}=1.2$ rad/s and consider 
two values of $R_{0}$; a rigid rotation model ($R_{0}=\infty$, referred to as UN100-rigid) and 
a differential rotation ($R_{0}=R_{\rm core}$, referred to as UN100-diff) model.
We note that the imposed rotation is moderately large (not rapid) because
$\Omega_{0}$ is much smaller than the Kepler value 
$(M_{\rm core}/R_{\rm core}^{3})^{1/2}\approx 5.2$ rad/s.

We assume axial and equatorial symmetries of the spacetime and 
the so-called Cartoon method~\cite{ShibataCa,AlcubierreCa} 
is adopted for integrating Einstein's equations.
In the current implementation, we use a fourth order Lagrange interpolation
scheme, which is necessary in the Cartoon method.

\begin{table}[t]
 \begin{center}
  \begin{tabular}{c|ccccc} \hline
     & $\Phi_{c} \le 0.02  $ & $  \le \Phi_{c} \le 0.044 $ & $  \le \Phi_{c} \le 0.09 $ 
     & $\le \Phi_{c} \le 0.2 $ & $\Phi_{c} \ge 0.2 $ \\
     \hline
 $\Delta x_{0}$ (km) & 4.0    & 2.0    & 1.0    & 0.5    & 0.25    \\
 $\delta$            & 0.0075 & 0.007  & 0.0065 & 0.006  & 0.0055  \\
 $N$                 & 332    & 428    & 542    & 668    & 812     \\
 $L$ (km)            & 5840   & 5370   & 5000   & 4450   & 3860    \\ \hline
 $\Delta x_{0}$ (km) &  5.5   & 2.85   & 1.45   & 0.7    & 0.35    \\
 $\delta$            & 0.0075 & 0.007  & 0.0065 & 0.006  & 0.0055  \\
 $N$                 & 294    & 380    & 484    & 610    & 752     \\
 $L$ (km)            & 5860   & 5360   & 4980   & 4370   & 3870    \\ \hline
  \end{tabular}
 \end{center}
\caption{Summary of the regridding procedure. The values of the minimum
 grid spacing $\Delta x_{0}$ (in units of km), 
 the non-uniform-grid factor $\delta$, and
 the grid number $N$ for each range of $\Phi_{c} = 1 -\alpha_c$ are
 listed for the finer and the coarser (lower table) resolutions.}\label{regrid}
\end{table}

In numerical simulations, we adopt a nonuniform grid, in which 
the grid spacing is increased according to the rule
\beq
d x_{j+1} = (1 + \delta) d x_{j}, \ \ \ \ d z_{l+1} = (1 + \delta) d z_{l},
\eeq
where $d x_{j} \equiv x_{j+1} - x_{j}$, $d z_{l} \equiv z_{l+1} - z_{l}$, 
and $\delta$ is a constant.
In addition, a regridding technique~\cite{Shibata02,Sekiguchi0507} 
is adopted to assign a sufficiently large number of grid points inside the
collapsing core, saving the CPU time efficiently.
The regridding is carried out whenever the characteristic radius
of the collapsing core, defined by~\cite{Shibata02} 
$\Phi_c \equiv 1 -\alpha_c~ (\Phi_c>0)$ 
where $\alpha_c$ is the central value of the lapse function,
decreases by a factor of $\sim 2$, and 
we set an infalling boundary condition at the outer boundary.

All the quantities on the new grid are calculated
using a fifth-order Lagrange interpolation. 
However, for the fluid quantities such as $\rho$ and $h$,
the fifth-order interpolation could fail because the
interpolation may give negative values of $\rho$ and $h-1$.
In case we have $\rho<0$ or $h<1$, we adopt the linear interpolation
to calculate the quantities on the new grid, based on
the prescription proposed by Ref.~\citen{Yamamoto08}.
In each regridding, we solve the Hamiltonian constraint
equation numerically. 

To check the convergence of numerical results, 
simulations are performed in two different grid resolutions.
Table \ref{regrid} summarizes the regridding parameters 
($N$ and $L$ are the number of the grid points and the computational domain, respectively) of each
level of the regridding procedure for finer (upper) and coarser (lower) 
resolutions. The numerical results in both grid resolutions agree well
except for the formation time of a BH and the stochastic behavior due to
connective and turbulent motions.

\begin{figure}[p]
\begin{center}
    (a)\includegraphics[scale=1.0]{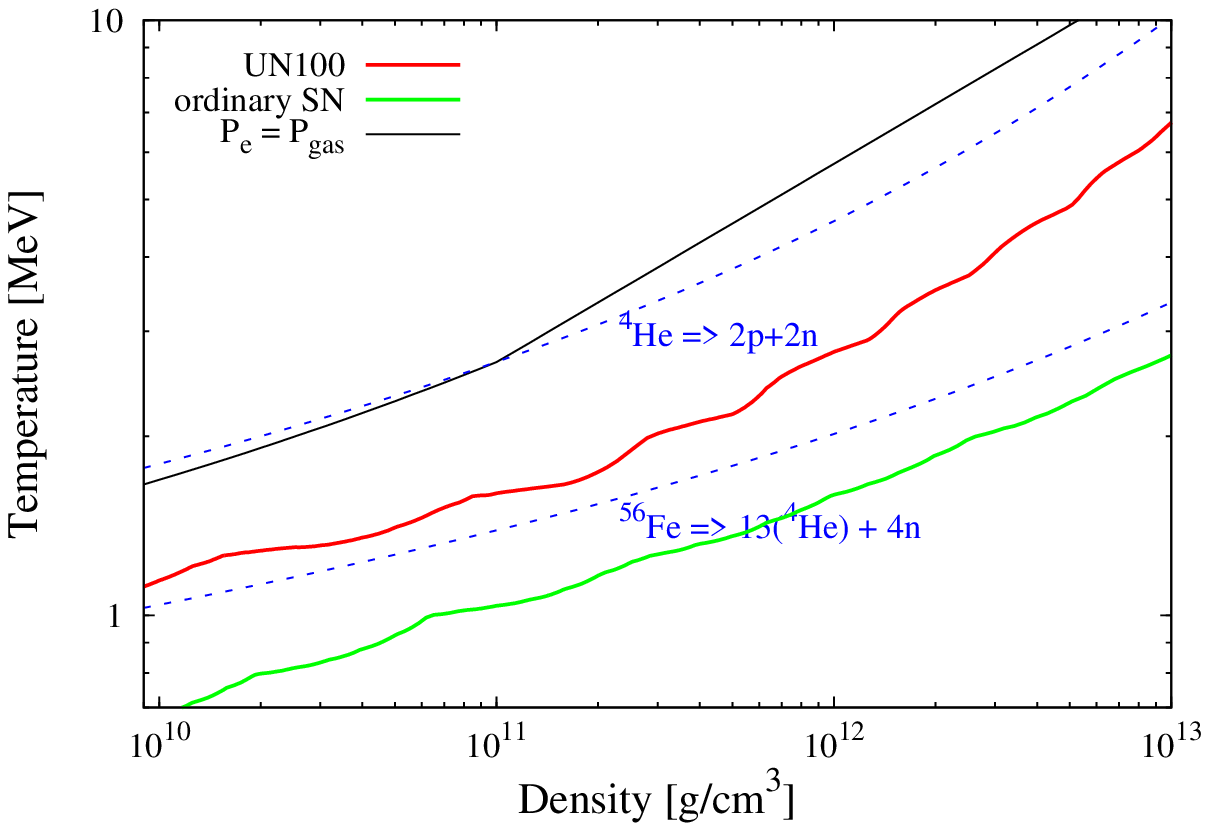}
    (b)\includegraphics[scale=1.0]{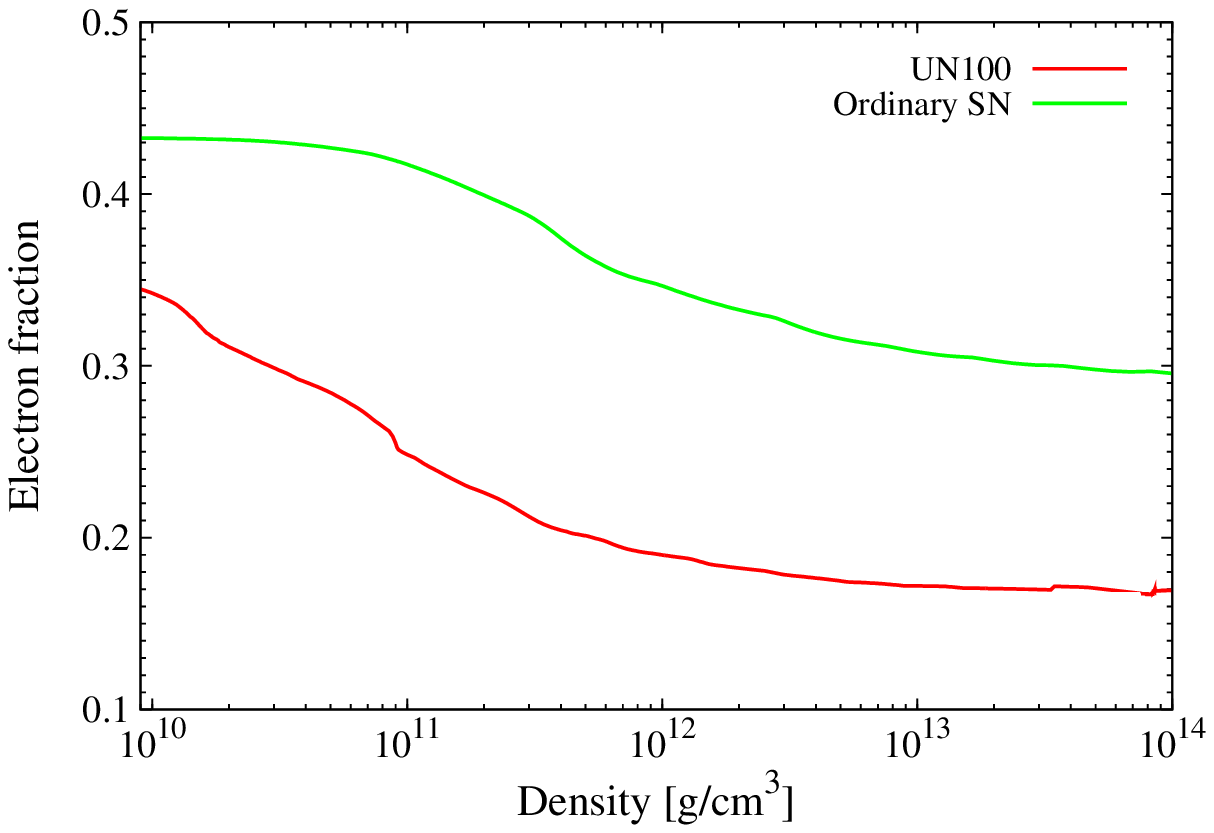}
\end{center}
  \caption{(a) Evolution paths of the central values of the rest-mass density and the temperature 
           in the $\rho$-$T$ plane. The red curve shows the evolution path
           of UN100-rigid.
           The black solid curve shows the boundary at which the condition 
           $P_{e}=P_{\rm gas}$ is satisfied ($P_{e}>P_{\rm gas}$ for the higher density side).
           The two blue dashed curves denote the values of $(\rho, T)$ with which
           $^{56}$Fe or $^{4}$He will be half by mass due to the photo-dissociation.
           An evolution path for an ordinary 
           supernova core~\cite{Sekiguchi2010} is shown together 
           for comparison (solid green curve).
           (b) Evolution paths of the central values of the rest-mass density and the electron fraction 
           in the $\rho$-$Ye$ plane.
\label{fig_SCC_rho-T-Ye}}
\end{figure}

\begin{figure}[t]
\begin{center}
    \includegraphics[scale=1.0]{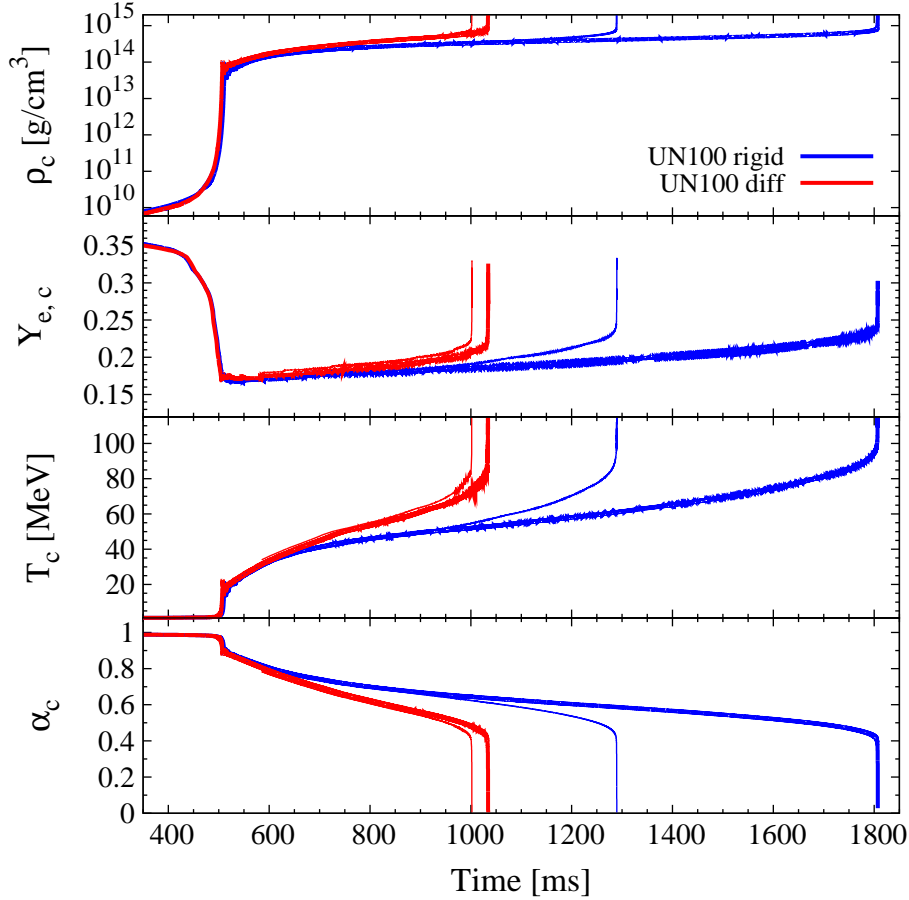}
\end{center}
  \caption{Time evolution of the central values of rest-mass density, electron fraction,
    temperature, and 
    the lapse function for UN100-rigid (blue curves) and UN100-diff (red curves).
    The results for the coarser grid resolution are shown together (thin curves).
    The collapsing core experiences a core bounce at $t\approx 510$ ms and
    collapse to a BH at $t\approx 1810$ ms.
\label{fig_SCC_central}}
\end{figure}

\begin{figure}[t]
\begin{center}
    \includegraphics[scale=1.6]{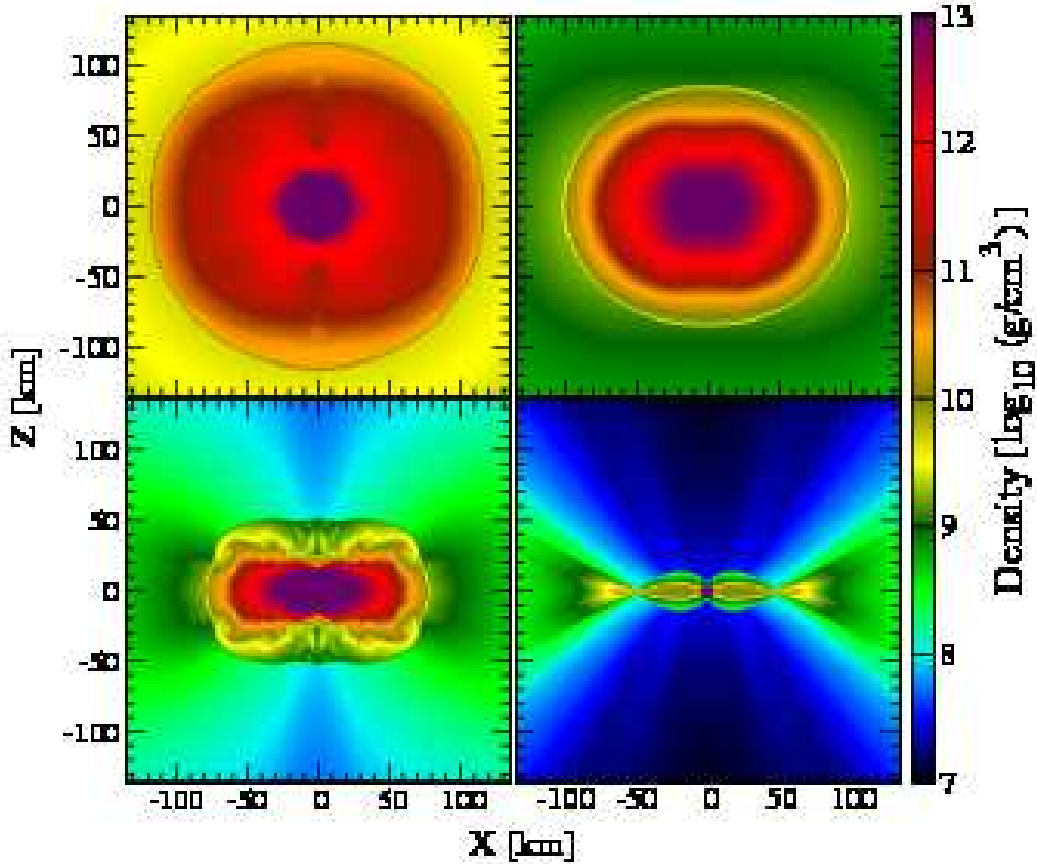}
\end{center}
  \caption{Contour profiles of the rest-mass density in the $x$-$z$ plane at
    $t=570$ ms, (top left), 645 ms (top right), 890 ms (bottom left), and 1150 ms (bottom right) 
    for UN100-diff.
\label{fig_SCC_con-diff}}
\end{figure}

\begin{figure}[p]
\begin{center}
    \includegraphics[scale=1.6]{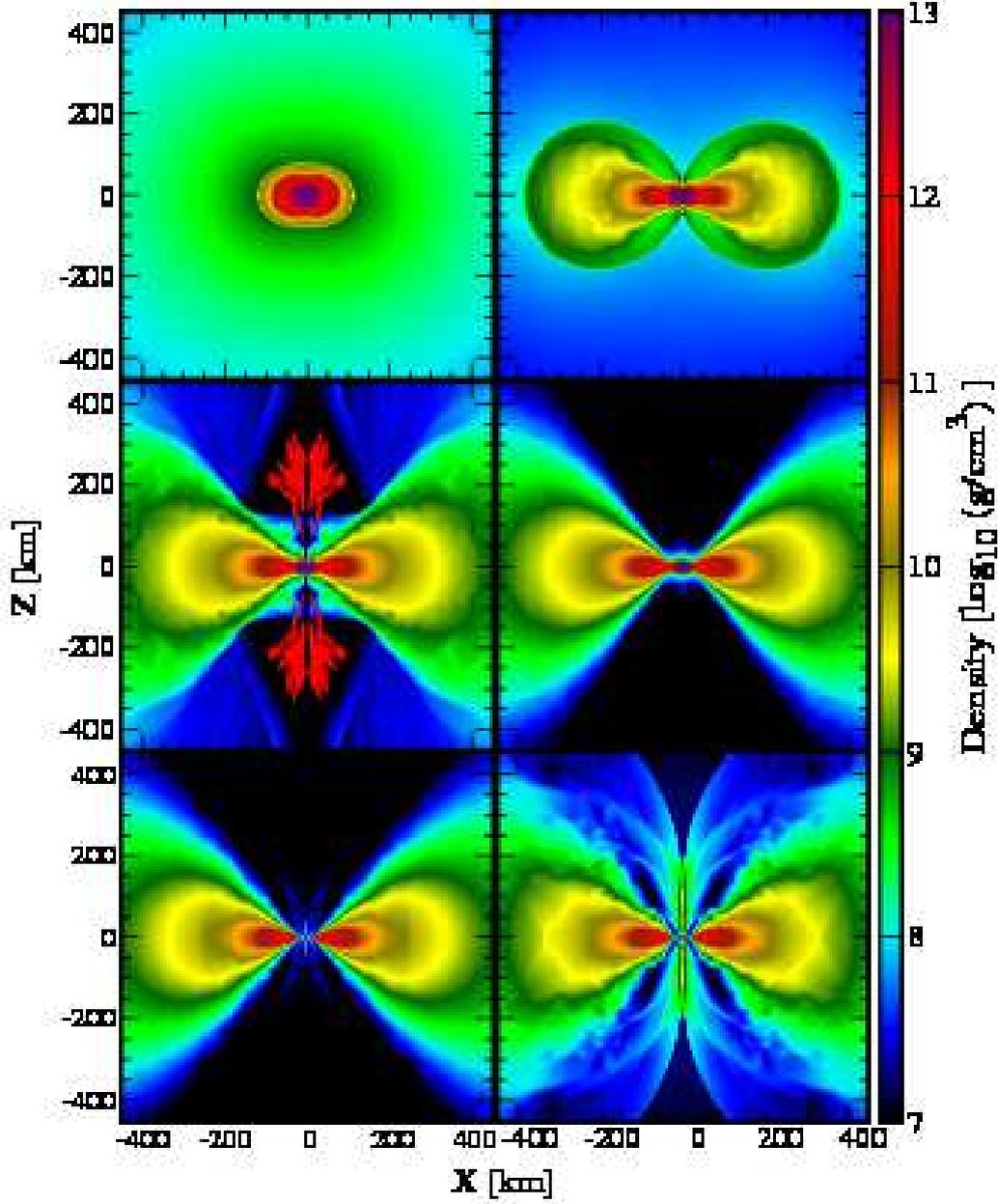}
\end{center}
  \caption{Contour profiles of the rest-mass density in the $x$-$z$ plane at
    $t=617$ ms, (top left), 863 ms (top right), 1256 ms (middle left), 1437 ms (middle right), 
    1822 ms (bottom left), and 2225 ms (bottom right) for UN100-rigid.
    In the middle left panel, outflow velocity vectors larger than $0.15c$ are plotted
    together (red arrows). 
    Note that the scale of the figure is different from that of Fig.~\ref{fig_SCC_con-diff}.
    In the bottom two panels, a BH is formed at the center.
\label{fig_SCC_con-rig-rho}}
\end{figure}

\begin{figure}[t]
\begin{center}
    \includegraphics[scale=1.6]{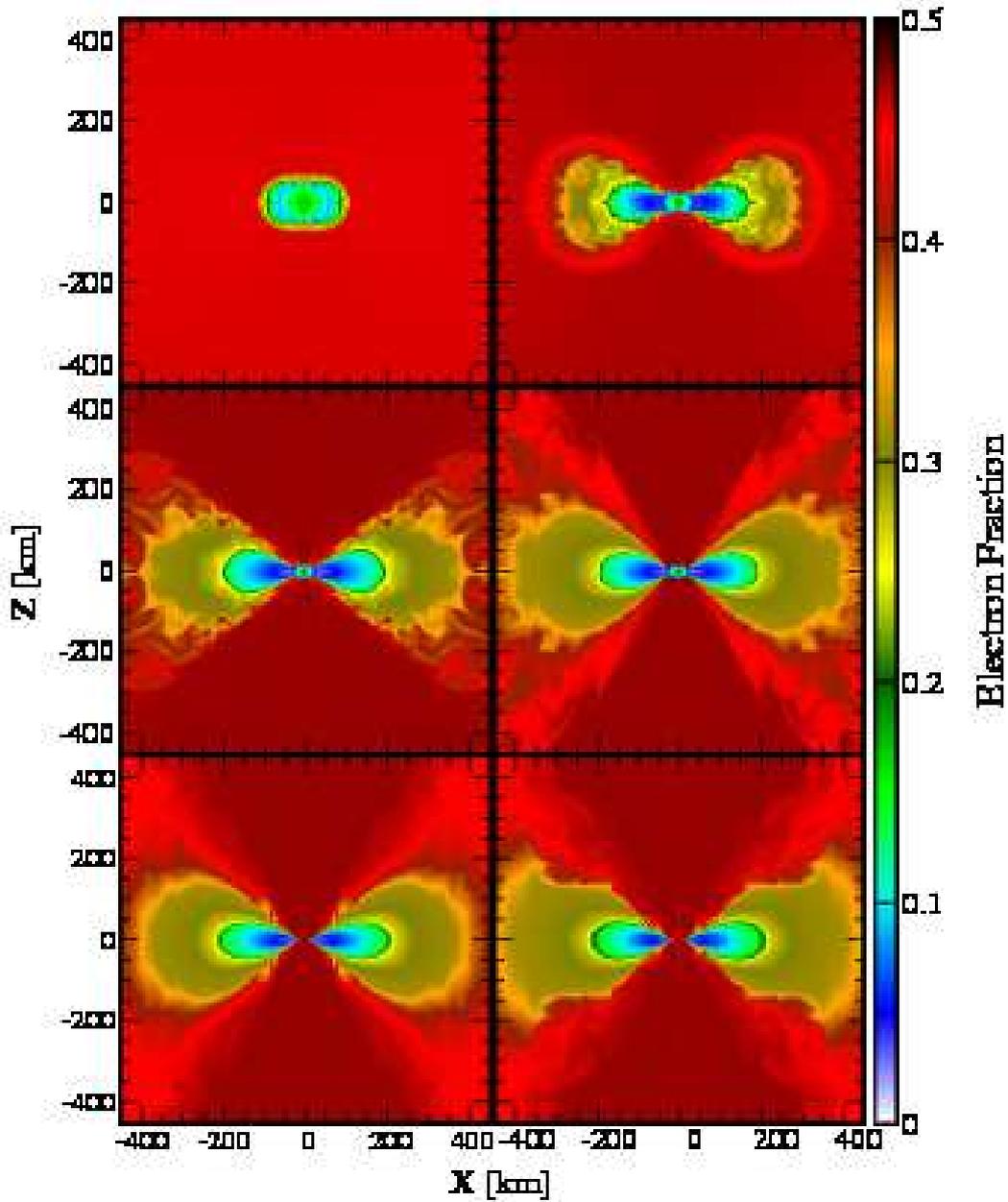}
\end{center}
  \caption{Contour profiles of the electron fraction in the $x$-$z$ plane at the same timeslices
    as Fig.~\ref{fig_SCC_con-rig-rho}.
\label{fig_SCC_con-rig-Ye}}
\end{figure}

\begin{figure}[t]
\begin{center}
    \includegraphics[scale=1.6]{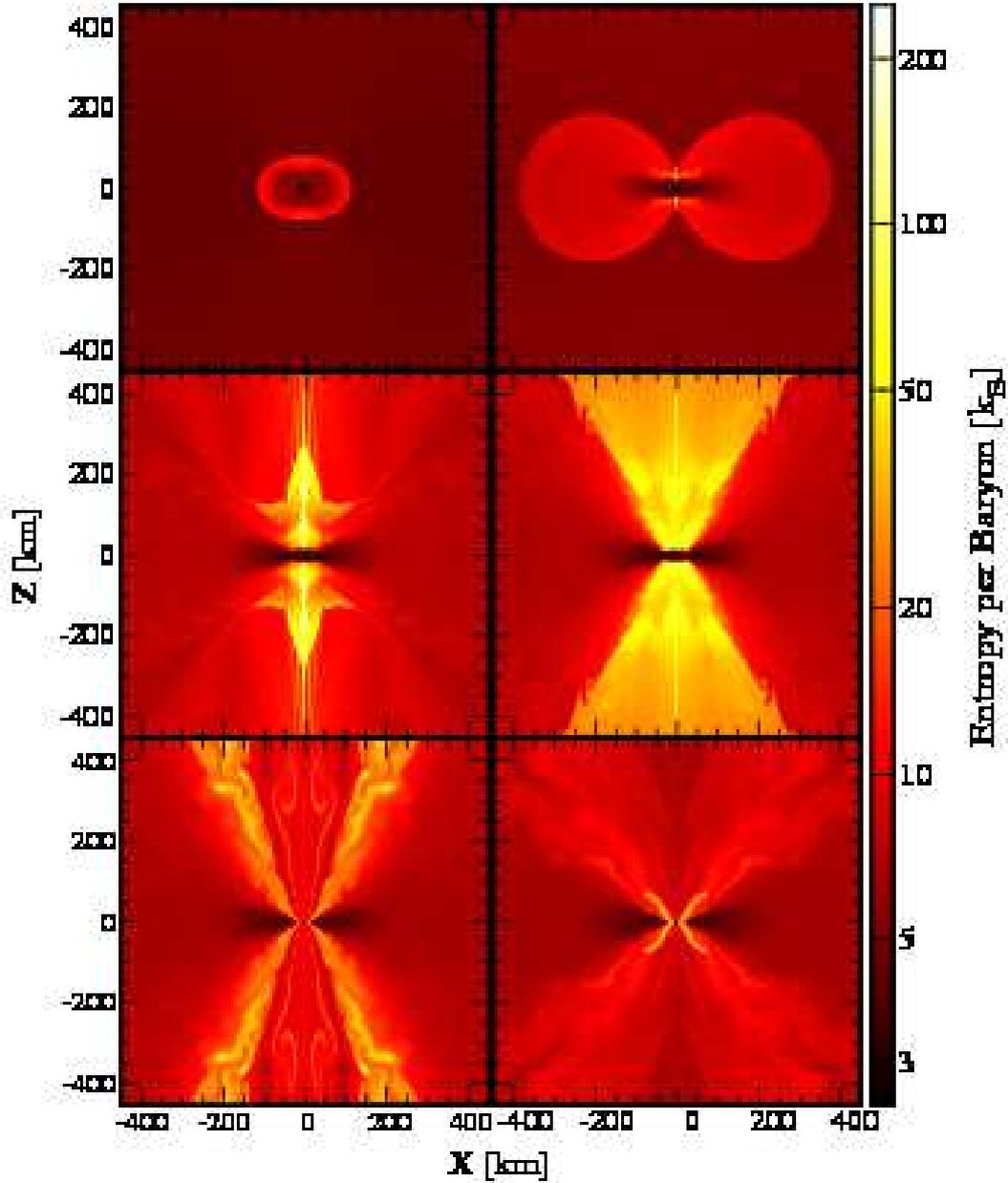}
\end{center}
  \caption{Contour profiles of the entropy per baryon in the $x$-$z$ plane at the same timeslices
    as Fig.~\ref{fig_SCC_con-rig-rho}.
\label{fig_SCC_con-rig-S}}
\end{figure}

\begin{figure}[t]
\begin{center}
    \includegraphics[scale=1.6]{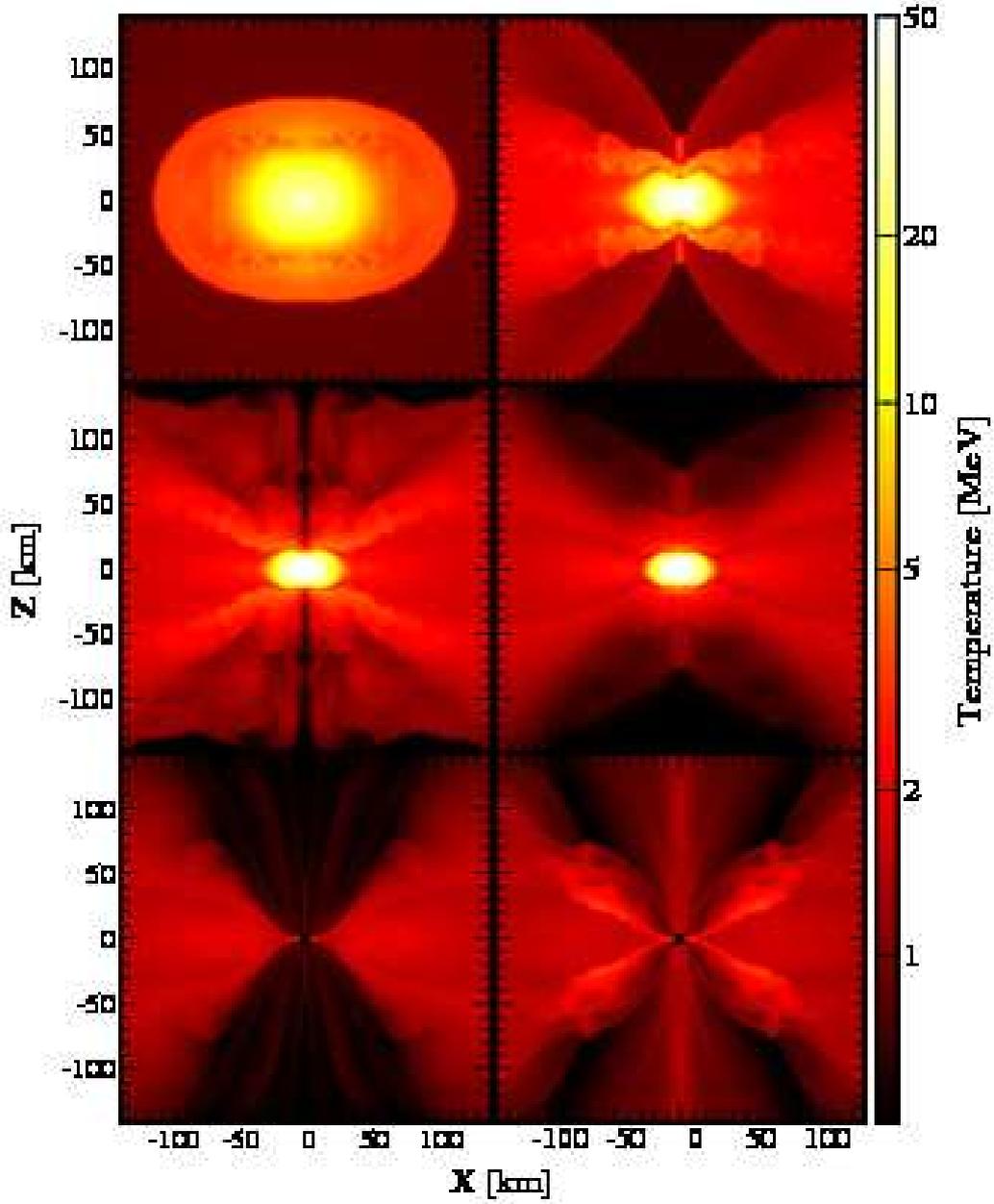}
\end{center}
  \caption{Contour profiles of the temperature in the $x$-$z$ plane at the same timeslices
    as Fig.~\ref{fig_SCC_con-rig-rho} but with a different (zooming) scale.
\label{fig_SCC_con-rig-T}}
\end{figure}

\begin{figure}[t]
\begin{center}
    \includegraphics[scale=1.6]{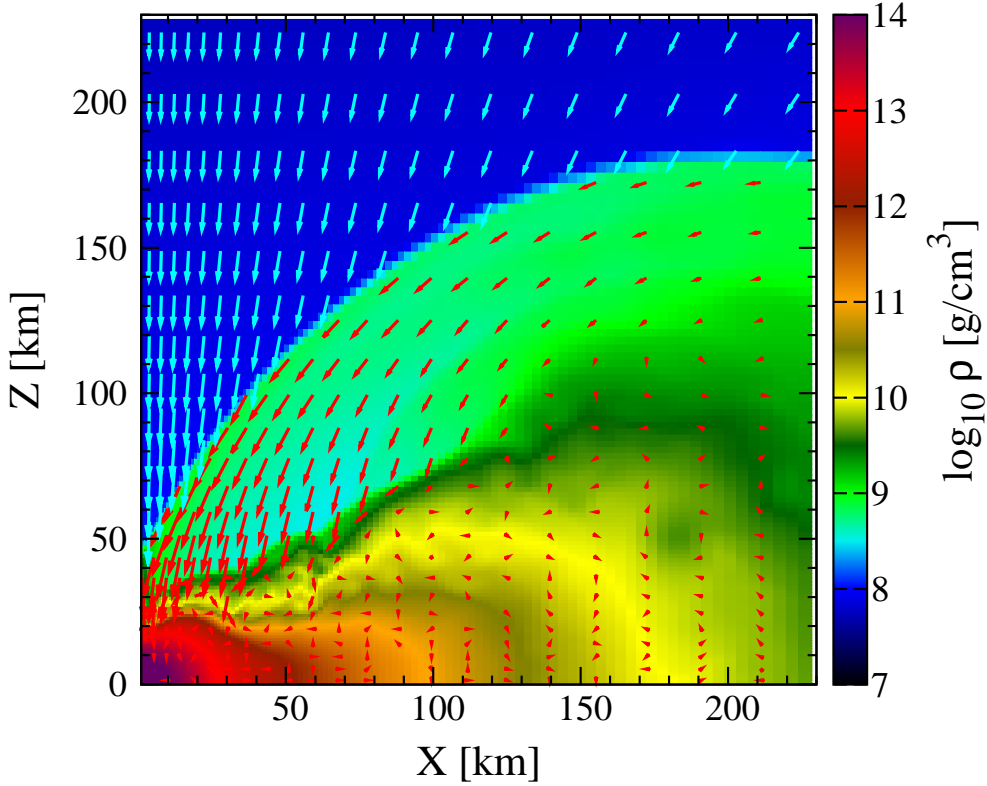}
\end{center}
  \caption{Velocity vector fields outside the shock surface (blue arrows) and inside the shock (red arrows) 
    together with a contour profile of the rest-mass density in the $x$-$z$ plane at $t=645$ ms.
    Only the velocity component perpendicular to the shock surface is dissipated at the shock.
    As a result, the infalling matter is accumulated in the central HMNS.
\label{fig_SCC_con-rig-accum}}
\end{figure}

\begin{figure}[t]
\begin{center}
    \includegraphics[scale=1.0]{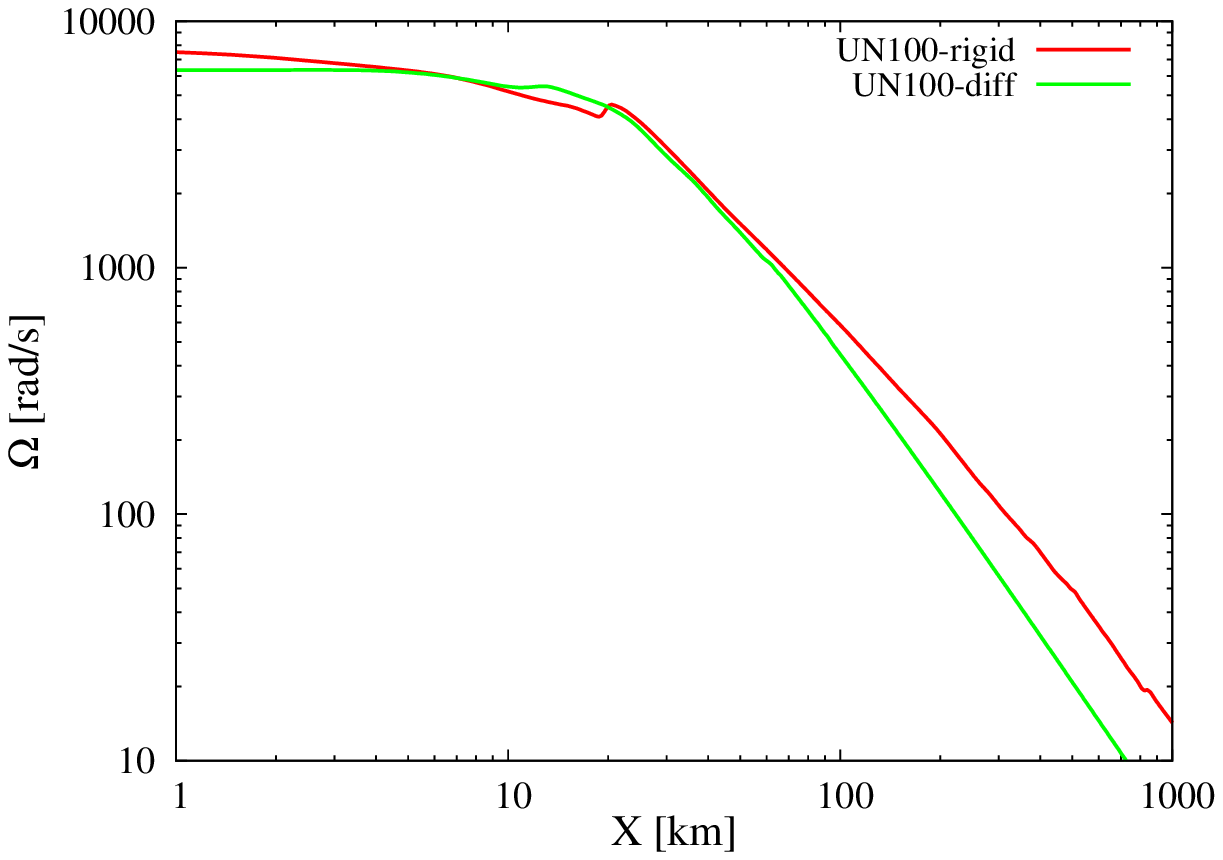}
\end{center}
  \caption{Profiles of the rotational angular velocity along the equator just before the BH formation 
    for UM100-diff (green curve) and UM100-rigid (red curve).
\label{fig_SCC_omega}}
\end{figure}

\subsection{Dynamical features}

Figure~\ref{fig_SCC_rho-T-Ye}(a) shows the evolution path (red curve) of central
values of the rest-mass density and the temperature in the $\rho$-$T$ plane for UN100-rigid.
As in the core collapse of an ordinary supernova for which the central 
value of entropy per baryon is $s/k_{B} \sim 1$,
gravitational collapse is triggered by the electron capture and 
the photo-dissociation of heavy nuclei. 
Because of the higher value of the entropy per baryon ($s/k_{B} \approx 4$),
the photo-dissociation is mainly responsible to the destabilization.
Note that a substantial amount of heavy nuclei are resolved into heliums 
by the photo-dissociation (see Fig.~\ref{fig_SCC_rho-T-Ye}):
The fraction of heavy nuclei in mass is $\approx 0.4$ and 0.2 for $\rho_{c}=10^{11}$ and 
$10^{12}$ g/cm$^{3}$, respectively.
Then the collapse in the early phase proceeds in a homologous manner.
As the collapse proceeds temperature increases, heliums
are resolved into free nucleons ($p$, $n$).

Figure~\ref{fig_SCC_rho-T-Ye}(b) shows the evolution path (red curve) of central
values of the rest-mass density and $Y_{e}$ in the $\rho$-$Y_{e}$ plane for UN100-rigid.
Because the temperature for the present models is higher than that for the ordinary supernova,
the electron capture on the free proton is enhanced due to the larger value of the free proton fraction,
and hence, the electron fraction for UN100 is by $\sim 0.1$ smaller than that for the ordinary supernova
in the collapse phase (compare the red and green curves in Fig.~\ref{fig_SCC_rho-T-Ye}(b)).

The time evolution of the central values of the rest-mass density, electron fraction,
temperature, and the lapse function for models UN100-rigid and UN100-diff 
is shown in Figure \ref{fig_SCC_central}. 
As in the collapse of the ordinary supernova core, the collapsing core 
experiences a bounce when the central density reaches the nuclear density 
$\rho_{\rm nuc}$ above which the pressure increases drastically due to the 
repulsive nuclear force, and then, shock waves are formed and launched. 
Because the electron fraction at the bounce is
small as $Y_{e}\approx 0.17$ and hence the core is neutron rich, the nuclear 
force starts playing a role at relatively low density, 
$\rho \sim 10^{14}$ g/cm$^{3}$, in Shen-EOS.
After the bounce, a HMNS, which is supported by
a significant rotation and thermal pressure, is formed.

The shock wave formed at the core bounce
propagates outward but eventually stalls at $r\approx 100$ km due the neutrino cooling and 
photodissociaion of heavy nuclei contained in the infalling matter 
(see the top panels of Figs.~\ref{fig_SCC_con-diff} and~\ref{fig_SCC_con-rig-rho}).
Then, a standing accretion shock is formed. As in the case of the ordinary supernova,
convection is activated between the HMNS and the standing shock.
However, it is not strong enough to push the standing shock outward.
The convection is stronger for the differentially rotating model, which rotates more slowly
than the rigidly rotating model.
This is likely to be due to the stabilizing effect of the epicyclic modes
which is stronger in UN100-rigid~\cite{Sekiguchi2011}. 
As the matter accretion proceeds, the central density and temperature 
increase gradually, and eventually, the HMNS collapses to a BH.
The formation time of the BH depends on the grid resolution (compare the thin and
thick curves in Fig.~\ref{fig_SCC_central}) in particular for UN100-rigid.
This is because the HMNS is close to the marginally stable configuration,
and hence, a small thermodynamical change results in a significant change in $\rho$.
In general, for a finer grid resolution, the lifetime
of HMNS increases. 
The longer lifetime for higher-resolution runs is a often-seen feature,
because the numerical dissipation is less severe for high resolution.
Note that the rotational profiles in the central region of UN100-rigid and
UN100-diff are very similar, and their evolution process agrees well with each other soon after the
core bounce. However, their evolution paths deviate as the matter in the outer region falls.

Dynamics of the system for the models UN100-diff and UN100-rigid in the later phase of the accretion onto 
the HMNS is qualitatively different. Figure~\ref{fig_SCC_con-diff} shows contour plots of
the rest-mass density in the $x$-$z$ plane at selected time slices for UN100-diff until
the HMNS collapses to a BH. 
Because of the accretion of the matter, the shock front of the standing accretion shock 
gradually recedes. Note that the shape of the shock wave is deformed by the rotation
to be spheroidal.
As we shall see below, this shows a remarkable contrast with the case of UN100-rigid
where the shock wave is deformed to be a torus-like shape.
When the shock wave stalls, negative gradients of the entropy per baryon 
and of the total-lepton (electron) fraction appear because neutrinos 
carry away both the energy and the lepton number, as in the collapse of an ordinary 
presupernova core (see the bottom left panel of Fig.~\ref{fig_SCC_con-diff}).
The HMNS finally collapses to a BH due to the mass accretion, and a geometrically thin
disk is formed around the BH. The system shows no violent time variability.

Figures \ref{fig_SCC_con-rig-rho}, \ref{fig_SCC_con-rig-Ye}, \ref{fig_SCC_con-rig-S}, and \ref{fig_SCC_con-rig-T}, respectively, show 
contour plots of the rest-mass density, the electron fraction, the entropy per baryon,
and the temperature in the $x$-$z$ plane at selected time slices for UN100-rigid. 
As in the model UN100-diff, the shock wave 
formed after the core bounce stalls at $r\sim 100$ km 
(the top left panel of Figs.~\ref{fig_SCC_con-rig-rho}--\ref{fig_SCC_con-rig-T}).
Due to the faster rotation of the outer region than in UN100-diff, 
the shock wave is deformed to be a torus-like configuration 
(the top right panel of Figs.~\ref{fig_SCC_con-rig-rho}--\ref{fig_SCC_con-rig-T}).
The formation of this torus-shaped shock is the key ingredient which characterizes the 
dynamics of UN100-rigid. 
At the shock, the kinetic energy associated with the motion perpendicular to the
shock surface is dissipated but that associated with the parallel component is preserved. 
In the model UN100-rigid, the shock front is highly deformed, and thus, the amount of the 
kinetic energy dissipated at the shock is not as large as that in UN100-diff. 
This implies that the infalling materials are eventually accumulated in the central region 
and their kinetic energy is dissipated at the surface of the HMNS. 
Figure~\ref{fig_SCC_con-rig-accum}, which displays the velocity field in the $x$-$z$ plane at a time slice,
clearly shows this mechanism.
During this process, oscillations of the HMNS are excited as the infalling matter hits it. 
Also, the shock waves gain the thermal energy via $PdV$ work and propagate outward.

Due to the accumulation of the matter onto the HMNS and the resulting shock heating,
the thermal energy is stored in the polar region of the HMNS, increasing the gas pressure, $P_{\rm gas}$,
there.
On the other hand, the ram pressure, $P_{\rm ram}$, of the infalling matter decreases with the
elapse of the time because its density decreases.
When the condition, $P_{\rm ram}<P_{\rm gas}$, is realized, 
outflows are launched from the polar surface of the HMNS, forming shocks 
(see the middle left panel of Figs.~\ref{fig_SCC_con-rig-rho}--\ref{fig_SCC_con-rig-T}).
It can be seen that the entropy around the rotational axis is significantly enhanced due to the shock 
heating associated with the outflows (e.g., see the middle right panel of Fig.~\ref{fig_SCC_con-rig-S}).

The outflows eventually lose the driving power by the neutrino cooling and matter again turns to
fall onto the polar region of the HMNS. 
Due to the continuous mass accretion, the HMNS eventually collapses to a BH surrounded by
a geometrically thick torus
(see the bottom left panel of Figs.~\ref{fig_SCC_con-rig-rho}--\ref{fig_SCC_con-rig-T}).
Note that in the present leakage scheme, neutrino heating is not taken into account and 
exploring the fate of the thermally driven outflows in the presence of 
the neutrino heating is an interesting subject. We plan to pursue this issue 
using a code based on the moment formalism~\cite{Thorne1981,SKSS2011} in the near future.

We found, as another novel feature of dynamics, that the BH-torus system shows a time variability
(see the bottom right panel of Figs.~\ref{fig_SCC_con-rig-rho}--\ref{fig_SCC_con-rig-T}).
This is reflected in the neutrino luminosities as we shall show in \S~\ref{SCC_Lum}.
Such a time variability has not been seen in UN100-diff. Reason for this will be explained as follows.

First, the infall timescale of the matter in the torus into the BH is longer for UN100-rigid 
due to the rapider rotation (in the outer region).
Also, the neutrino cooling timescale for the torus will be longer for UN100-rigid 
because of the larger optical depth due to the higher density and temperature.
Furthermore, the heating rate due to the mass accretion in the central region is larger for
UN100-rigid due to the mass accumulation mechanism.
Due to these reasons, the energy deposition by the accretion cannot be valanced by these cooling 
mechanisms: $\dot{Q}^{+}_{\rm acc} > \dot{Q}^{-}_{\rm infall} + \dot{Q}^{-}_{\nu}$. 
Then the torus will expand lowering
the optical depth which results in the enhancement of the neutrino cooling rate 
$\dot{Q}^{-}_{\nu}$.
Because of the strong dependence of neutrino opacities and cooling rate on the temperature, 
a slight change in the shock configuration may result in a huge loss of the thermal energy 
by neutrino emission. 
Here, note that the shock heated matter is partially 
supported by the pressure gradient due to the moderate (not very rapid) rotation. 
Therefore, if some materials lose their thermal energy, they will drop into the BH
like an avalanche.
The modulation of the shock configuration which triggers the above dropping appears to come from
the Kelvin-Helmholtz instability,
developed at the interface between the torus and the accumulating flows.

In the model UN100-diff, by contrast, $\dot{Q}^{+}_{\rm acc}$ is smaller due to the absence of 
the accumulation mechanism,
and $\dot{Q}^{-}_{\rm infall}$ and $\dot{Q}^{-}_{\nu}$ are larger due to the slower rotation.
As a result, the above energy balance will be satisfied without expansion of the disk, and thus,
there is no violent time variability.

It is remarkable that the above qualitative differences in dynamics between UN100-diff and UN100-rigid 
stem from a small difference in the initial angular velocity profile in the outer region.
Figure~\ref{fig_SCC_omega} compares profiles of the rotational angular velocity along the equator 
just before the BH formation for UM100-diff and UM100-rigid. The rotational profiles 
of the HMNS ($r\lesssim 40$ km) are similar in the central region, and in the outer 
region, the difference at most by a factor of 2--3. 
This result shows that the final outcome depends strongly on the rotational 
profile of progenitor stars.

\begin{figure}[p]
\begin{center}
    (a)\includegraphics[scale=1.00]{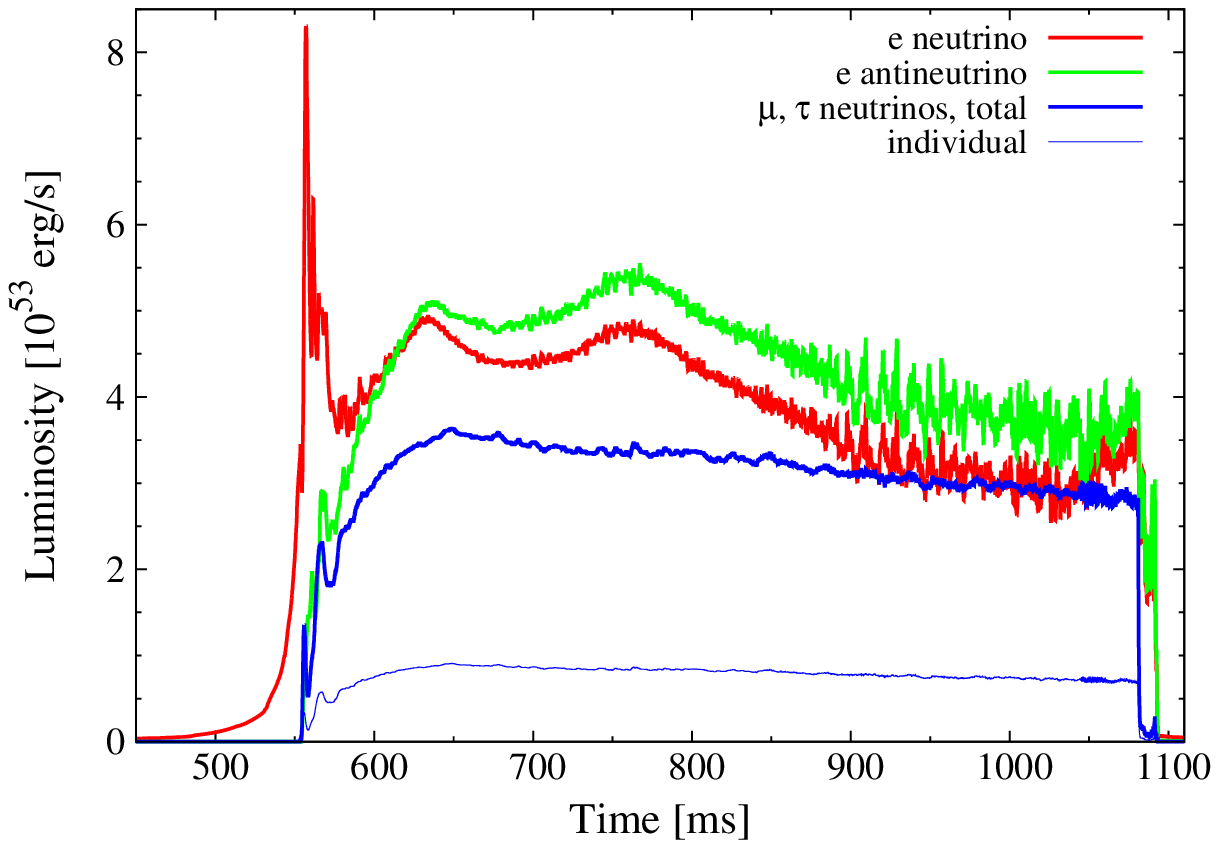}
    (b)\includegraphics[scale=1.00]{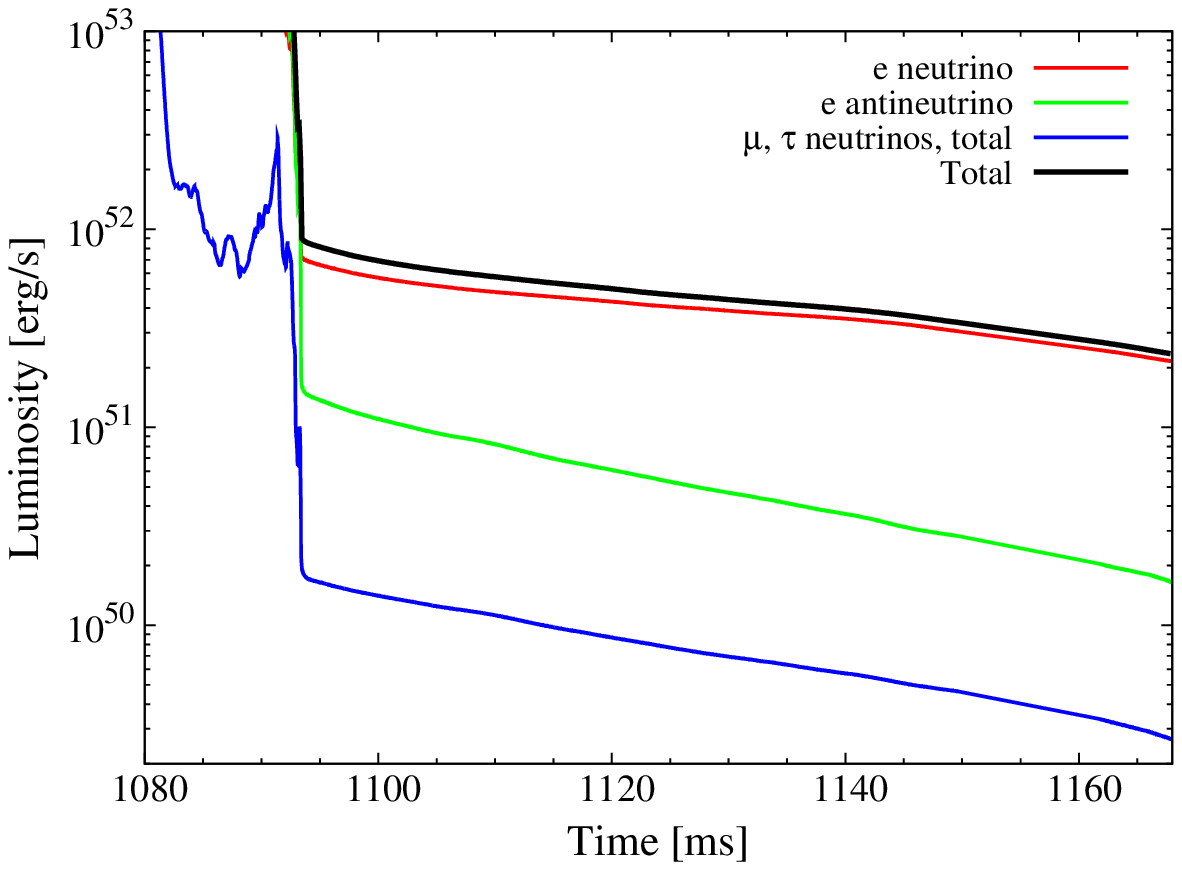}
\end{center}
  \caption{Time evolution of neutrino luminosities for UN100-diff (a) before the BH formation 
    and (b) after the BH formation. The red, green, and blue
    thick curves correspond to the luminosities of $\nu_{e}$, $\bar{\nu}_{e}$, total of $\mu$ and $\tau$
    pair neutrinos, respectively. The thin blue curve in the upper panel shows the luminosity of 
    {\it individual} $\mu$/$\tau$ neutrino (namely the quarter of thick blue curve).
    The thick black curve in the lower panel shows the total neutrino luminosity.
\label{fig_SCC_lum-diff}}
\end{figure}

\begin{figure}[p]
\begin{center}
    (a)\includegraphics[scale=1.00]{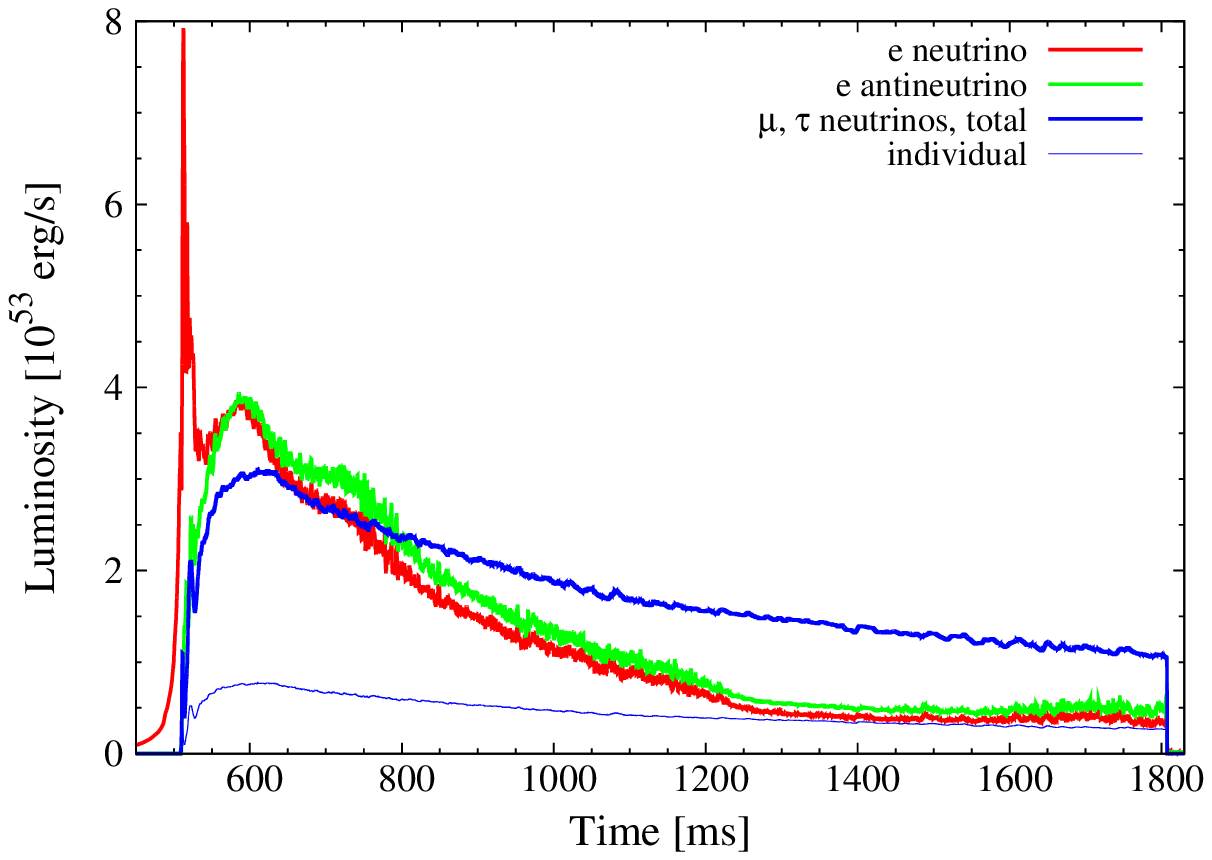}
    (b)\includegraphics[scale=1.00]{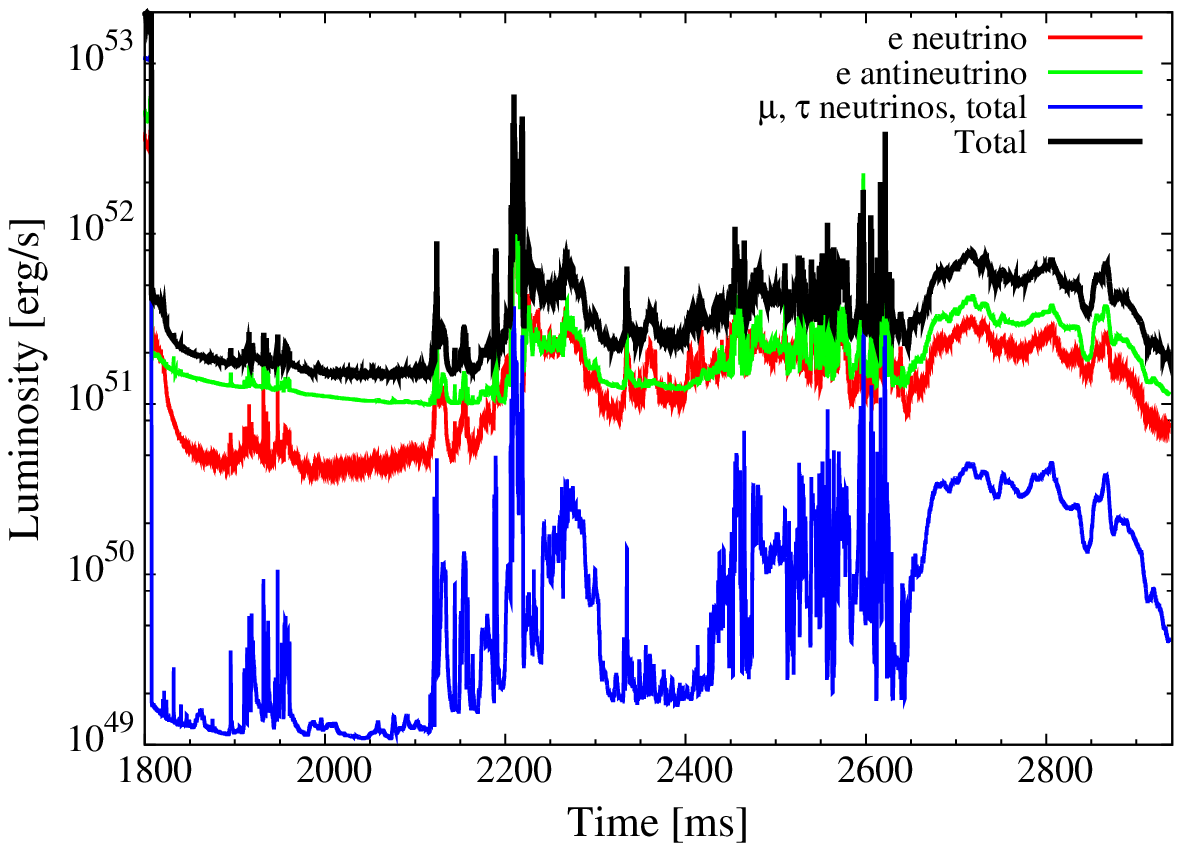}
\end{center}
  \caption{Time evolution of neutrino luminosities for UN100-rigid (a) before the BH formation 
    and (b) after the BH formation. Meanings of all curves are the
    same as those of Fig.~\ref{fig_SCC_lum-diff}.
\label{fig_SCC_lum-rig}}
\end{figure}

\subsection{Neutrino Luminosity and Gravitational Waves}\label{SCC_Lum}

Figures~\ref{fig_SCC_lum-diff}(a) and \ref{fig_SCC_lum-diff}(b) plot the time evolution of neutrino luminosities
for UN100-diff. 
In the prebounce phase, electron neutrinos are dominantly emitted and the emissivity of 
electron anti-neutrinos is much smaller. This is because the electrons are (mildly) 
degenerate blocking the inverse $\beta$-decay 
and also the positron fraction, which is responsible for the anti-neutrino emission, is small. 
Soon after the core bounce, the so-called neutrino burst occurs at the time that 
the shock wave passes through the neutrino-sphere, as in the collapse of an ordinary supernova core.

After the neutrino burst, the emission of electron anti-neutrinos is enhanced and
their luminosity becomes larger that of electron neutrinos.
This property is different from that in the ordinary supernova~\cite{Bethe1990,SpheGR},
and explained as follows. During the post neutrino burst phase, a large number of positrons 
are produced because the degeneracy parameter becomes low as 
$\eta_{e} \sim 1$ due to the high temperature of  $T \gtrsim 20$ MeV, which is
higher than the temperature in the ordinary supernova $T\sim 5$ MeV~\cite{Bethe1990}.
Then, because the neutron fraction $X_{n}$ is much larger than the proton fraction $X_{p}$,
the positron capture on neutrons occurs more efficiently, and hence, 
the electron anti-neutrino luminosity becomes larger than the electron neutrino luminosity.
This dominant emission of electron anti-neutrinos are also found for a HMNS formed after 
the BNS merger (see \S~\ref{Sec_ResultBNS}) and in a BH-torus system formed in the collapse 
of a more massive core~\cite{Sekiguchi2011} with $s=8k_B$. Both systems have a higher temperature and
a lower electron fraction than the collapse of the ordinary supernova core, as in the
present case.

The luminosity of $\mu$ and $\tau$ neutrinos is smaller than
that of electron neutrinos and anti-neutrinos. This is simply due to the absence of
the neutrino production channel mediated by the charged weak current.
At later phases in the HMNS evolution (800 ms $\lesssim t \lesssim$ 1100 ms), 
the electron neutrino and anti-neutrino luminosities show weak time variability. 
This is due to the convective activities that occur near the neutrino sphere.
The $\mu$ and $\tau$ neutrino luminosity does not show the variability because
they are mainly emitted by the hot central regions that do not suffer from the
convection.

Soon after the BH formation at $t\approx 1090$ ms, 
neutrino luminosities decrease drastically because the main 
neutrino-emission region is swallowed into the BH.
After that, the geometrically thin accretion disk emits $\sim 10^{51}$--$10^{52}$ ergs/s
by neutrinos in its early evolution phase with the duration $\sim 100$ ms 
and the luminosities decrease monotonically in time. 
We do not find any enhancement of the neutrino luminosities after 
the BH formation in our simulation time. 
In this phase, electron neutrinos are dominantly emitted because the disk is 
at a lower temperature of $T\lesssim$ a few MeV,
and hence, there are less positrons.
Also, the number of the target neutrons are smaller because the disk is composed mainly of 
proton-rich matter in the outer neutrino emission region.

Figures~\ref{fig_SCC_lum-rig}(a) and \ref{fig_SCC_lum-rig}(b) show the time evolution of neutrino luminosities for UN100-rigid.
The features of the evolution is similar to those for UN100-diff
before the neutrino burst ($t\lesssim 700$ ms).
After that time, the luminosities of electron neutrinos and anti-neutrinos gradually decrease.
This is because the optical depth (diffusion time) gets larger (longer) as the torus grows.
The luminosities of electron neutrinos and anti-neutrinos show only weak time variability, 
reflecting the weaker convective activity in UN100-rigid.

At $t\approx 800$ ms, the {\it total} luminosity of $\mu$ and $\tau$ neutrinos 
($L_{\nu_{\mu}}+L_{\bar{\nu}_{\mu}}+L_{\nu_{\tau}}+L_{\bar{\nu}_{\tau}}$) becomes larger than 
the electron neutrino and anti-neutrino luminosities. This is partly due to a 
very high temperature of the neutrino sphere which enhances pair neutrino production 
processes, as well as a smaller optical depth along the rotational axis: 
Thermal neutrinos from the hot HMNS will be almost directly {\it seen} due to the low 
density along the rotational axis. 
Indeed, after $t\gtrsim 1300$ ms, all flavor of neutrinos and anti-neutrinos are almost 
equally emitted, indicating the dominant emission of thermal neutrinos from the very hot HMNS in the
neutrino luminosity. 
This feature is not seen in the BNS merger (see \S~\ref{Sec_ResultBNS}) and  
the collapse of the more massive stellar core~\cite{Sekiguchi2011}.
This is because the continuous mass accretion is absent in the BNS merger and
the HMNS is quickly collapses to a BH in the collapse of the more massive stellar core.
At the final phase in the fallback collapse of an ordinary core 
(a failed supernova)~\cite{Sumiyoshi07},
an enhancement of the emission of $\nu_{x}$ is also seen.

Note that the above result implies that observational signals of neutrinos could depend on the viewing
angle: If we would see the system from the direction along the rotational axis, we might see
a brighter emission of neutrinos with a higher average energy from the hot HMNS, while we would
see neutrino emissions from the torus if we see the system from the direction along the
equator.
In the leakage scheme adopted in this paper, unfortunately, we cannot investigate such 
an angle dependence of neutrino luminosities.

The neutrino luminosities show a precipitation when the BH is formed at $t\approx 1805$ ms, 
as in the case of UN100-diff. The total neutrino luminosity emitted from
the torus around the BH amounts to $L_{\nu, {\rm tot}}\sim 10^{51}$ ergs/s.
A remarkable property in UN100-rigid is that this luminosity is maintained for $\gtrsim 1$ s.
In addition, by contrast with the case of UN100-diff, the neutrino luminosities show a violent
time variability. Such a long-term high luminosity and a time variability may be associated with 
the time variability that LGRBs show.

\begin{figure}[t]
\begin{center}
    \includegraphics[scale=1.0]{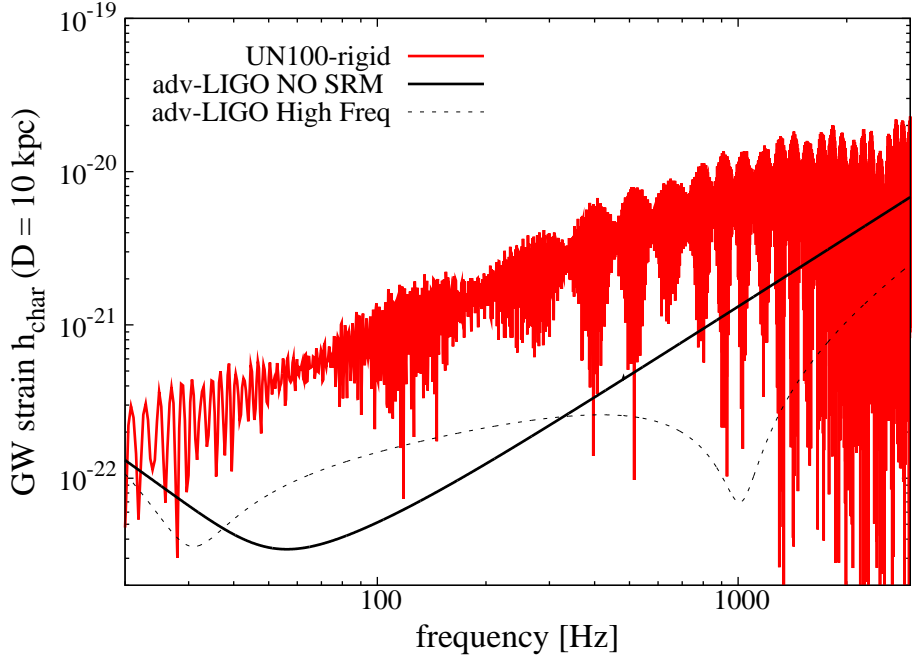}
\end{center}
  \caption{The spectrum of the characteristic gravitational-wave strain for UN100-rigid.
    The noise amplitudes of Advanced LIGO for a version in which 
    no signal recycling mirror is used (NO SRM) and a design with a narrow-band tuning 
    at 1kHz (High Freq) are shown together.
\label{fig_SCC_GW}}
\end{figure}

Figure~\ref{fig_SCC_GW} plots the spectra of the characteristic gravitational-wave
strain~\cite{FH98}, 
\beq
h_{\rm char}(f) \equiv \sqrt{
\frac{2}{\pi^{2}} \frac{G}{c^{3}}\frac{1}{D^{2}}\frac{dE}{df}},
\eeq \label{hchar}
where $D$ is the distance to the source and
\beq
\frac{dE}{df} = \frac{8\pi^{2}}{15}\frac{c^{3}}{G} f^{2}
\left| \tilde{A}_{2}(f) \right|^{2}
\eeq 
is the energy power spectra of the gravitational radiation.
$\tilde{A}_{2}(f)$ is the Fourier transform of $A_{2}$,
\beq
\tilde{A}_{2}(f) = \int A_{2}(t) e^{2\pi i f t} dt
\eeq
with $A_{2}$ being the $+$-mode of gravitational waves with $l=2$ and $m=0$,
\beq
h_+^{\rm quad} = {\ddot I_{zz}(t_{\rm ret}) - \ddot I_{xx}(t_{\rm ret})
\over D}\sin^2\theta \equiv \frac{A_{2}(t)}{D}\sin^{2}\theta,
\label{quadr}
\eeq
where $I_{ij}$ denotes a quadrupole moment~\cite{SS03}, $\ddot I_{ij}$ 
its second time derivative, and $t_{\rm ret}$ a retarded time.
Because the strain of gravitational waves has a
broad-band spectrum, they appear to be emitted
primarily by a long-term stochastic motion of
the infalling material and of the matter in the
HMNS, as in the collapse of ordinary cores~\cite{Kotake}.
Note that $h_{\rm char}$ includes only gravitational waves from the matter contribution and 
not from the anisotropic neutrino emissions.
We also show the noise amplitudes of Advanced LIGO for a version
in which no signal recycling mirror is used (NO SRM) and of a version of
a specially designed narrow-band Advanced LIGO (High Freq) together~\cite{LIGOnoise}. 

The effective amplitude of gravitational waves observed in the most optimistic 
direction is $h_{\rm char}\sim 10^{-20}$ for an hypothetical event at a distance of 10 kpc, which is 
larger than that for the ordinary supernova~\cite{Kotake}.
The frequency at the peak amplitude is 1--2 kHz.
The reason for this larger gravitational-wave amplitude is that
the HMNS in the present model has a larger mass and 
the gravitational-wave signal is accumulated
during the long-term convective activity, for which the
duration $\gtrsim 1$ s is longer than that for the ordinary supernova.
Figure~\ref{fig_SCC_GW} shows that with NO-SRM Advanced LIGO the signal of gravitational
waves will be detected with a signal-to-noise ratio ${\rm S/N}\sim 10$ for $D=10$ kpc.
For High-Freq Advanced LIGO, the detection may be done with ${\rm S/N}\sim 100$ for $D=10$ kpc.
Even for $D=50$ kpc (the distance to the Large Magellanic Cloud), the detection will be possible
with ${\rm S/N}\gtrsim 20$.

For an event of $D\lesssim 50$ kpc, a large number of neutrinos will be also detected by 
water-Cherenkov neutrino detectors such as Super-Kamiokande and future Hyper-Kamiokande
(see also a discussion in \S~\ref{BNS_Dyn}).

\subsection{Possible association to Gamma-ray bursts}\label{pairGRB}

\begin{figure}[t]
\begin{center}
    \includegraphics[scale=1.2]{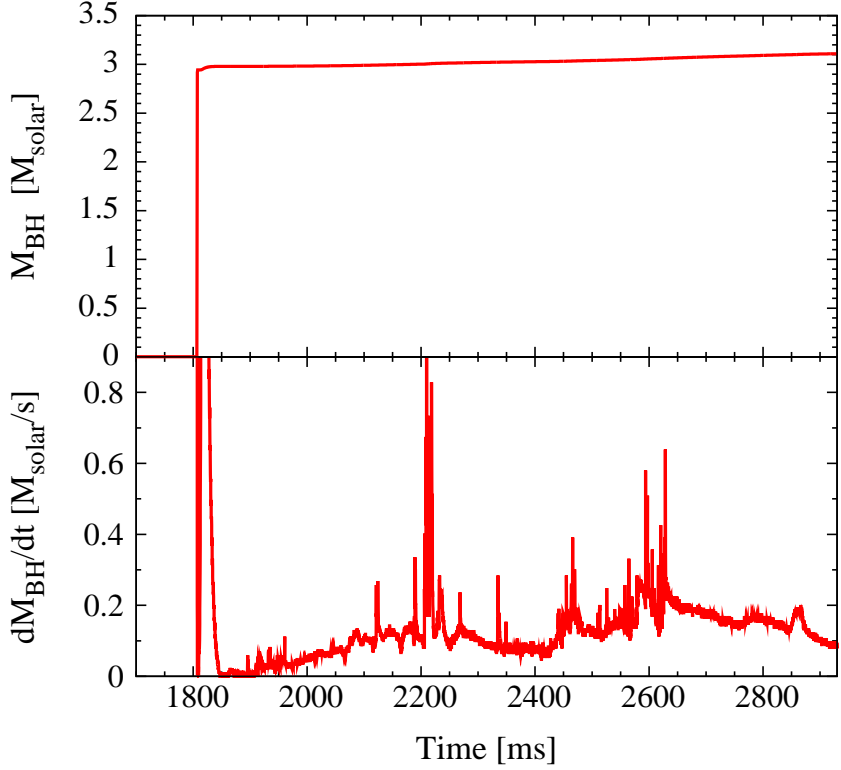}
\end{center}
  \caption{Time evolution of the irreducible mass of the BH (the upper panel) 
    and the mass accretion rate (the lower panel) for UN100-rigid.
\label{fig_SCC_MBH}}
\end{figure}

Before closing this section, we give an order estimate of the energy deposition rates
($\dot{E}_{\nu\bar{\nu}}$) 
by the neutrino pair-annihilation which is one of the possible processes to drive relativistic
jets required to produce LGRBs.
Note that the energy from the neutrino pair annihilation should be deposited
in a baryon-poor region in order to generate highly relativistic outflows.
The funnel region near the rotational axis above the torus formed in UN100-rigid is
a promising place for this purpose.

According to the estimate in Ref.~\citen{Beloborodov08}, 
the deposition rate in the BH-torus system would be proportional to 
$\dot{M}^{9/4}M_{\rm BH}^{-3/2}$.
In this estimation, the neutrino luminosity is assumed to be originated from a
viscous heating. In our present simulations, the neutrino luminosity is determined by 
the infalling rate of the material which experiences the shock heating
at the surface of the torus to increase the thermal energy of the torus. 
However, the dependence of the pair-annihilation rate on the mass infall rate 
$\dot{M}$ is essentially the same for thick torus phase.
Figure~\ref{fig_SCC_MBH} shows the irreducible mass of the BH (the upper panel) 
and the mass accretion rate (the lower panel).
Due to this strong dependence on $\dot{M}$, the energy deposition
by the neutrino pair-annihilation would be important only for a phase 
in which $L_{\nu, {\rm tot}} \gtrsim 10^{51}$ ergs/s 
(see Figs.~\ref{fig_SCC_lum-diff}(b) and \ref{fig_SCC_lum-rig}(b)).
Figure~\ref{fig_SCC_lum-rig}(b) shows that for UN100-rigid, the duration of the neutrino
emission (in the BH phase) with $L_{\nu, {\rm tot}} \gtrsim 10^{51}$ ergs/s is longer than 1 s,
and thus, a long-term energy deposition for a LGRB may be explained. 
Taking into account the dependence of the neutrino pair annihilation rate 
on the geometry of the torus~\cite{Mochkovitch93,Beloborodov08,Liu10,Zalamea11},
$\dot{E}_{\nu\bar{\nu}}$ would be given by~\cite{Beloborodov08}, 
\beqn
\dot{E}_{\nu\bar{\nu}} &\sim& 10^{48} \,{\rm ergs/s} 
\left(\frac{100\,{\rm km}}{R_{\rm fun}}\right)
\left(\frac{0.1}{\theta_{\rm fun}}\right)^{2}
\left(\frac{E_{\nu}+E_{\bar{\nu}}}{10\,{\rm MeV}}\right) \nonumber \\
&& \ \ \ \ \ \ \ \ \ \ \ \times \left(\frac{L_{\nu}}{10^{51}\,{\rm ergs/s}}\right)
\left(\frac{L_{\bar{\nu}}}{10^{51}\,{\rm ergs/s}}\right)
\sin^{2} \Theta,
\eeqn
where $R_{\rm fun}$ and $\theta_{\rm fun}$ are the characteristic radius and 
the opening angle of the funnel region.
$\Theta$ denotes the collision angle of the neutrino pair.
Thus a low-luminosity LGRB could be explained.

In the HMNS phase, by contrast, the neutrino luminosity is huge 
as $L_{\nu} \gtrsim 10^{53}$ ergs/s (see Figs.~\ref{fig_SCC_lum-diff}(a) and \ref{fig_SCC_lum-rig}(a)), 
and hence, the deposition rate would be very large as
\beqn
\dot{E}_{\nu\bar{\nu}} &\sim& 3\times 10^{52} \,{\rm ergs/s} 
\left(\frac{100\,{\rm km}}{R_{\rm fun}}\right)
\left(\frac{0.1}{\theta_{\rm fun}}\right)^{2}
\left(\frac{E_{\nu}+E_{\bar{\nu}}}{30\,{\rm MeV}}\right) \nonumber \\
&& \ \ \ \ \ \ \ \ \ \ \ \ \ \ \ \ \times \left(\frac{L_{\nu}}{10^{53}\,{\rm ergs/s}}\right)
\left(\frac{L_{\bar{\nu}}}{10^{53}\,{\rm ergs/s}}\right)
\sin^{2} \Theta.
\eeqn
If the outflows launched due to the mass accumulation mechanism can
penetrate the stellar envelope, a system composed of a long-lived HMNS and a 
geometrically thick torus may be a promising candidate of the central engine of
LGRBs a relatively short duration.


\section{Binary neutron star merger}\label{Sec_ResultBNS}

Coalescence of binary neutron stars (BNS) is one of the most promising
sources for next-generation kilo-meter-size gravitational-wave
detectors~\cite{LIGO,VIRGO,LCGT}, and also a possible candidate for the 
progenitor of SGRBs~\cite{GRB-BNS1,GRB-BNS2}. Motivated by these facts, numerical 
simulations have been extensively performed for the merger of BNS in the 
framework of full general relativity in the past decade since the first 
success~\cite{SU00} in 2000 (see also, e.g., Refs.~\citen{Duez} for reviews).

BNS evolve due to the gravitational radiation reaction and eventually
merge. Before the merger sets in, each neutron star is cold (i.e.,
thermal energy of constituent nucleons is much smaller than their
Fermi energy), because the thermal energy inside the neutron stars is
significantly reduced by neutrino and photon emissions~\cite{ST83}
in the long-term inspiral phase (typically $\gtrsim 10^8$~years~\cite{Lorimer}) 
until the
merger.  By contrast, after the merger sets in, shocks are generated
by hydrodynamic interactions. In particular, when a HMNS is formed 
in the merger, spiral arms are developed in its
envelope and continuous heating occurs due to the collision between
the HMNS and spiral arms (e.g., Refs.~\citen{STU,KSST,LR}).  Newtonian
simulations indeed suggest that by this process the maximum
temperature increases to $\sim 30$--50~MeV, and hence, copious
neutrinos are
emitted~\cite{Ruffert,Rosswog,Setiawan,Dessart09}.  Thus,
to accurately explore the merger process, the evolution of the hot
HMNS, and possible subsequent formation of a BH with a physical
modeling, numerical-relativity simulations have to be performed
incorporating microphysical processes such as neutrino emission and
equation of state (EOS) based on a theory for the high-density and
high-temperature nuclear matter.  However, such simulations have not
been done in full general relativity until quite recently (but
see Ref.~\citen{OJM} for a work in an approximate general relativistic
gravity with finite-temperature EOS).  Incorporation of microphysical
processes is in particular important for exploring the merger
hypothesis of SGRB because it may be driven through pair annihilation
of neutrino-antineutrinos pairs~\cite{GRB-BNS1,GRB-BNS2}.

In this section, we review our first results of numerical-relativity
simulations for the BNS merger presented in Refs.~\citen{SKKS1,SKKS2}, which
are performed incorporating both a finite-temperature
EOS~\cite{Shen1998,Shen2011} and neutrino cooling~\cite{Sekiguchi2010}. In the
following, we summarize the possible outcome formed after the merger,
criteria for the formation of HMNS and BH, thermal properties of the
HMNS and torus surrounding the formed BH, and neutrino luminosity and
gravitational waveforms from the HMNS and in the BH formation.

\subsection{Initial condition and grid setting}

\begin{figure}[t]
\begin{center}
    \includegraphics[scale=1.1]{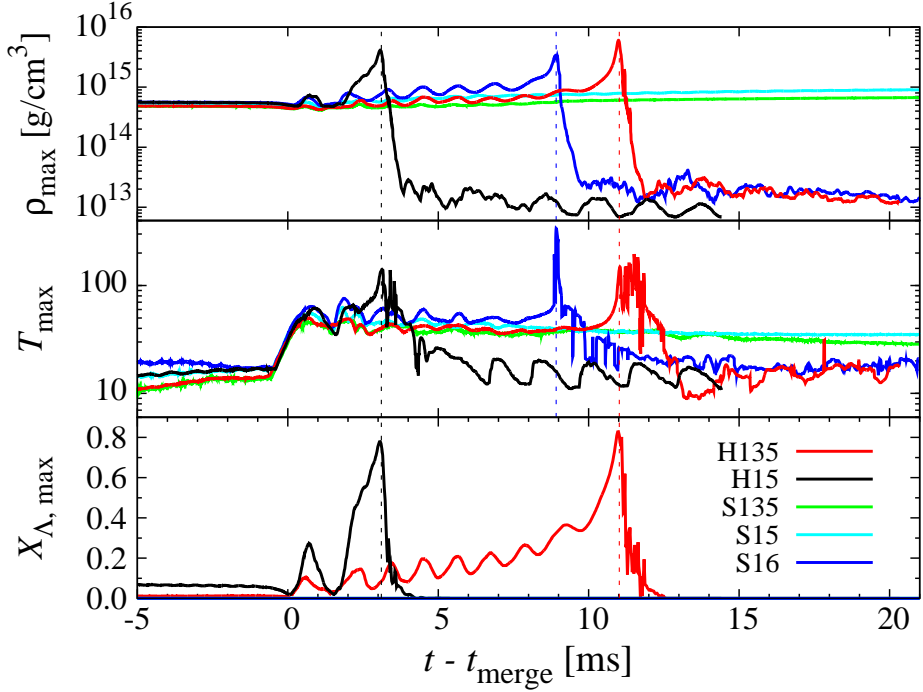}
\end{center}
\caption{Maximum rest-mass density, maximum matter temperature, and
maximum hyperon fraction in mass as functions of time for all the
models.  $t_{\rm merge}$ denotes the onset time of the merger.  The
dashed vertical line shows the time at which a BH is formed for models
S16, H135, and H15. 
\label{fig_BNS_max}}
\end{figure}

\begin{figure}[t]
\begin{center}
    \includegraphics[scale=1.1]{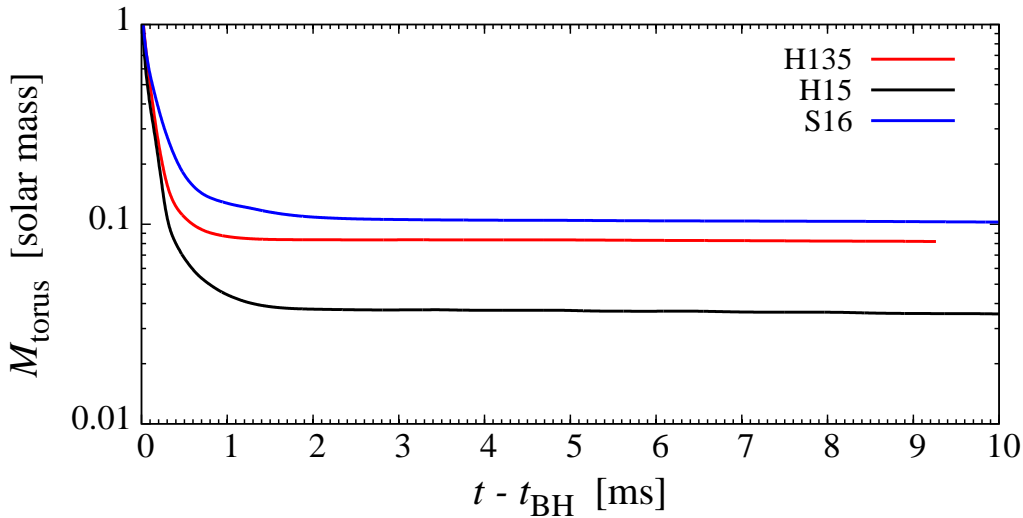}
\end{center}
\caption{The rest mass of a torus surrounding the formed BH as a 
function of time for models H135, H15, and S16. $t_{\rm BH}$ denotes
the time at the formation of the BH.
\label{fig_BNS_Mtorus}}
\end{figure}

\begin{figure}[t]
\vspace{-2mm}
\begin{center}
    \includegraphics[scale=1.8]{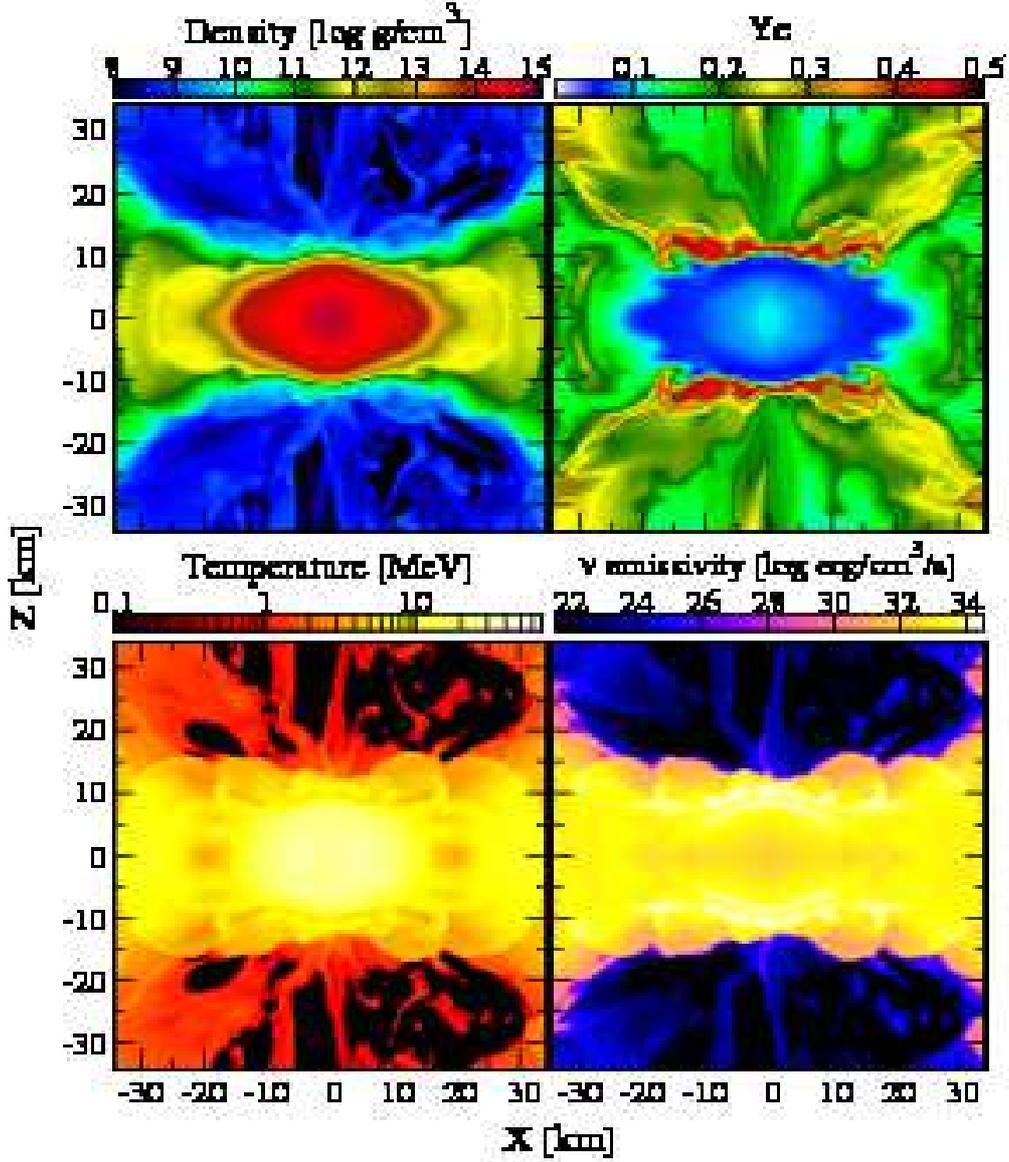}
\end{center}
\caption{
Contour maps in the $x$-$z$ plane of the rest-mass density (top left), the electron fraction (top right),
the temperature (bottom left), and the total neutrino emissivity (bottom right) 
at $t \approx 16.7$~ms after the onset of the merger for model S135. 
\label{fig_BNS_conSxz}}
\end{figure}

\begin{figure}[t]
\vspace{-2mm}
\begin{center}
    \includegraphics[scale=1.8]{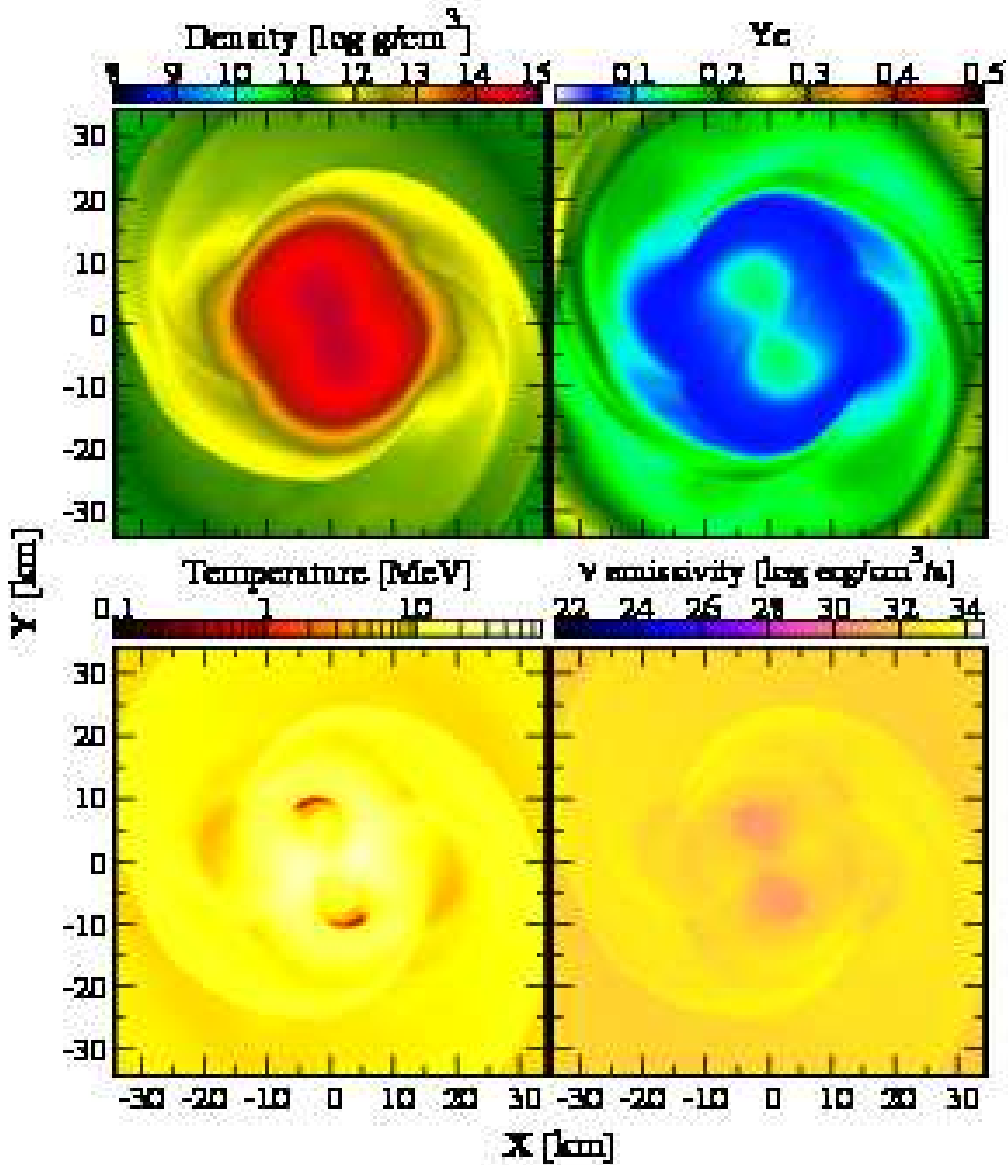}
\end{center}
\caption{
The same figure as Fig.~\ref{fig_BNS_conSxz} but in the $x$-$y$ plane.
\label{fig_BNS_conSxy}}
\end{figure}

\begin{figure}[t]
\vspace{-2mm}
\begin{center}
    \includegraphics[scale=1.6]{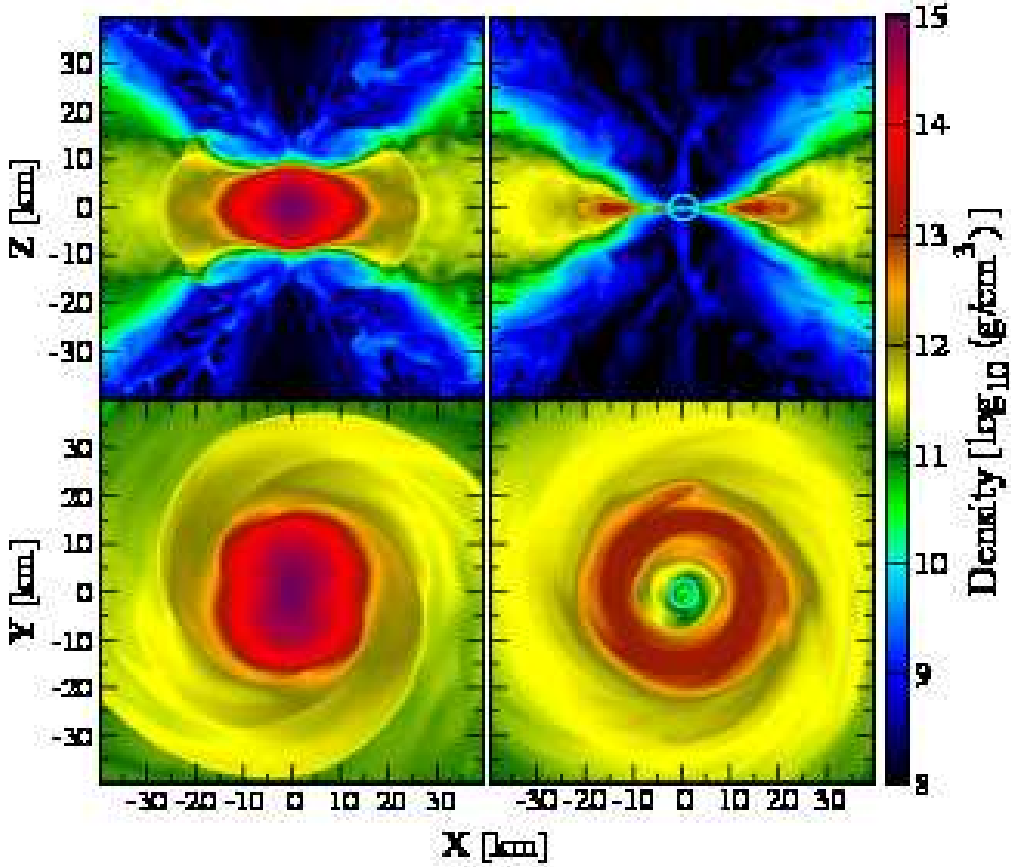}
\end{center}
\caption{
Contour maps of the rest-mass density for a HMNS phase at $t \approx 17.5$~ms after the merger
(left panels) and that for a BH phase at $t \approx 26.8$~ms after the onset of the merger for model H135. 
The upper and lower panels show the configuration in the $x$-$y$ and
$x$-$z$ planes, respectively. The blue circle of the right panels shows the location of
the apparent horizon.
\label{fig_BNS_conHrho}}
\end{figure}

\begin{figure}[t]
\vspace{-2mm}
\begin{center}
    \includegraphics[scale=1.6]{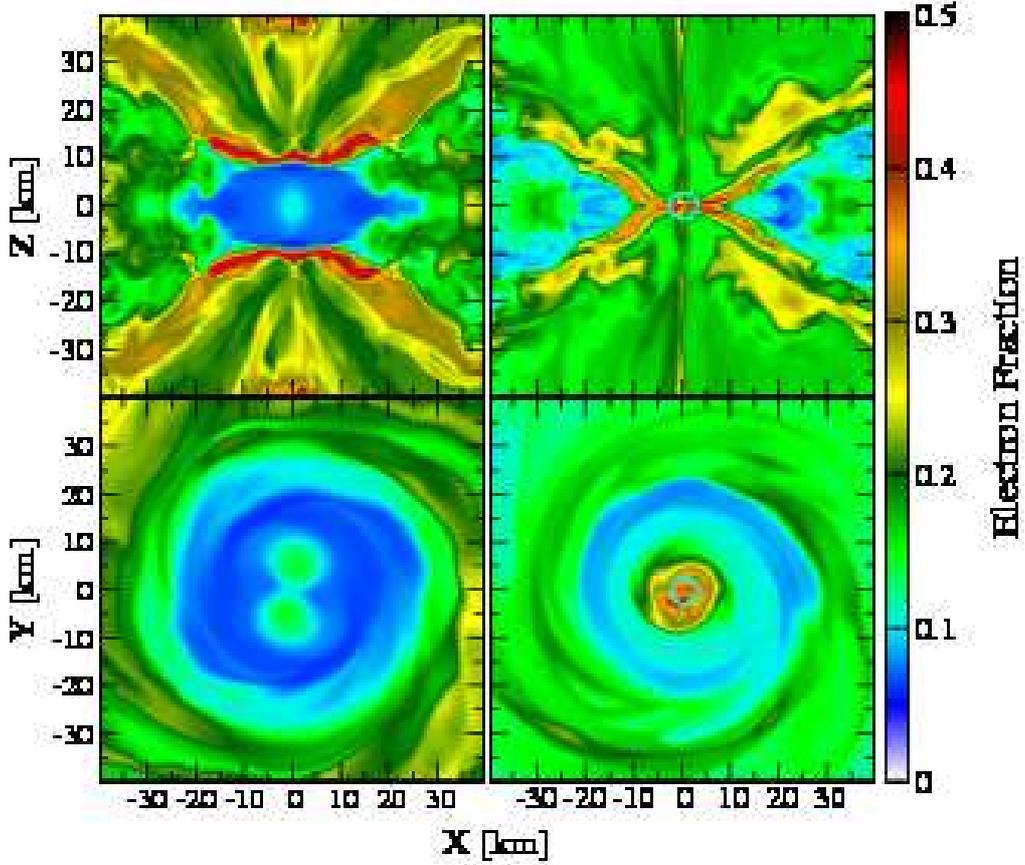}
\end{center}
\caption{
The same figure as Fig.~\ref{fig_BNS_conHrho} but for the electron fraction.
\label{fig_BNS_conHYe}}
\end{figure}

\begin{figure}[t]
\vspace{-2mm}
\begin{center}
    \includegraphics[scale=1.6]{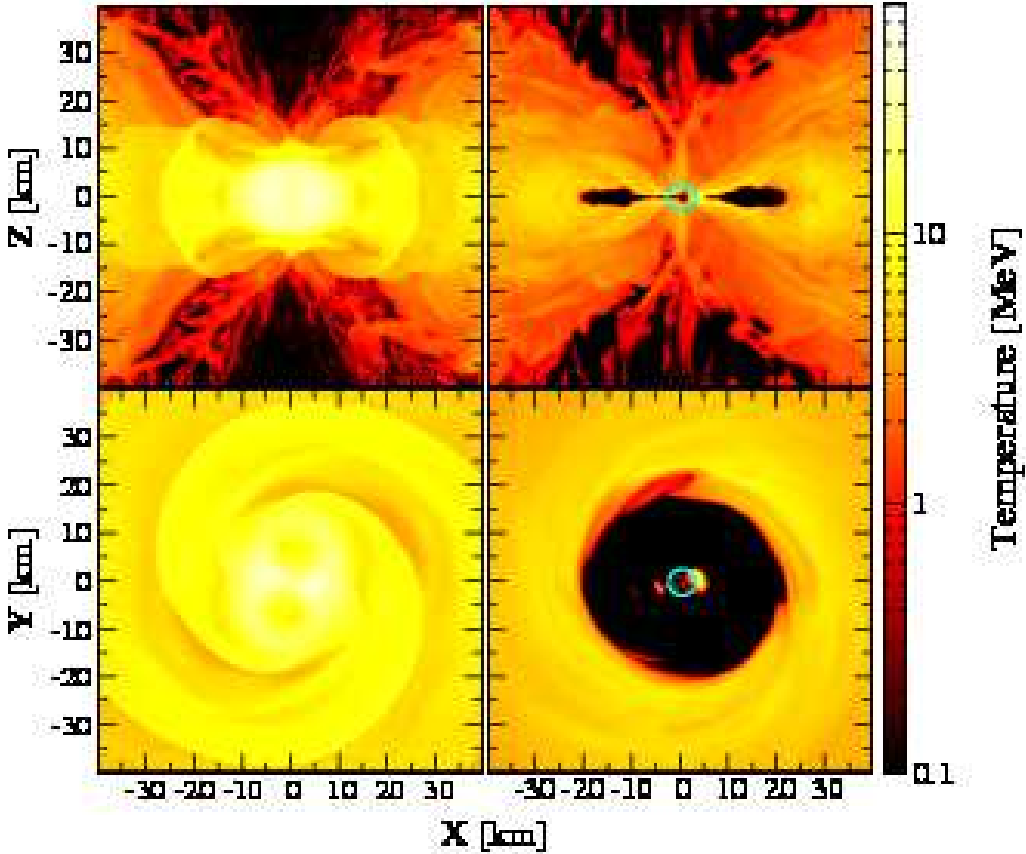}
\end{center}
\caption{
The same figure as Fig.~\ref{fig_BNS_conHrho} but for the temperature.
\label{fig_BNS_conHT}}
\end{figure}

\begin{figure}[t]
\vspace{-2mm}
\begin{center}
    \includegraphics[scale=1.6]{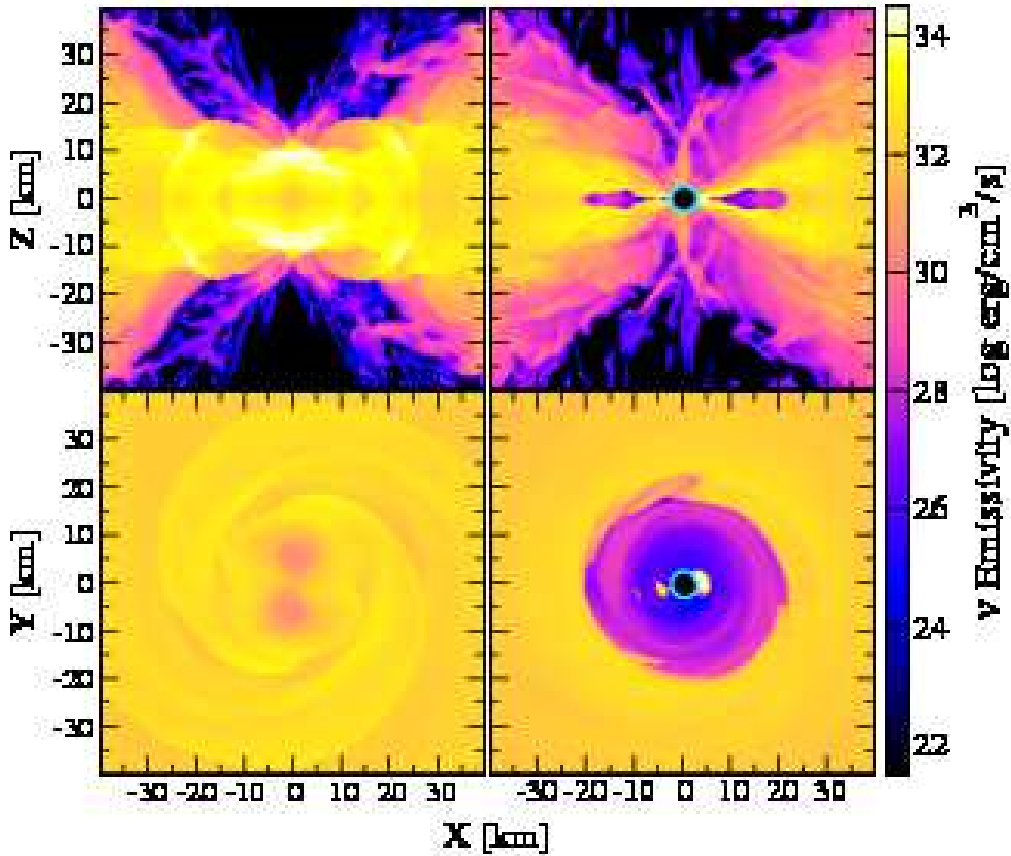}
\end{center}
\caption{
The same figure as Fig.~\ref{fig_BNS_conHrho} but for the total neutrino emissivity.
\label{fig_BNS_conHdQ}}
\end{figure}

\begin{figure}[t]
\begin{center}
    \includegraphics[scale=1.1]{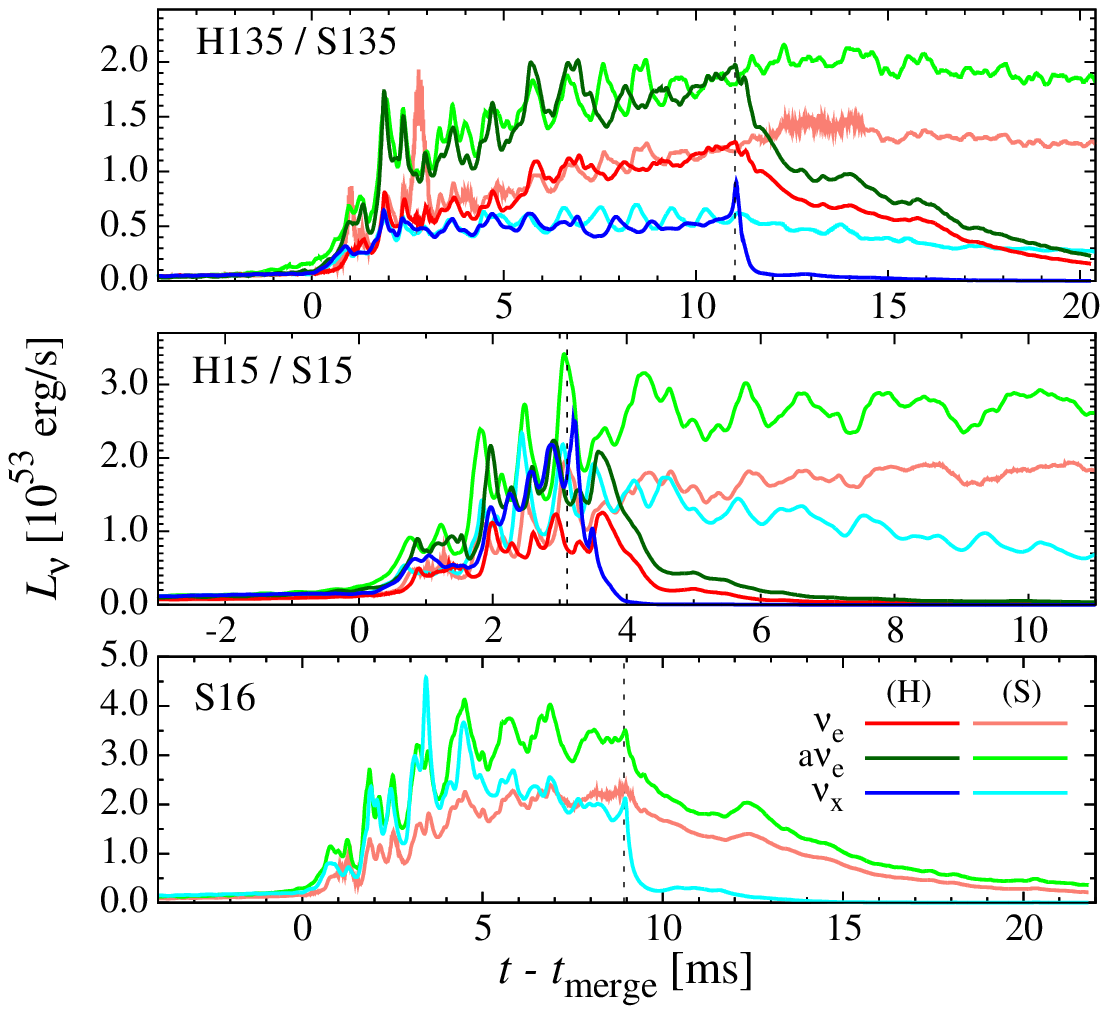}
\end{center}
\caption{Neutrino luminosities for three flavors for all the models.
The upper, middle, and lower panels show the results for $M_{\rm
    NS}=1.35$, 1.5, and $1.6M_{\odot}$, respectively.  The meaning of
    the dashed line is the same as in Fig.~\ref{fig_BNS_max}.
\label{fig_BNS_lum}}
\end{figure}

 In Refs.~\citen{SKKS1,SKKS2}, we focused only on the merger of equal-mass
BNS, because the mass difference for the observed BNS is not very large
\cite{Stairs,Lorimer}. To date, we have performed simulations for 5 models: For Shen-EOS, 
we employed three masses for each neutron star: $M_{\rm NS}=1.35$, 1.5,
and $1.6M_{\odot}$ ($M_{\rm NS}$ is the gravitational mass of a
neutron star in isolation).  We refer to each model as models S135,
S15, and S16, respectively.  For Hyp-EOS, we employed $M_{\rm NS}=1.35$
and $1.5M_{\odot}$, and refer to two models as H135 and H15, respectively.  
The simulations
were performed with the initial condition of about 3--4 orbits before
the onset of the merger, until the system relaxes to a quasi-stationary
state. Quasi-equilibrium states of BNS were prepared as the initial
conditions, as in Refs.~\citen{STU,KSST}, using the LORENE library~\cite{lorene}.

There are two possible fates~\cite{STU} of BNS: If its total
mass $M$ is larger than a critical mass $M_c$, a BH will be formed
soon after the onset of the merger, while a differentially rotating
HMNS will be formed for $M < M_c$.  The value of $M_c$ depends
strongly on the EOS.  Because Shen-EOS is quite stiff, $M_c$ is much
larger than the typical total mass of BNS, $\sim
2.7M_{\odot}$~\cite{Stairs,Lorimer}, as shown in Ref.~\citen{SKKS1} and below.
Thus, with this EOS, the HMNS will be the frequent outcomes, as in the
cases of stiff EOS with which $M_{\rm
max}>2M_{\odot}$~\cite{hotoke}. By contrast, Hyp-EOS is not stiff in
particular for a high-density range. Thus, a BH is often formed with
this EOS, although a HMNS could be a transient outcome soon after the
onset of the merger~\cite{SKKS2}.

Numerical simulations were performed preparing a non-uniform grid as in
Ref.~\citen{KSST}. The inner domain was composed of a finer uniform grid and
the outer domain of a coarser nonuniform grid. The grid resolution in
the inner zone is chosen so that the major diameter of each neutron
star in the inspiral orbit was covered by 60 and 80 grid points for
low- and high-resolution runs, respectively: We always performed
simulations for both grid resolutions to confirm that the convergence,
sufficient to draw a scientific conclusion on the final outcome,
gravitational waveforms, and neutrino luminosities, is approximately
achieved.  Outer boundaries are located in a local wave zone (at
$\approx 560$--600~km along each coordinate axis which is longer than
gravitational wavelength in the inspiral phase).  During the
simulations, we checked the conservation of the baryon rest-mass, total
gravitational mass (Arnowitt-Deser-Misner mass plus radiated energy of
gravitational waves), and total angular momentum (including that
radiated by gravitational waves), and found that the errors are within
0.5\%, 1\%, and 3\%, respectively, for the high-resolution runs within
the physical duration $\approx 30$~ms.

\subsection{Merger and subsequent evolution}\label{BNS_Dyn}

Figure~\ref{fig_BNS_max} plots the maximum rest-mass
density, $\rho_{\rm max}$, maximum matter temperature, $T_{\rm max}$,
and maximum hyperon fraction in mass $X_{\Lambda,{\rm max}}$ as
functions of $t-t_{\rm merge}$ where $t_{\rm merge}$ is the onset time
of the merger. For $t< t_{\rm merge}$, $\rho_{\rm max}$ is
approximately constant besides a small decline due to tidal
elongation, while for $t \gtrsim t_{\rm merge}$, it gradually increases
because a HMNS is formed at least temporarily irrespective of models,
and subsequently contracts due to the dissipation of the angular momentum 
by the gravitational-wave emission. Thus, $\rho_{\rm max}$ increases in 
the gravitational radiation time scale. The subsequent
evolution process depends on the mass and EOS. For models S135 and
S15, the degree of non-axial symmetry of the HMNS becomes low enough at
$t-t_{\rm merge} \sim 20$~ms that the emissivity of gravitational
waves is significantly reduced.  Because no dissipation process except
for the neutrino cooling is present, the HMNS will be alive at least for
the cooling time scale before collapsing to a BH (see below).  For
models S16, H135, and H15, the HMNS collapse to a BH 
at $t-t_{\rm merge} \lesssim 10$ ms after
the gradual contraction due to the gravitational-wave emission and a
massive disk of $\approx 0.03$--$0.1 M_{\odot}$ is formed around the
BH. It should be noted that for Hyp-EOS, a BH is formed at $t-t_{\rm
merge} \sim 10$~ms even with the total mass $2.7M_{\odot}$. This is
due to the softening effect by the appearance of $\Lambda$ hyperons: 
See the bottom panel of Fig.~\ref{fig_BNS_max}, which shows that $X_{\Lambda, {\rm max}}$
increases steeply just before the HMNS collapses to a BH.

The evolution of $T_{\rm max}$ plotted in Fig.~\ref{fig_BNS_max} shows that
the HMNS formed just after the merger are hot with $T_{\rm max} \sim
50$--70~MeV (much higher than that in the ordinary supernova and as high as
that in the HMNS formed after the collapse of the massive stellar core; cf.
\S~\ref{Sec_ResultSCC}). 
Such a high temperature is achieved due to the liberation
of the kinetic energy of the orbital motion at the collision of two
neutron stars. For the case that a long-lived HMNS is formed,
subsequently, $T_{\rm max}$ decreases due to the neutrino cooling,
with the maximum luminosity 3--$10 \times 10^{53}$~ergs/s (see
Fig.~\ref{fig_BNS_lum}), but relaxes to a high value with $25$--50~MeV when
the HMNS relaxes to a quasi-steady state.  Around the HMNS, spiral
arms are formed and shock heating continuously occurs when the spiral
arms hit the HMNS (see Figs.~\ref{fig_BNS_conSxz}--\ref{fig_BNS_conHdQ} for snapshots). 
Due to this process and because of the long neutrino-cooling time scale, the
temperature (and thermal energy) does not significantly decrease in
$\sim 100$~ms: We estimated the cooling time scale as $E_{\rm
th}/L_{\nu}\sim 2$--3~s where $E_{\rm th}$ is the total thermal energy of
the HMNS.

For the case that a BH is eventually formed, the maximum temperature
raises significantly to $\gtrsim 100$~MeV just before the BH
formation. This is simply due to the adiabatic compression effect.
For models S16, H135, and H15, a torus surrounding the BH is
subsequently formed. The typical maximum density and temperature of the torus are
$\sim 10^{13}~{\rm g/cm^3}$ and 20~MeV, respectively, with the
mass $\approx 0.03$--$0.1M_{\odot}$ (see Fig.~\ref{fig_BNS_Mtorus}). 
This mass has a correlation with
the lifetime of the HMNS; for the longer lifetime (e.g., for model S16), the torus mass is
larger (compare the mass of the torus in Fig.~\ref{fig_BNS_Mtorus}). 
The reason is that during the evolution
of the HMNS which is deformed in a non-axisymmetric manner, the angular 
momentum is transported from the inner to the
outer region via the hydrodynamic torque associated with its
non-axisymmetric structure.  Thus, the longer lifetime helps increasing
the mass element of a sufficiently large specific angular momentum
which can escape falling into the formed BH.  This fact implies that
stiff EOS are favored for the formation of a massive torus. 

Figures~\ref{fig_BNS_conSxz} and \ref{fig_BNS_conSxy} plot the contour maps of 
the rest-mass density, electron fraction, matter temperature, and total neutrino luminosity 
of a HMNS for model S135 at $t-t_{\rm merge} \approx 16.7$~ms in the $x$-$z$ and $x$-$y$ planes,
respectively, at which it already relaxed to a
semi-final quasi-steady state. This shows that the HMNS is weakly
spheroidal and the temperature is high ($T\sim 30$ MeV) in its outer region. 
The neutrino luminosity is also high in its outer region, in particular,
near the polar surface. With the fact that the rest-mass density is
relatively small near the rotation axis above the polar surface, this
is a favorable feature for the merger hypothesis of SGRB; pair
annihilation of neutrinos and anti-neutrinos could supply a large
amount of thermal energy which may drive a fire ball along the
rotation axis. 
As in \S~\ref{pairGRB}, the neutrino pair annihilation rate is estimated as
\beqn
\dot{E}_{\nu\bar{\nu}} &\sim& 10^{51} \,{\rm ergs/s} 
\left(\frac{50\,{\rm km}}{R_{\rm fun}}\right)
\left(\frac{0.3}{\theta_{\rm fun}}\right)^{2}
\left(\frac{E_{\nu}+E_{\bar{\nu}}}{20\,{\rm MeV}}\right) \nonumber \\
&& \ \ \ \ \ \ \ \ \ \ \  \times \left(\frac{L_{\nu}}{10^{53}\,{\rm ergs/s}}\right)
\left(\frac{L_{\bar{\nu}}}{10^{53}\,{\rm ergs/s}}\right)
\sin^{2} \Theta,
\eeqn
which would be sufficient for driving SGRBs.
The pair annihilation efficiency has been
approximately estimated in the previous
works~\cite{Ruffert,Rosswog,Setiawan,Dessart09} and our result is
consistent with these works.

Possible reasons that HMNS are formed are; (i) it is rapidly rotating with
the period $\sim 1$~ms, and hence, the centrifugal force increases the
possible mass that can be sustained; (ii) because it is hot, the thermal
energy enhances the pressure. We find that the rotational velocity
with the period $\sim 1$~ms does not play a substantial role.
Exploring in detail Shen-EOS for the high density tells us that the effect
of the thermal energy is significant and can increase $M_{\rm max}$ by
$\sim 20$--30\% for a high-temperature state with $T \gtrsim 20$~MeV.
This indicates that the HMNS will alive before collapsing to a BH for a long
cooling time $\gtrsim 1$~s.  At the time when the HMNS collapse to a BH,
it will be close to a spherical configuration with low temperature due
to long-term gravitational-wave and neutrino emissions.  Thus,
observable signals from the late-time collapse will not be remarkable.

Figure~\ref{fig_BNS_lum} plots neutrino luminosities as functions of time for
three flavors ($\nu_{e}$, $\bar{\nu}_{e}$, and sum of $\nu_{x}$). 
It is found that electron anti-neutrinos are dominantly emitted
for any model. The reason for this is as follows: The HMNS has a high
temperature, and hence, electron-positron pairs are efficiently
produced from thermal photons, in particular in its envelope.  
Neutrons efficiently capture positrons to emit anti-neutrinos
whereas electrons are not captured by protons as frequently as
positrons because the proton fraction is much smaller.
This hierarchy in the neutrino luminosities was reported 
also in Refs.~\citen{Ruffert,Rosswog}.

Soon after the BH formation for models S16, H135, and H15, $\mu/\tau$
neutrino luminosity steeply decreases because high-temperature regions
are swallowed into the BH, while luminosities of electron neutrinos
and anti-neutrinos decrease only gradually because these neutrinos are
emitted via charged-current processes from the massive accretion
disk. We here note that magnetic fields, which are not taken into
account in the present simulations, could be amplified significantly
in the accretion disk~\cite{LR} and may play a role in the late
evolution of the BH-disk system.

The {\em anti-neutrino} luminosity for the long-lived HMNS is $L_{\bar
\nu} \sim 1.5$--$3\times 10^{53}$~ergs/s with a small time variability. 
It is by a factor of $\sim 1$--5 larger than that from proto-neutron
stars formed after supernovae~\cite{SpheGR}, while it is comparable to the luminosities
for UN100-rigid model.  The averaged neutrino energy is
$\epsilon_{\bar\nu} \sim 20$--30~MeV.  The sensitivity of
water-Cherenkov neutrino detectors such as Super-Kamiokande and future
Hyper-Kamiokande have a good sensitivity for such high-energy
neutrinos in particular for electron anti-neutrinos~\cite{SM2009}.  The
detection number for electron anti-neutrinos is approximately estimated
by $\sigma \Delta T L_{\bar\nu}/(4\pi D^2 \epsilon_{\bar\nu})$ where
$\sigma$ is the total cross section of the detector against target
neutrinos, $\Delta T$ is the lifetime of the HMNS, and $D$ is the
distance to the HMNS.  For a one-Mton detector such as Hyper-Kamiokande, the
expected detection number is $\gtrsim 10$ for $D \lesssim 5$~Mpc with $\Delta
T\sim 2$--3~s, based on an analysis of Ref.~\citen{SM2009}.  Thus, if the
BNS merger fortunately happens within $D \sim 5$~Mpc, neutrinos from
the HMNS may be detected and its formation may be confirmed.  Note
that gravitational waves from the HMNS will be simultaneously detected
for such a close event (see below), reinforcing the confirmation of
the HMNS formation.

\begin{figure}[p]
\begin{center}
    \includegraphics[scale=1.1]{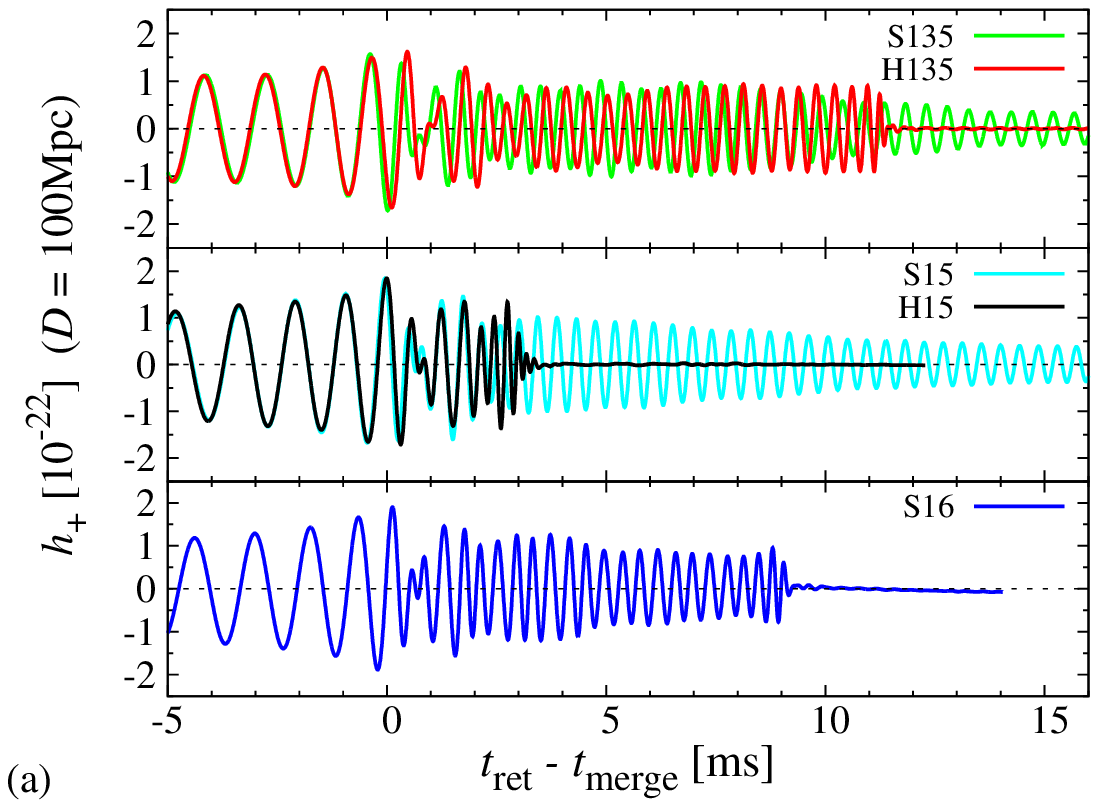}
    \includegraphics[scale=1.01]{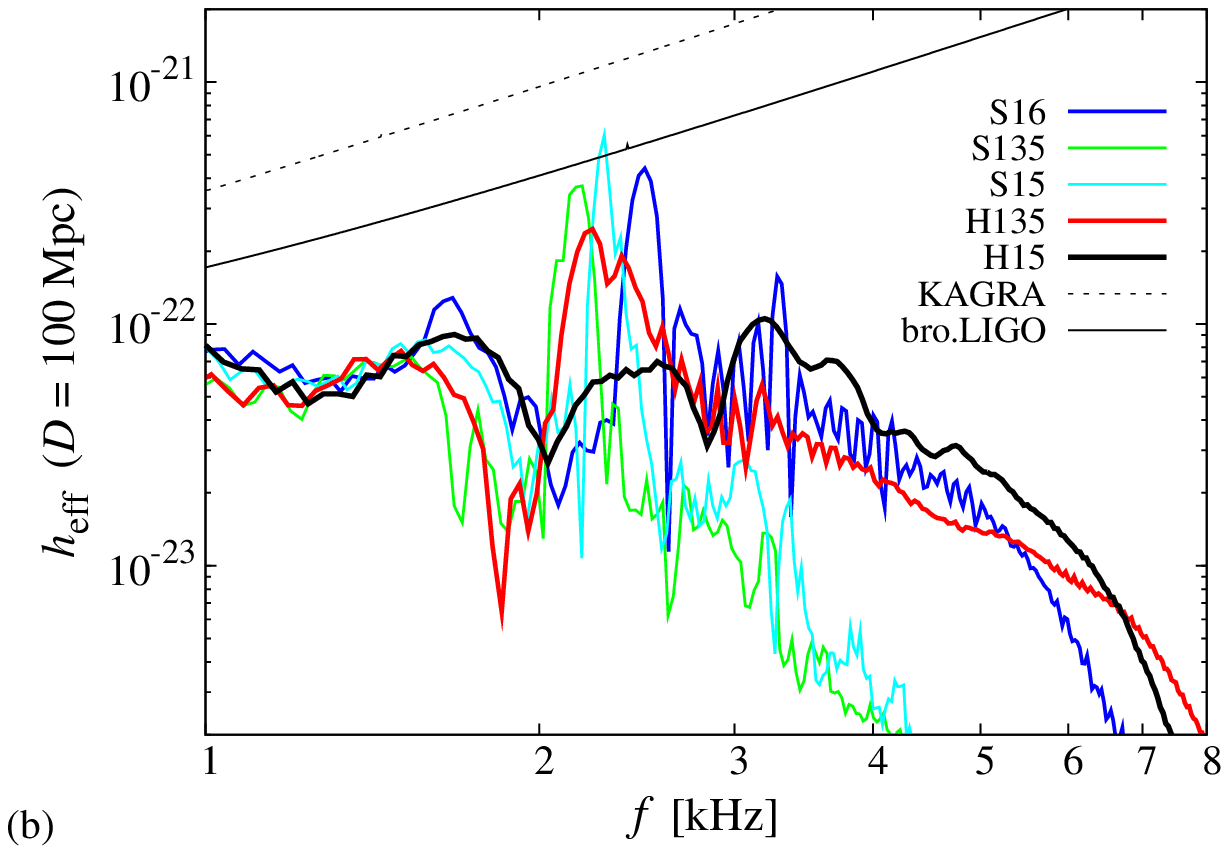}
\end{center}
\vspace{-2mm}
\caption{
(a) Gravitational waves observed along the axis perpendicular to the
orbital plane for the hypothetical distance to the source $D=100$~Mpc
for all the models. (b) The effective amplitude of gravitational waves
as a function of frequency for $D=100$~Mpc.  The noise amplitudes of
a broadband configuration of Advanced LIGO
(bro.~LIGO) and KAGRA are shown together.
\label{fig_BNS_GW1}}
\end{figure}

\begin{figure}[th]
\begin{center}
  \includegraphics[scale=1.1]{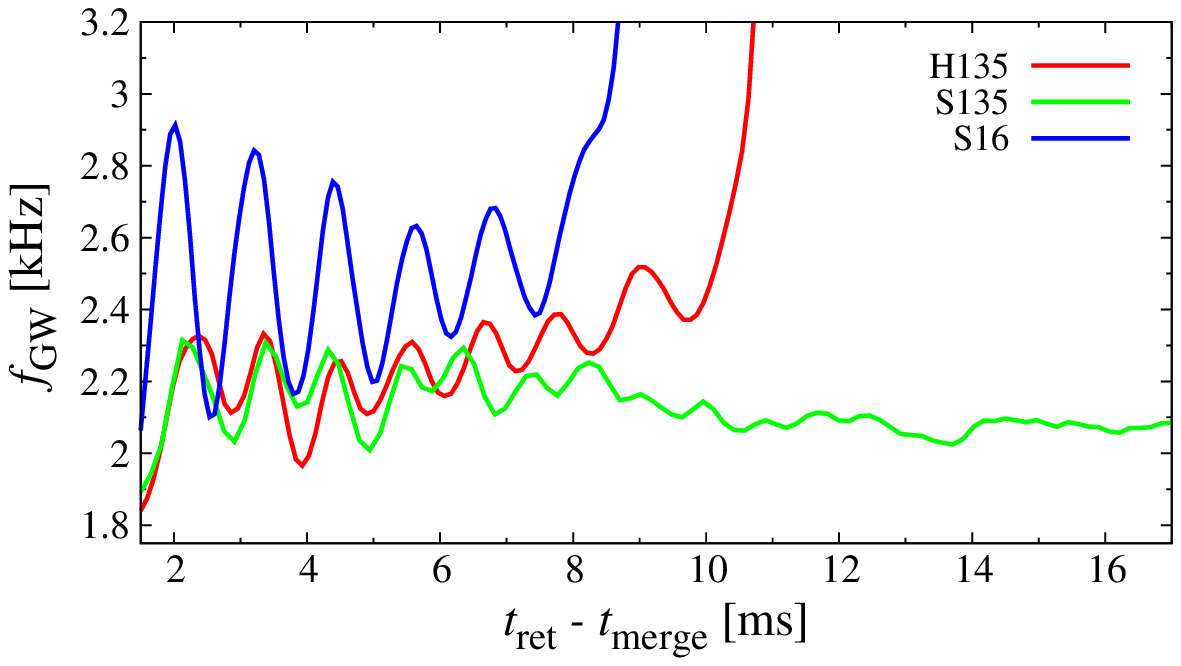}
\end{center}
 \vspace{-3mm}
  \caption{$f_{\rm GW}(t)$ in the HMNS evolution phase, smoothed by a weighted
  spline, for models H135, S135, and S16.\label{fig_BNS_GW2}}
\end{figure}

\subsection{Gravitational waves}

Figure \ref{fig_BNS_GW1}(a) plots the plus mode ($h_{+}$) of gravitational
waves as a function of $t_{\rm ret}-t_{\rm merge}$ where $t_{\rm ret}$
is the retarded time, $t_{\rm ret}=t-D-2M {\rm log}(D/M)$ ($M=2M_{\rm
NS}$). Gravitational waves are extracted from the metric through the
outgoing component of the complex Weyl scalar, $\Psi_4$, in the local wave
zone. The waveforms are composed of the so-called chirp waveform,
which is emitted when the BNS is in an inspiral motion (for $t_{\rm
ret} \lesssim t_{\rm merge}$), and the merger waveform (for $t_{\rm ret}
\gtrsim t_{\rm merge}$). Gravitational waves from the inspiral phase 
(for $t_{\rm ret} \lesssim t_{\rm merge}$) agree well with each other for
the models with Hyp-EOS and Shen-EOS for the same mass. On the other
hand, quasi-periodic gravitational waves from the HMNS (for $t_{\rm
ret} \gtrsim t_{\rm merge}$) show several differences.  First, the amplitude of
quasi-periodic gravitational waves damps steeply at the BH formation
for H135 and H15. This is because the HMNS collapse to a BH before
relaxing to a stationary spheroid.  Second, the characteristic
gravitational-wave frequency, $f_{\rm GW}$, {\em increases} with time
for Hyp-EOS models, while it is approximately constant for Shen-EOS
models with $f_{\rm peak} \approx 2.0$--2.5~kHz which depends weakly on
$M$.  These facts are clearly observed in the effective amplitude (see
Fig.~\ref{fig_BNS_GW1}(b)) defined by $h_{\rm eff}(f) \equiv 0.4 f |h(f)|$
where $h(f)$ is the Fourier transform of $h_+ -i h_{\times}$ with
$h_{\times}$ being the cross mode and the factor 0.4 comes from taking
the average in terms of the random direction to the source and rotational axis of
the HMNS.  Reflecting a shorter lifetime of the HMNS in Hyp-EOS
models, the peak amplitude of $h_{\rm eff}(f)$ is smaller, in
particular for H15 where the HMNS survives only for a short period
$\sim 3$~ms.  Reflecting the shift of the characteristic frequency,
the prominent peak in $h_{\rm eff}$ for Hyp-EOS models (H135 and H15)
is broadened.  The reason for this is described as follows in more
detail.

In the case that hyperons are absent, the HMNS slightly contract
during their evolution simply due to the angular momentum loss
(weakening centrifugal force).  By contrast, in the case that hyperons
are present, $X_{\Lambda}$ increases with the contraction of the HMNS,
resulting in the relative reduction of the pressure. As a result, the
HMNS contracts by a larger fraction.  Recent studies showed that
$f_{\rm GW}$ is associated with the frequency of an $f$-mode which is
approximately proportional to $\sqrt{M_{\rm H}/R_{\rm H}^{3}}$ where
$M_{\rm H}$ and $R_{\rm H}$ are the mass and radius of the
HMNS~\cite{oscillation}. This indicates that $f_{\rm GW}$ should
increase with time.  To see that this is indeed the case, we show
$f_{\rm GW} (\equiv d\phi_{\rm NP}/dt)$ calculated from $\Psi_{4}
\equiv |Psi_{4}| e^{i\phi_{\rm NP}}$ in the HMNS phase for H135, S135, and
S16 in Fig.~\ref{fig_BNS_GW2}.  It is clearly seen that the mean value of
$f_{\rm GW}$ is approximately constant for Shen-EOS models; $f_{\rm
GW} \approx 2.1$ and 2.5 kHz for S135 and S16, respectively.  By
contrast, $f_{\rm GW}$ for H135 increases with time (from $f_{\rm
GW}\approx 2.0$ kHz at $t_{\rm ret}-t_{\rm merge}=2$ ms to $\approx
2.5$ kHz at $t_{\rm ret}-t_{\rm merge}=10$ ms) as the HMNS becomes
compact.

Figure \ref{fig_BNS_GW1}(b) shows that if a HMNS with the lifetime $\gtrsim
10$~ms is formed, the effective amplitude $h_{\rm eff} \sim 4$--$6
\times 10^{-22}$ at $D=100$~Mpc for Shen-EOS models and $2 \times
10^{-22}$ at $D=100$~Mpc for H135 with $f_{\rm peak} \approx
2.0$--2.5~kHz which depends weakly on $M$ and EOS. For the detection,
Hyp-EOS is obviously unfavored. By contrast, for Shen-EOS, the maximum
amplitude for a hypothetical distance of 100 Mpc is as high as the
sensitivity curve of a specially-designed version of advanced
gravitational-wave detectors such as broadband LIGO~\cite{LIGOnoise}, which has a good
sensitivity for a high-frequency band. This suggests that
gravitational waves from the HMNS oscillations may be detected with
${\rm S/N}=5$ if $D \lesssim 20$~Mpc. If the source is located in an optimistic
direction, the detection with ${\rm S/N}=5$ may be possible for $D=50$~Mpc.

\section{Summary} \label{Sec_Summary}

We have described our latest results of numerical-relativity simulations
of rotating stellar core collapses to a BH and BNS mergers,
performed incorporating a finite-temperature EOS and neutrino
cooling effects. The following is the summary of our latest findings and 
prospects for the near future.  

\subsection{Stellar core collapse}

We presented our latest results of axisymmetric simulations of very
massive stellar core collapsing to a system composed of a rotating BH and 
surrounding disk/torus in full general relativity.
The simulation were performed taking into account of the
microphysical processes and the neutrino cooling. 
Because progenitor models of LGRBs suggested in the literatures~\cite{Fryer07} propose a
possibility that they may have an entropy higher than that of ordinary supernova cores,
we employed a model of a presupernova core with a high entropy of $s/k_B = 4$ calculated by
Umeda and Nomoto~\cite{UN2008} together with hypothetical two rotational profiles 
(UN100-rigid and UN100-diff).

As in the collapse of ordinary supernova cores, the gravitational collapse sets in due to 
the photo-dissociation of heavy nuclei and the electron capture.
The collapsing core eventually experiences a core bounce, forming a shock wave. 
Then a HMNS, which is supported by the centrifugal force
and the thermal pressure is, formed. 
The neutrino luminosity in this HMNS phase is larger than than that in the ordinary
supernova as $\gtrsim 5\times 10^{53}$ ergs/s.
The HMNS eventually collapses to a BH 
irrespective of the initial rotational profiles. 
However, the dynamics and the geometry of
the final outcome depend strongly on the degree of the initial rotation.

For the model UN100-rigid, the shock wave formed at the core bounce is deformed to
be a torus-like shape. Then the infalling materials are accumulated in the central
region after they pass through the oblique shock formed at the tours-shaped outer region
of the HMNS. As a result, the thermal energy 
is efficiently stored at the surface of the HMNS due to the dissipation of the
kinetic energy of the accumulated materials, driving outflows.
After the collapse of the HMNS, a torus is formed around the BH.
We found that the torus shows a time variability.
The total neutrino luminosity emitted from
the torus around the BH amounts to $L_{\nu, {\rm tot}}\sim 10^{51}$--$10^{52}$ ergs/s, which lasts
for a long duration of $\gtrsim 1$ s.
Associated with the time variability of the BH-torus system,
the neutrino luminosities also show a violent time variability. 
Such a long-term high luminosity with the time variation may be related to 
the time variability that LGRBs show.
For the model UN100-diff, by contrast, a geometrically thin disk is formed
around the BH and the BH-disk system shows essentially no time variability.
Remarkably, the above differences in the dynamics and the outcome stem
from a small difference in the initial rotational profile.

We also calculated the characteristic gravitational-wave strain $h_{\rm char}$ 
for UN100-rigid. The effective amplitude is as large as $\sim 10^{-20}$ at $f \sim 1$ kHz for
a hypothetical event occurred at a distance of 10 kpc. 
This shows a possibility that we may observe multi-messenger information,
namely, gravitational waves, neutrinos, and electromagnetic radiation from 
such a nearby event and may obtain a clue to understand the stellar core collapse
and the central engine of LGRBs.

\subsection{Binary neutron star merger}

We showed that for a stiff, purely nucleonic EOS, a
HMNS is the canonical outcome and a BH is not promptly formed after the
onset of the merger as long as the total mass of the system is smaller
than $3.2M_{\odot}$. The primary reason is that the thermal pressure plays
an important role for sustaining the HMNS.  We further showed that the
lifetime of the formed HMNS with mass $\lesssim 3M_{\odot}$ would be much longer
than its dynamical time scale, i.e., $\gg 10$~ms, and will be
determined by the time scale of the subsequent neutrino cooling.  The neutrino luminosity
in the early evolution phase of the HMNS was shown to be high as 
$\sim 3$--$10\times 10^{53}$~ergs/s.  The effective amplitude of
gravitational waves averaged over the random source direction and orbital plane
inclination is $h_{\rm eff}=4$--$6 \times 10^{-22}$ at $f_{\rm
peak}=2.1$--2.5~kHz for a hypothetical source distance of $D=100$~Mpc.
If the BNS merger happens at a relatively short source distance $\sim
20$~Mpc or is located in an optimistic direction with $D \sim 50$~Mpc,
such gravitational waves may be detected by advanced
gravitational-wave detectors with ${\rm S/N}=5$, and the HMNS formation will be
confirmed.

For an EOS in which effects of hyperons are taken into account, the
EOS becomes softer than the purely nucleonic EOS. With this EOS, a BH
is often formed in a short time scale after the onset of the merger,
although a HMNS could be a transient outcome with a short lifetime
$\lesssim 10$~ms. Because the EOS becomes soft during the evolution of the
HMNS, the compactness significantly changes in a short time scale in
this EOS. This is well reflected in gravitational waveforms and their
spectra.  Specifically, the characteristic frequency changes with
time.  This effect reduces the amplitude at a peak frequency of
gravitational waves in the Fourier space, and make a feature
unfavorable for the detection of gravitational waves.  Roughly
speaking, the allowed distance for the detection of gravitational
waves from the HMNS is by a factor of 2 smaller than that in the
nucleonic EOS for the same mass of BNS.

\subsection{Future prospects}

\subsubsection{Massive stellar core collapse}

As mentioned in \S~\ref{Sec_ResultSCC}, we did not take into account effects of 
the neutrino heating in the numerical simulations to date.
Recently, we have developed~\cite{SKSS2011} a formulation for numerical simulations of 
general relativistic radiation transfer based on Thorne's moment formalism~\cite{Thorne1981}.
Based on this formalism, we have already performed general relativistic radiation 
magnetohydrodynamics (GRRMHD) simulations~\cite{SS2012} for the evolution of a system 
composed of a BH and a surrounding torus with 
a simplified treatment of microphysics, as a step toward a more physical modelling. 
Furthermore, we have succeeded in implementing a code which can solve the neutrino transfer 
with a detailed microphysics (in preparation).
Using this code, we plan to perform simulations of the stellar core collapse to 
explore a supernova explosion mechanism and the formation of a BH in full general
relativity. 

Throughout the study of the massive stellar collapse in this article,
we assume the axial symmetry in the simulations. However, this
assumption would be invalid if non-axisymmetric instabilities set in.
For example, in Ref.~\citen{Kiuchi2011}, it has been shown that the
BH-torus systems could exhibit the so-called Papaloizou-Pringle
instability~\cite{PP}. Also, for a rapidly rotating HMNS,
non-axisymmetric instabilities such as a bar-mode instability could set
in~\cite{SS05,Rampp98,Ott3D,Simon,LLR}.  Once these instabilities turn on, the torus/HMNS may
deform to be a highly non-axisymmetric structure. This will enhance
the angular momentum transport in the torus and HMNS, and the evolution
processes of these systems may be modified.  We plan to perform a
three dimensional simulation to explore if the non-axisymmetric
instabilities set in and play an important role for the collapsar
models. 

\subsubsection{Binary neutron star merger}

To date, we performed the BNS merger simulations for the case of the equal-mass binaries.
As a straightforward extension of the previous studies, we plan to perform simulations 
for unequal-mass binaries, for which the merger dynamics, gravitational waveforms, and the mass of the 
disk will be modified. We also plan to perform merger simulations of BHNS binaries.
It has been reported (see Ref.~\citen{Kyutoku2011} and references therein) that a massive disk of 
$M_{\rm disk}\gtrsim 0.1M_{\odot}$ can be formed for stiff EOSs even when the BH rotates moderately 
($a_{\rm BH} \gtrsim 0.5 $). Such a system is a promising candidate of the central engine of SGRBs.
Only simulations with detailed microphysics will enable a quantitative study for this merger
hypothesis of SGRB.

In the study of the BNS merger simulation, we totally ignore the effect of magnetic fields. 
Recent studies on magnetized BNS merger showed that the magnetic fields in the merger 
and post-merger phases have an impact on the dynamics of the torus formed around the BH~\cite{LR}. 
This is because the angular velocity inside the torus has a steep gradient. 
Thus, the magnetic field is subject to the amplification via magnetic winding and/or magneto-rotational 
instability~\cite{Balbus}. 
This condition holds in the HMNS because it has strong and rapid differential rotation as discussed in 
\S~4.2. 
We plan to incorporate magneto-hydrodynamics in our code and explore its impact on the evolution 
of HMNSs.

\section*{Acknowledgments}
We thank H.~Umeda for providing us the presupernova model (UN100) adopted 
in this work. 
Numerical simulations were performed on SR16000
at YITP of Kyoto University, on SX9 and XT4 at CfCA of NAOJ, and 
on the NEC SX-8 at RCNP in Osaka University. 
This work was supported by Grant-in-Aid for Scientific Research (21018008,
21105511, 21340051, 21684014, 22740178, 23740160), Grant-in-Aid on Innovative
Area (20105004), and HPCI Strategic Program of Japanese MEXT.

\appendix

\section{Electron and positron captures}\label{A_EPC} \label{EleCap}

In this section, we briefly summarize our methods of handling  electron and positron
captures based on Ref.~\citen{Fuller85}, and give the explicit
forms of $\gamma^{\rm ec}_{\nu_e}$, $\gamma^{\rm pc}_{\bar{\nu}_e}$,
$Q^{\rm ec}_{\nu_e}$, and $Q^{\rm pc}_{\bar{\nu}_e}$ in 
Eqs. (\ref{gnlocal}), (\ref{galocal}), (\ref{Qnlocal}), and (\ref{Qalocal}),
for completeness

\subsection{The electron and positron capture rates 
$\gamma^{\rm ec}_{\nu_e}$ and $\gamma^{\rm pc}_{\bar{\nu}_e}$}
The 'net' electron fraction is written as $Y_{e} = Y_{-} - Y_{+}$ 
where $Y_{-}$ ($Y_{+}$) denotes the number of electrons (positrons) 
per baryon including pair electrons. 
Then the electron-neutrino number emission rate by the electron capture
and the electron-anti-neutrino number emission rate by the positron capture
are given by 
\beqn
&& \gamma^{\rm local}_{\nu_e} = -\dot{Y}_{-} = 
  -(\dot{Y}_{-}^{f} +\dot{Y}_{-}^{h}), \\
&& \gamma^{\rm local}_{\bar{\nu}_e} = -\dot{Y}_{+} = 
  -(\dot{Y}_{+}^{f} +\dot{Y}_{+}^{h}),
\eeqn
where the electron and positron capture rates are decomposed into two
parts, capture on by free nucleons (with the superscript $f$) and on heavy
nuclei (with the superscript $h$). 
In the following, we will present the explicit forms of 
$\dot{Y}_{-}^{f}$, $\dot{Y}_{+}^{f}$, $\dot{Y}_{-}^{h}$, and
$\dot{Y}_{+}^{h}$.

\subsection{Capture on free nucleons $\dot{Y}^{f}$}

The electron capture rate (including the contribution of the inverse
reaction of the neutrino capture) on free nucleons ($\dot{Y}_{-}^{f}$) is
given by 
\beq
\dot{Y}_{-}^{f} = X_{n}\lambda^{\nu_{e} {\rm c},f} 
-X_{p}\lambda^{{\rm ec},f},
\label{dot_Yef}
\eeq
where $\lambda^{{\rm ec},f}$ is the specific 
electron capture rate on free protons, 
$\lambda^{\nu_{e} {\rm c},f}$ is the specific electron-neutrino capture
rate on free neutrons, 
and  $X_{p}$ and $X_{n}$ are the mass fraction of free protons and neutrons,
respectively. 
Based on a balance argument\cite{Fuller85}, 
one can show that $\lambda^{\nu_{e} {\rm c},f}$
is related to $\lambda^{{\rm ec},f}$ by 
\beq 
\lambda^{\nu_{e} {\rm c},f} = 
\exp \left(
\eta_{\nu_e} -\eta_{e} - \frac{\delta m}{k_{B}T}
\right)
\lambda^{{\rm ec},f},
\label{ecnc}
\eeq
where $\eta_{\nu_ e}$ and $\eta_{e}$ are the chemical potentials of
electron neutrinos and electrons in units of $k_{B}T$ and 
$\delta m = (m_{n} - m_{p})c^{2}$.
Furthermore, we use the following Saha's relation for non-degenerate free nucleons,
\beq
X_{n} \approx X_{p} 
\exp \left( \eta_{n}-\eta_{p}+\frac{\delta m}{k_{B}T} \right), 
\eeq
where $\eta_{n}$ and $\eta_{p}$ are the chemical potentials of
free neutrons and protons in units of $k_{B}T$.
Then we obtain
\beq
\dot{Y}_{-}^{f} = \left[
\exp \left(\eta_{\nu_e}-\eta_{e} + \eta_{n}-\eta_{p} \right)
-1 \right] X_{p}\lambda^{{\rm ec},f}.
\eeq

The positron capture rate 
(including the contribution of the inverse reaction) on free nucleons is
similarly given by   
\beq
\dot{Y}_{+}^{f} = X_{p}\lambda^{\bar{\nu}_{e} {\rm c},f} -X_{n}\lambda^{{\rm pc},f}
= 
\left[ \exp 
\left(\eta_{\bar{\nu}_e} +\eta_{e} + \eta_{p}-\eta_{n} \right)
- 1 \right] X_{n} \lambda^{{\rm pc},f}, \label{dot_Ypf}
\eeq
where $\eta_{\bar{\nu}_e}$ is the chemical potential of
electron-anti-neutrinos in units of $k_{B}T$, 
$\lambda^{\rm pc}$ is the specific positron capture rate on free
neutrons, and $\lambda^{\bar{\nu}_{e} {\rm c},f}$ is the specific
electron-anti-neutrino capture rate on free protons.

\subsection{Capture on heavy nuclei $\dot{Y}^{h}$}

The electron capture rate (including the contribution of the inverse
reaction of the neutrino capture) on a heavy nucleus of mass number 
$A$ ($\dot{Y}_{-}^{h}$) is given by\cite{Fuller85}
\beq
\dot{Y}_{-}^{h} = \frac{X_{D}}{A}\lambda^{\nu_{e} {\rm c}, h} 
                 -\frac{X_{P}}{A}\lambda^{{\rm ec}, h}, 
\label{dot_Yeh}
\eeq
where $\lambda^{{\rm ec},h}$ is the specific electron capture rate on the parent
nucleus (mass fraction $X_{P}$),
$\lambda^{\nu_{e} {\rm c},h}$ is the specific electron-neutrino capture rate on the
daughter nucleus (mass fraction $X_{D}$),
and $A$ is the atomic mass of the parent and daughter nuclei.
In the present simulations, we set $X_{D} = X_{P} = X_{A}$.
Then, under the assumption of a nuclear statistical equilibrium,
one may approximate the capture rate on heavy nuclei as\cite{Fuller85},
\beq
\dot{Y}_{-}^{h} 
\approx
\left[  
\exp \left(
\eta_{\nu_e} - \eta_{e} + \eta_{n}-\eta_{p}
\right) - 1 
\right] \frac{X_{A}}{A} \lambda^{{\rm ec},h}. 
\label{dot_Yeh3}
\eeq

Similarly, the positron capture rate (including the contribution of the
inverse reaction) on heavy nuclei ($\dot{Y}_{+}^{h}$) is given by 
\beq
\dot{Y}_{+}^{h} = \frac{X_{D}}{A}\lambda^{\bar{\nu}_{e} {\rm c}, h} 
                 -\frac{X_{P}}{A}\lambda^{{\rm pc}, h}
\approx
\left[  \exp \left(
\eta_{\bar{\nu}_e} + \eta_{e} + \eta_{p}-\eta_{n}
\right) - 1
\right]\frac{X_{A}}{A}\lambda^{{\rm pc},h} .
\label{dot_Yph}
\eeq

\subsection{The specific capture rate $\lambda$}

The specific electron and positron capture rates on 
free nucleons and on heavy nuclei are written 
in the same form as\cite{Fuller85}
\beqn
&& 
\lambda^{{\rm ec}, f} = 
\frac{\ln 2}{\langle ft \rangle _{\rm eff}^{{\rm ec},f}} I^{{\rm ec},f},
\ \ \ \ \ 
\lambda^{{\rm pc}, f} = 
\frac{\ln 2}{\langle ft \rangle _{\rm eff}^{{\rm pc},f}} I^{{\rm pc},f}, 
\\ 
&&
\lambda^{{\rm ec}, h} = 
\frac{\ln 2}{\langle ft \rangle _{\rm eff}^{{\rm ec},h}} I^{{\rm ec},h},
\ \ \ \ \ 
\lambda^{{\rm pc}, h} = 
\frac{\ln 2}{\langle ft \rangle _{\rm eff}^{{\rm pc},h}} I^{{\rm pc},h}, 
\eeqn
where $I^{{\rm ec},f}$ and $I^{{\rm pc},f}$ are the phase space factors
for the electron and positron captures on free electrons,
and $I^{{\rm ec},h}$ and $I^{{\rm pc},h}$ are those on heavy nuclei.
$\langle ft \rangle _{\rm eff}$'s are the effective
$ft$-values introduced by Fuller et al.\cite{Fuller85}, 
which is essentially the same as the square of 
the nuclear transition matrix.

The phase space factors are given by
\beqn
\!\!\!\!\!\!\!\!\!
I^{{\rm ec},f} \!&=&\!
\left( \frac{k_{B}T}{m_{e}c^{2}} \right)^{5}
\int_{\eta_{0}}^{\infty} 
\eta^{2}(\eta + \zeta^{{\rm ec},f})^{2}
\frac{1}{1+e^{\eta - \eta_{e}}}
\left[
1- \frac{1}{1+e^{\eta -\eta_{\nu_e} +\zeta^{{\rm ec},f} }}
\right]
d\eta, \label{def_Ie}
\\
\!\!\!\!\!\!\!\!\!
I^{{\rm pc},f} \!&=&\! 
\left( \frac{k_{B}T}{m_{e}c^{2}} \right)^{5}
\int_{\eta_{0}}^{\infty} 
\eta^{2}
(\eta + \zeta^{{\rm pc},f})^{2}
\frac{1}{1+e^{\eta + \eta_{e}}}
\left[
1- \frac{1}{1+e^{\eta -\eta_{\bar{\nu}_e} +\zeta^{{\rm pc},f} }}
\right]
d\eta, \label{def_Ip}
\\
\!\!\!\!\!\!\!\!\!
I^{{\rm ec},h} \!&=&\!
\left( \frac{k_{B}T}{m_{e}c^{2}} \right)^{5}
\int_{\eta_{0}}^{\infty} 
\eta^{2}(\eta + \zeta^{{\rm ec},h})^{2}
\frac{1}{1+e^{\eta - \eta_{e}}}
\left[
1- \frac{1}{1+e^{\eta -\eta_{\nu_e} +\zeta^{{\rm ec},h} }}
\right]
d\eta, \label{def_Ieh}
\\
\!\!\!\!\!\!\!\!\!
I^{{\rm pc},h} \!&=&\! 
\left( \frac{k_{B}T}{m_{e}c^{2}} \right)^{5}
\int_{\eta_{0}}^{\infty} 
\eta^{2}
(\eta + \zeta^{{\rm pc},h})^{2}
\frac{1}{1+e^{\eta + \eta_{e}}}
\left[
1- \frac{1}{1+e^{\eta -\eta_{\bar{\nu}_e} +\zeta^{{\rm pc},h} }}
\right]
d\eta, \label{def_Iph}
\eeqn 
where $\zeta^{{\rm ec},f}$, $\zeta^{{\rm pc},f}$, 
$\zeta^{{\rm ec},h}$, and $\zeta^{{\rm pc},h}$ are the nuclear mass-energy
differences for the electron and positron captures in units of $k_{B}T$.
The superscripts 'f' and 'h' again denote free nucleons and heavy nuclei. 
The nuclear mass-energy differences for the capture on free nuclei are given by
\beq
\zeta^{{\rm ec},f} 
= -\zeta_{n}^{{\rm pc},f}
\approx 
\eta_{p} - \eta_{n}.
\eeq
We follow Fuller et al.~\cite{Fuller85} for the nuclear
mass-energy differences in the capture on heavy nuclei:
In the case of $N<40$ or $Z>20$ (referred to as 'unblocked' case),
we set
\beq
\zeta^{{\rm ec},h} 
= -\zeta^{{\rm pc},h}
\approx 
\eta_{p} - \eta_{n}.
\eeq
In the case of $N\ge40$ or $Z\le20$ (referred to as 'blocked' case),
on the other hand, we set
\beqn
\zeta^{{\rm ec},h} 
&\approx&
\eta_{p}-\eta_{n} - \frac{5\,{\rm MeV}}{k_{B}T}, \\
\zeta^{{\rm pc},h} 
&\approx& 
-\eta_{p}+\eta_{n} + \frac{5\, {\rm MeV}}{k_{B}T}.
\eeqn
Then, the threshold value of the electron and positron captures is given by
$\eta_{0} = m_{e}c^{2}/(k_{B}T)$ for $\zeta > -m_{e}c^{2}/(k_{B}T)$
and $\eta_{0} = |\zeta|$ for $\zeta < -m_{e}c^{2}/(k_{B}T)$ where
we have dropped the superscripts 'ec', 'pc', '$f$', and '$h$' in $\zeta$
for simplicity.

The effective $ft$-value of the electron or positron capture on free nuclei
is given by (e.g. Ref.~\citen{Fuller85}
\beq
\log_{10} \langle ft \rangle_{\rm eff}^{{\rm ec},f} =
\log_{10} \langle ft \rangle_{\rm eff}^{{\rm pc},f} \approx 3.035.
\eeq
We follow Fuller et al.\cite{Fuller85} for the effective $ft$-value of
the capture on heavy nuclei, who proposed to use 
\beqn
&&
\log_{10} \langle ft \rangle_{\rm eff}^{{\rm ec},h} \approx \left\{
\begin{array}{ccc}
3.2 & {\rm unblocked} & \eta_{e}<|\zeta^{{\rm ec},h}|  \\
2.6 & {\rm unblocked} & \eta_{e}>|\zeta^{{\rm ec},h}|  \\
2.6 + \frac{25.9}{T_{9}} & {\rm blocked} &
\end{array}
\right. , \label{eft1} \\
&&
\log_{10} \langle ft \rangle_{\rm eff}^{{\rm pc},h} \approx \left\{
\begin{array}{ccc}
3.2 & {\rm unblocked} & \eta_{e}<|\zeta^{{\rm pc},h}|  \\
2.6 & {\rm unblocked} & \eta_{e}>|\zeta^{{\rm pc},h}|  \\
2.6 + \frac{25.9}{T_{9}} & {\rm blocked} &
\end{array}
\right. , \label{eft2}
\eeqn
where $T_{9} = T/(10^{9}K)$. 
In this expression, the thermal unblocking effect\cite{CW84}
is readily taken into account.
In the thermal unblocking, it costs $\approx 5.13$ MeV to remove a neutron
from a filled orbital 1$f_{5/2}$ and place it in the
$gd$-shell\cite{Fuller85}. 

\subsection{Energy emission rates 
$Q^{\rm ec}_{\nu_e}$ and $Q^{\rm pc}_{\bar{\nu}_e}$} 

The neutrino energy emission rates associated with 
the electron and positron captures in units of $m_{e}c^{2}$ s$^{-1}$ 
are given by\cite{Fuller85}
\beq
\pi^{\rm ec} = \ln 2 \frac{J^{\rm ec}}
{\langle ft \rangle_{\rm eff}^{\rm ec}}, \ \ \ \ \ \ 
\pi^{\rm pc} = \ln 2 \frac{J^{\rm pc}}
{\langle ft \rangle_{\rm eff}^{\rm pc}},
\label{pi_epc}
\eeq
where the phase space factors are given by
\beqn
J^{\rm ec} &=& \left(\frac{k_{B}T}{m_{e}c^{2}}\right)^{6}
\int_{\eta_{0}}^{\infty}
\eta^{2}(\eta + \zeta^{\rm ec})^{3}
\frac{1}{1+e^{\eta - \eta_{e}}}
\left[ 1- 
\frac{1}{1-e^{\eta -\eta_{\nu_e} + \zeta^{\rm ec}}}\right]d\eta 
\label{J_ec},\\
J^{\rm pc} &=& \left(\frac{k_{B}T}{m_{e}c^{2}}\right)^{6}
\int_{\eta_{0}}^{\infty}
\eta^{2}(\eta + \zeta^{\rm pc})^{3}
\frac{1}{1+e^{\eta + \eta_{e}}}
\left[ 1- 
\frac{1}{1-e^{\eta -\eta_{\bar{\nu}_e} + \zeta^{\rm pc} }}\right]d\eta .
\label{J_pc}
\eeqn
In Eqs. (\ref{pi_epc})--(\ref{J_pc}), we have dropped the superscripts
'$f$' and '$h$' in $\pi^{\rm ec}$, $\pi^{\rm pc}$, $J^{\rm ec}$, 
$J^{\rm pc}$, $\langle ft \rangle_{\rm eff}^{\rm ec}$, 
$\langle ft \rangle_{\rm eff}^{\rm pc}$, $\zeta^{\rm ec}$, and
$\zeta^{\rm pc}$ for simplicity.

The average energy of the electron neutrinos produced by electron 
and positron captures is defined, in units of $m_{e}c^{2}$, as
\beq
\langle \epsilon_{\nu_e} \rangle^{\rm ec} = 
\frac{J^{\rm ec}}{I^{\rm ec}}, \ \ \ \ \ \ 
\langle \epsilon_{\bar{\nu}_e} \rangle^{\rm pc} = 
\frac{J^{\rm ec}}{I^{\rm pc}}.
\eeq
Then, the local neutrino energy emission rates by the electron and
positron captures per unit volume is given by 
\beqn
Q_{\nu e}^{\rm ec} &=& \frac{\rho}{m_{u}}
\left[\,
  X_{p} \langle \epsilon_{\nu_e} \rangle^{{\rm ec},f} 
  \lambda^{{\rm ec}, f}
+ \frac{X_{A}}{A} \langle \epsilon_{\nu_e} \rangle^{{\rm ec},h}
  \lambda^{{\rm ec}, h}
\right], \\
Q_{\bar{\nu} e}^{\rm pc} &=& \frac{\rho}{m_{u}}
\left[\,
 X_{n} \langle \epsilon_{\bar{\nu}_e} \rangle^{{\rm pc},f} 
  \lambda^{{\rm pc}, f}
+\frac{X_{A}}{A} \langle \epsilon_{\bar{\nu}_e} \rangle^{{\rm pc},h}
 \lambda^{{\rm pc}, h}\, 
\right].
\eeqn

\section{Neutrino pair processes} \label{A_nupair}

In this section, we briefly summarize our method of handling the pair processes of
the neutrino emission and give the explicit forms of 
$\gamma^{\rm pair}_{\nu_{e}\bar{\nu}_e}$, 
$\gamma^{\rm plas}_{\nu_{e}\bar{\nu}_e}$, 
$\gamma^{\rm Brems}_{\nu_{e}\bar{\nu}_e}$, 
$\gamma^{\rm pair}_{\nu_{x}\bar{\nu}_x}$, 
$\gamma^{\rm plas}_{\nu_{x}\bar{\nu}_x}$, 
$\gamma^{\rm Brems}_{\nu_{x}\bar{\nu}_x}$, 
$Q^{\rm pair}_{\nu_{e}\bar{\nu}_e}$, 
$Q^{\rm plas}_{\nu_{e}\bar{\nu}_e}$, 
$Q^{\rm Brems}_{\nu_{e}\bar{\nu}_e}$, 
$Q^{\rm pair}_{\nu_{x}\bar{\nu}_x}$, 
$Q^{\rm plas}_{\nu_{x}\bar{\nu}_x}$, and
$Q^{\rm Brems}_{\nu_{x}\bar{\nu}_x}$
for completeness.

\subsection{Electron-positron pair annihilation}

We follow Cooperstein et al. \cite{Cooperstein86} for computing the rate of
neutrino emission by the electron-positron pair annihilation.
The number emission rate of $\nu_{e}$ or $\bar{\nu}_{e}$ by the
electron-positron pair annihilation can be written as 
\beq
\gamma_{\nu_{e} \bar{\nu}_{e}}^{\rm pair} =
\frac{m_{u}}{\rho}
\frac{C^{\rm pair}_{\nu_{e}\bar{\nu}_e}}{36\pi^{4}} 
\frac{\sigma_{0}c  }{m_{e}^{2}c^{4}}
\frac{(k_{B}T)^{8} }{(\hbar c)^{6}}
F_{3}(\eta_{e})F_{3}(-\eta_{e})
\langle {\rm block} \rangle_{\nu_{e}\bar{\nu}_{e}}^{\rm pair},
\eeq
where $\sigma_{0} \approx 1.705 \times 10^{-44}$cm$^{-2}$ and 
$C^{\rm pair}_{\nu_{e}\bar{\nu}_{e}} = (C_{V}-C_{A})^{2}+(C_{V}+C_{A})^{2}$ 
with $C_{V} = \frac{1}{2} + 2\sin^{2}\theta_{W}$ and $C_{A} = \frac{1}{2}$.
The Weinberg angle is given by $\sin^{2}\theta_{W} \approx 0.23$. 
Using the average energy of neutrinos produced by the pair annihilation,
\beq
\langle \epsilon_{\nu_{e}\bar{\nu}_{e}} \rangle^{\rm pair} =
\frac{k_{B}T}{2}\left(
\frac{F_{4}(\eta_{e})}{F_{3}(\eta_{e})} +
\frac{F_{4}(-\eta_{e})}{F_{3}(-\eta_{e})}
\right),
\eeq
the blocking factor $\langle {\rm block}
\rangle_{\nu_{e}\bar{\nu}_{e}}^{\rm pair}$ is 
evaluated as
\beq
\langle {\rm block} \rangle_{\nu_e \bar{\nu}_{e}}^{\rm pair} \approx
\left[
1+\exp \left(
\eta_{\nu_e} - 
\frac{\langle \epsilon_{\nu_{e}\bar{\nu}_{e}} \rangle^{\rm pair}}
{k_{B}T}
\right)
\right]^{-1} 
\left[
1+\exp \left(
\eta_{\bar{\nu}_e} - 
\frac{\langle \epsilon_{\nu_{e}\bar{\nu}_{e}} \rangle^{\rm pair}}
{k_{B}T}
\right)
\right]^{-1} .
\eeq
The associated neutrino energy emission rate by the pair annihilation is
given by 
\beq
Q_{\nu_{e}\bar{\nu}_{e}}^{\rm pair} =
\frac{\rho}{m_{u}}\gamma_{\nu_{e}\bar{\nu}_{e}}^{\rm pair}
\langle \epsilon_{\nu_{e}\bar{\nu}_{e}} \rangle^{\rm pair}.
\eeq

Similarly, the number emission rate of $\nu_{x}$ or $\bar{\nu}_{x}$ by the
electron-positron pair annihilation and the associated energy emission
rate are given by
\beqn
&&
\gamma_{\nu_{x} \bar{\nu}_{x}}^{\rm pair} =
\frac{m_{u}}{\rho}
\frac{C^{\rm pair}_{\nu_{x}\bar{\nu}_{x}}}{36\pi^{4}} 
\frac{\sigma_{0}c  }{m_{e}^{2}c^{4}}
\frac{(k_{B}T)^{8} }{(\hbar c)^{6}}
F_{3}(\eta_{e})F_{3}(-\eta_{e})
\langle {\rm block} \rangle_{\nu_{x}\bar{\nu}_{x}}^{\rm pair}, \\
&&
Q_{\nu_{x}\bar{\nu}_{x}}^{\rm pair} = 
\frac{\rho}{m_{u}}\gamma_{\nu_{x}\bar{\nu}_{x}}^{\rm pair}
\langle \epsilon_{\nu_{x}\bar{\nu}_{x}} \rangle^{\rm pair},
\eeqn
where $C_{\nu_{x}\bar{\nu}_{x}} = (C_{V}-C_{A})^{2}+(C_{V}+C_{A}-2)^{2}$.
The average neutrino energy and the blocking factor are given by
\beq
\langle \epsilon_{\nu_{x}\bar{\nu}_{x}} \rangle^{\rm pair} =
\langle \epsilon_{\nu_{e}\bar{\nu}_{e}} \rangle^{\rm pair},
\eeq
and 
\beq
\langle {\rm block} \rangle_{\nu_x \bar{\nu}_{x}}^{\rm pair} \approx
\left[
1+\exp \left(
\eta_{\nu_x} - 
\frac{\langle \epsilon_{\nu_{x}\bar{\nu}_{x}} \rangle^{\rm pair}}
{k_{B}T}
\right)
\right]^{-1} 
\left[
1+\exp \left(
\eta_{\bar{\nu}_x} - 
\frac{\langle \epsilon_{\nu_{x}\bar{\nu}_{x}} \rangle^{\rm pair}}
{k_{B}T}
\right)
\right]^{-1},
\eeq
where $\eta_{\bar{\nu}_x} = \eta_{\nu_x}$ because they are produced only by
the pair processes.

\subsection{Plasmon decay}

We follow Ruffert et al. \cite{Ruffert} for computing the pair creation rate of neutrinos
by the decay of transversal plasmons.
The number emission rate of $\nu_{e}$ or $\bar{\nu}_{e}$ can be written as
\beq
\gamma_{\nu_{e}\bar{\nu}_{e}}^{\rm plas} = 
\frac{m_{u}}{\rho}
\frac{C_{V}^{2}}{192 \pi^{3} \alpha_{\rm fine}}
\frac{\sigma_{0}c}{m_{e}^{2}c^{4}}
\frac{(k_{B}T)^{8}}{(\hbar c)^{6}}
\gamma_{p}^{6}e^{-\gamma_{p}}(1+\gamma_{p})
\langle {\rm block} \rangle_{\nu_{e}\bar{\nu}_{e}}^{\rm plas} ,
\eeq
where $\alpha_{\rm fine}\approx 1/137$ is the fine-structure constant and
$\gamma_{p} \approx 2\sqrt{(\alpha_{\rm fine}/9\pi) (\pi^{2}+3\eta_{e})}$.
The blocking factor is approximately given by
\beq
\langle {\rm block} \rangle_{\nu_{e}\bar{\nu}_{e}}^{\rm plas} \approx
\left[
1+\exp \left(
\eta_{\nu_e} - 
\frac{\langle \epsilon_{\nu_{e}\bar{\nu}_{e}} \rangle^{\rm plas}}
{k_{B}T}
\right)
\right]^{-1}
\left[
1+\exp \left(
\eta_{\bar{\nu}_e} - 
\frac{\langle \epsilon_{\nu_{e}\bar{\nu}_{e}} \rangle^{\rm plas}}
{k_{B}T}
\right)
\right]^{-1},
\eeq
where
\beq
\langle \epsilon_{\nu_{e}\bar{\nu}_{e}} \rangle^{\rm plas} =
\frac{k_{B}T}{2}\left(
2 + \frac{\gamma_{p}^{2}}{1+1\gamma_{p}}
\right)
\eeq
is the average energy of neutrinos produced by the plasmon decay.
The associated neutrino energy emission rate is given by
\beq
Q_{\nu_{e}\bar{\nu}_{e}}^{\rm plas} =
\frac{\rho}{m_{u}}\gamma_{\nu_{e}\bar{\nu}_{e}}^{\rm plas}
\langle \epsilon_{\nu_{e}\bar{\nu}_{e}} \rangle^{\rm plas}.
\eeq

Similarly, the number emission rate of $\nu_{x}$ or $\bar{\nu}_{x}$ by the
plasmon decay and the associated energy emission rate are given by
\beqn
&&
\gamma_{\nu_{x}\bar{\nu}_{x}}^{\rm plas} = 
\frac{m_{u}}{\rho}
\frac{(C_{V}-1)^{2}}{192 \pi^{3} \alpha_{\rm fine}}
\frac{\sigma_{0}c}{m_{e}^{2}c^{4}}
\frac{(k_{B}T)^{8}}{(\hbar c)^{6}}
\gamma_{p}^{6}e^{-\gamma_{p}}(1+\gamma_{p})
\langle {\rm block} \rangle_{\nu_{x}\bar{\nu}_{x}}^{\rm plas} , \\
&&
Q_{\nu_{x}\bar{\nu}_{x}}^{\rm plas} =
\frac{\rho}{m_{u}}\gamma_{\nu_{x}\bar{\nu}_{x}}^{\rm plas}
\langle \epsilon_{\nu_{x}\bar{\nu}_{x}} \rangle^{\rm plas}, 
\eeqn
where the average neutrino energy is 
$\langle \epsilon_{\nu_{x}\bar{\nu}_{x}} \rangle^{\rm plas} =
 \langle \epsilon_{\nu_{e}\bar{\nu}_{e}} \rangle^{\rm plas}$
and the blocking factor is given by
\beq
\langle {\rm block} \rangle_{\nu_x \bar{\nu}_{x}}^{\rm pair} \approx
\left[
1+\exp \left(
\eta_{\nu_x} - 
\frac{\langle \epsilon_{\nu_{x}\bar{\nu}_{x}} \rangle^{\rm plas}}
{k_{B}T}
\right)
\right]^{-1} 
\left[
1+\exp \left(
\eta_{\bar{\nu}_x} - 
\frac{\langle \epsilon_{\nu_{x}\bar{\nu}_{x}} \rangle^{\rm plas}}
{k_{B}T}
\right)
\right]^{-1}.
\eeq

\subsection{Nucleon-nucleon bremsstrahlung}

We follow Burrows et al. \cite{Burrows06} for computing 
the pair creation rate of neutrinos by the nucleon-nucleon bremsstrahlung
radiation. They derived the neutrino energy emission
rate associated with the pair creation of $\nu_{x}$ or $\bar{\nu}_{x}$
by the nucleon-nucleon bremsstrahlung radiation without the blocking
factor as
\beq
Q_{\nu_{x}\bar{\nu}_{x}}^{\rm Brems, 0} =
3.62 \times 10^{5} \zeta^{\rm Brems} 
\left( X_{n}^{2} + X_{p}^{2} + \frac{28}{3}X_{n}X_{p} \right)
\rho^{2}
\left( \frac{k_{B}T}{m_{e}c^{2}} \right)^{4.5} 
\langle \epsilon_{\nu_{x}\bar{\nu}_{x}} \rangle^{\rm Brems}, 
\eeq 
where $\zeta^{\rm Brems} \sim 0.5$ is a correction factor and the
average energy is 
\beq
\langle \epsilon_{\nu_{x}\bar{\nu}_{x}} \rangle^{\rm Brems} 
\approx 4.36k_{B}T.
\eeq
To obtain the 'blocked' neutrino energy emission rate
we multiply the blocking factor,
\beq
\langle {\rm block} \rangle_{\nu_{x}\bar{\nu}_{x}}^{\rm Brems} \approx
\left[
1+\exp \left(
\eta_{\nu_x} - 
\frac{\langle \epsilon_{\nu_{x}\bar{\nu}_{x}} \rangle^{\rm Brems}}
{k_{B}T}
\right)
\right]^{-1}
\left[
1+\exp \left(
\eta_{\bar{\nu}_x} - 
\frac{\langle \epsilon_{\nu_{x}\bar{\nu}_{x}} \rangle^{\rm Brems}}
{k_{B}T}
\right)
\right]^{-1},
\eeq
to give
\beq
Q_{\nu_{x}\bar{\nu}_{x}}^{\rm Brems} = 
Q_{\nu_{x}\bar{\nu}_{x}}^{\rm Brems, 0}
\langle {\rm block} \rangle_{\nu_{x}\bar{\nu}_{x}}^{\rm Brems}.
\eeq
The number emission rate of $\nu_{x}$ or $\bar{\nu}_{x}$ is readily
given by
\beq
\gamma_{\nu_{x}\bar{\nu}_{x}}^{\rm Brems} =
3.62 \times 10^{5} \zeta^{\rm Brems} 
\left( X_{n}^{2} + X_{p}^{2} + \frac{28}{3}X_{n}X_{p} \right)
m_{u}\rho
\left( \frac{k_{B}T}{m_{e}c^{2}} \right)^{4.5} 
\langle {\rm block} \rangle_{\nu_{x}\bar{\nu}_{x}}^{\rm Brems}. 
\eeq 

Noting that the weak interaction coefficients of the
bremsstrahlung radiation are\cite{Itoh1996}
$(1-C_{V})^{2} + (1-C_{A})^{2}$ for the pair
creation of $\nu_{x}\bar{\nu}_{x}$ and 
$C_{V}^{2} + C_{A}^{2}$ for the pair
creation of $\nu_{e}\bar{\nu}_{e}$,
the number emission rate and the associated energy emission rate
for $\nu_{e}$ or $\bar{\nu}_{e}$ are written as
\beqn
&&
\gamma_{\nu_{e}\bar{\nu}_{e}}^{\rm Brems} = 
\frac{C_{V}^{\ 2} + C_{A}^{\ 2}}
     {(1-C_{V})^{2} +(1-C_{A})^{2}}
 \gamma_{\nu_{x}\bar{\nu}_{x}}^{\rm Brems}, \\
&&
Q_{\nu_{e}\bar{\nu}_{e}}^{\rm Brems} = 
\frac{C_{V}^{\ 2} + C_{A}^{\ 2}}
     {(1-C_{V})^{2} + (1-C_{A})^{2}}
 Q_{\nu_{x}\bar{\nu}_x}^{\rm Brems}. 
\eeqn

\section{Neutrino diffusion rates} \label{A_nudiff}

We follow Ref.~\citen{Rosswog} for computing the diffusive neutrino-number emission
rate $\gamma_{(\nu)}^{\rm diff}$ and the associated energy emission rate
$Q_{(\nu)}^{\rm diff}$ in Eqs. (\ref{Q_leak}) and (\ref{g_leak}). 
An alternative definition of the diffusion rates are found in Ref.~\citen{Ruffert}.

\subsection{Neutrino diffusion rates}
To calculate the neutrino diffusion rates 
$\gamma_{(\nu)}^{\rm diff}$ and $Q_{(\nu)}^{\rm diff}$, we first
define the neutrino diffusion time.
In this paper, we consider cross sections for scattering on nuclei ($\sigma_{\nu A}^{\rm sc}$), 
and on free nucleons ($\sigma_{\nu p}^{\rm sc}$ and $\sigma_{\nu n}^{\rm sc}$), as well as that
for absorption on free nucleons ($\sigma_{\nu n}^{\rm ab}$ and $\sigma_{\nu p}^{\rm ab}$).

Ignoring the higher-order correction terms in neutrino energy $E_{\nu}$, 
these neutrino cross sections can be written in general as
\beq
\sigma(E_{\nu}) = E_{\nu}^{2}\tilde{\sigma} , \label{crosssect}
\eeq
where $\tilde{\sigma}$ is a 'cross section' in which $E_{\nu}^{2}$
dependence is factored out.
In practice, the cross sections contain the correction terms which cannot be expressed
in the form of Eq. (\ref{crosssect}). We take account of these correction terms,
approximating neutrino-energy dependence on temperature according to
\beq
E_{\nu} \approx k_{B}T\frac{F_{3}(\eta_{\nu})}{F_{2}(\eta_{\nu})}.
\eeq

The opacity is written as 
\beq
\kappa(E_{\nu}) = \sum \kappa_{i}(E_{\nu}) 
= E_{\nu}^{2} \sum \tilde{\kappa}_{i} = E_{\nu}^{2} \tilde{\kappa},
\eeq
and the corresponding optical depth is calculated by
\beq
\tau (E_{\nu}) = \int \kappa (E_{\nu}) ds 
= E_{\nu}^{2}\int \tilde{\kappa} ds 
= E_{\nu}^{2} \tilde{\tau}.
\eeq
Then, we define the neutrino diffusion time by
\beq
T_{\nu}^{\rm diff} (E_{\nu}) \equiv
a^{\rm diff} \frac{\Delta x(E_{\nu})}{c}\tau(E_{\nu})
=
E_{\nu}^{2} a^{\rm diff} \frac{\tilde{\tau}^{2}}{c\tilde{\kappa}} 
= E_{\nu}^{2} \tilde{T}_{\nu}^{\rm diff},
\eeq
where the distance parameter $\Delta x (E_{\nu})$ is given by
\beq
\Delta x (E_{\nu}) = \frac{\tau (E_{\nu})}{\kappa (E_{\nu})}. 
\eeq
Note that $\tilde{T}_{\nu}^{\rm diff}$ can be calculated only using
matter quantities.
Here, $a^{\rm diff}$ is a parameter which controls how many neutrinos
diffuse outward and we chose it to be $3$ following Ref.~\citen{Ruffert}.
For a larger value of $a^{\rm diff}$, the corresponding neutrino emission
rate due to diffusion becomes smaller.

Finally, we define the neutrino diffusion rates by
\beqn
&& \gamma_{(\nu)}^{\rm diff} \equiv \frac{m_u}{\rho}
\int \frac{n_{\nu}(E_{\nu})}{T_{\nu}^{\rm diff}(E_{\nu})}dE_{\nu}
= \frac{1}{a^{\rm diff}} \frac{m_u}{\rho} 
  \frac{4\pi c g_{\nu}}{(hc)^{3}}
  \frac{\tilde{\kappa}}{\tilde{\tau}^{2}} T F_{0}(\eta_{\nu}), \\
&& Q_{(\nu)}^{\rm diff} \equiv 
\int \frac{E_{\nu}n_{\nu}(E_{\nu})}{T_{\nu}^{\rm diff}(E_{\nu})}dE_{\nu}
= \frac{1}{a^{\rm diff}}\frac{4\pi c g_{\nu}}{(hc)^{3}}
  \frac{\tilde{\kappa}}{\tilde{\tau}^{2}} T^{2} F_{1}(\eta_{\nu}).
\eeqn

%
\subsection{Summary of cross sections} \label{CS}

In this subsection, we briefly summarize the cross sections adopted in the
present neutrino leakage scheme.

\subsubsection{Neutrino nucleon scattering}

The total $\nu$-$p$ scattering cross section $\sigma_{p}$ for all
neutrino species is given by
\beq
\sigma_{\nu p}^{\rm sc} =
\frac{\sigma_{0}}{4}\left(\frac{E_{\nu}}{m_{e}c^{2}}\right)^{2} 
\left[(C_{V}-1)^{2} + 3g_{A}^{2} (C_{A}-1)^{2} \right] W^{\rm sc}_{p}(E_{\nu}),
\eeq
where $g_{A}$ is
the axial-vector coupling constant $g_{A} \approx -1.26$.
$W^{\rm sc}_{p}$ is the correction for the proton recoil.
We use the exact expression derived by Horowitz~\cite{Horowitz2002} for 
high neutrino energies $E_{\nu}/m_{p}c^{2} \gtrsim 0.01$.
However, the exact expression has a behavior which is inconvenient to 
treat numerically (such as $0/0$). Thus we adopt expanded forms of the
exact expression in low neutrino energies $E_{\nu}/m_{p}c^{2} \lesssim 0.01$, which give
\beqn
W^{\rm sc}_{p}(E_{\nu})     &\approx&  
1 - 1.524 \frac{E_{\nu}}{m_{p}c^{2}} + 1.451 \left(\frac{E_{\nu}}{m_{p}c^{2}}\right)^{2} ,\\
W^{\rm sc}_{p}(E_{\bar{\nu}}) &\approx&  
1 - 6.874 \frac{E_{\bar{\nu}}}{m_{p}c^{2}} + 29.54 \left(\frac{E_{\bar{\nu}}}{m_{p}c^{2}}\right)^{2} ,
\eeqn
for neutrinos and anti-neutrinos, respectively.
Note that in the case of black hole formation, the neutrino energy becomes
large and this correction becomes important.
On the other hand, the total $\nu -n$ scattering cross section
$\sigma_{n}$ is
\beq
\sigma_{\nu n}^{\rm sc} =
\frac{\sigma_{0}}{16}\left(\frac{E_{\nu}}{m_{e}^{2}c^{2}}\right)^{2} 
\left[ 1 + 3g_{A}^{2} \right] W^{\rm sc}_{n}.
\eeq
We evaluate $W^{\rm sc}_{n}$ by the same method as for $W^{\rm sc}_{p}$. The expanded forms
in the low neutrino energies are
\beqn
W^{\rm sc}_{n}(E_{\nu})    &\approx&  
1 - 0.7659 \frac{E_{\nu}}{m_{n}c^{2}} -1.3947 \left(\frac{E_{\nu}}{m_{n}c^{2}}\right)^{2} , \\
W^{\rm sc}_{n}(E_{\bar{\nu}}) &\approx&  
1 - 7.366 \frac{E_{\bar{\nu}}}{m_{n}c^{2}} + 33.25 \left(\frac{E_{\bar{\nu}}}{m_{n}c^{2}}\right)^{2} ,
\eeqn
for neutrinos and anti-neutrinos, respectively.

\subsubsection{Coherent scattering of neutrinos on nuclei}


The differential cross section for the $\nu$-$A$ neutral current
scattering is written as~\cite{Burrows06}
\beq
\frac{d\sigma_{A}^{\rm sc}}{d\Omega} = \frac{\sigma_{0}}{64 \pi}
\left(\frac{E_{\nu}}{m_{e}c^{2}}\right)^{2}
A^{2}\left[{\cal W}{\cal C}_{\rm FF} + {\cal C}_{\rm LOS}\right]^{2}
\langle {\cal S}_{\rm ion} \rangle
(1+ \cos\theta),
\eeq
where $\theta$ is the azimuthal angle of the scattering and
\beq
{\cal W} = 1- \frac{2Z}{A}(1-2\sin^{2}\theta_{W}).
\eeq
$\langle {\cal S}_{\rm ion} \rangle$, ${\cal C}_{\rm LOS}$, and 
${\cal C}_{FF}$ are correction factors due to 
the Coulomb interaction among the nuclei,\cite{Horowitz97}
due to the electron polarization,\cite{Leinson88}
and due to the finite size of heavy nuclei\cite{FsizeN}.
Because it is known that the correction factor ${\cal C}_{\rm LOS}$ is
important only for low-energy neutrinos~\cite{Burrows06}, 
we consider only $\langle {\cal S}_{\rm ion} \rangle$ and $C_{FF}$.

The correction factor due to the Coulomb interaction among the nuclei
is given by
\beq
\langle {\cal S}_{\rm ion} \rangle = 
\frac{3}{4}\int_{-1}^{1}d\cos \theta (1+\cos\theta) (1- \cos \theta) S_{\rm ion}.
\eeq
Itoh et al.\cite{Itoh2004} presented a detailed fitting formula for the
correction factor. However, the fitting formula is so complicated that we
use a simple approximation based on Ref.~\citen{Itoh75}, in which 
\beq
S_{\rm ion} \approx \frac{(q a_{I})^{2}}{3\Gamma + f(\Gamma)(q a_{I})^{2}},
\eeq
where $q = (2E_{\nu}/\hbar c) \sin (\theta /2)$,
$a_{I} = (4\pi n_{A}/3)^{-1/3}$ is the ion-sphere radius, $n_{A}$ is
the number density of a nucleus,
$\Gamma = (Ze)^{2}/(a_{I}k_{B}T)$ is the conventional parameter that 
characterizes the strongness of the Coulomb interaction, and 
$f(\Gamma)$ is given by\cite{Itoh2004}
\beq
f(\Gamma) \approx 0.73317 - 0.39890 \Gamma + 0.34141 \Gamma^{1/4}
+ 0.05484 \Gamma^{-1/4}.
\eeq
The integration approximately gives for 
$x \equiv E_{\nu}a_{I}/(\hbar c) < 1$
\beq
\langle {\cal S}_{\rm ion} \rangle \approx
 \frac{1}{6  }\frac{1        }{\Gamma    }x^{2}
-\frac{1}{30 }\frac{f(\Gamma)}{\Gamma^{2}}x^{4}
+\frac{1}{135}\frac{(f(\Gamma))^{2}}{\Gamma^{3}}x^{6}
-\frac{1}{567}\frac{(f(\Gamma))^{3}}{\Gamma^{4}}x^{8}
+\frac{1}{2268}\frac{(f(\Gamma))^{4}}{\Gamma^{5}}x^{10}.
\eeq
To use this expression for the case of 
$x \ge 1$, 
we set the maximum value as $ \langle {\cal S}_{\rm ion} \rangle = 
{\rm max} (1, \langle {\cal S}_{\rm ion} \rangle)$ where 
$\langle {\cal S}_{\rm ion} \rangle =1 $ corresponds to the
case without the correction.


\subsubsection{Absorption on free neutrons}
The total cross section of the absorption of electron neutrinos on free neutrons
is given by~\cite{Burrows06}
\beq
\sigma_{n}^{\rm ab} = \sigma_{0} \left(\frac{1+3g_{A}^{2}}{4}\right)
\left(\frac{E_{\nu} + \Delta_{\rm np}}{m_{e}c^{2}}\right)^{2}
\left[1- \left(\frac{m_{e}c^{2}}{E_{\nu}+\Delta_{\rm
np}}\right)\right] W^{\rm ab}_{n},
\eeq
where $\Delta_{\rm np} = m_{n}c^{2}-m_{p}c^{2}$, and
$W^{\rm ab}_{n}$ is the correction for weak magnetism and recoil of neutron.
We use the exact expression derived by Horowitz~\cite{Horowitz2002}.
By contrast with the corrections in the scattering cross section on free nucleons,
the corrections in the absorption do not show the bad behavior at low neutrino energies.
Similarly, the total cross section of the absorption of electron anti-neutrinos on 
free protons is given by~\cite{Burrows06}
\beq
\sigma_{p}^{\rm ab} = \sigma_{0} \left(\frac{1+3g_{A}^{2}}{4}\right)
\left(\frac{E_{\bar{\nu}} - \Delta_{\rm np}}{m_{e}c^{2}}\right)^{2}
\left[1- \left(\frac{m_{e}c^{2}}{E_{\bar{\nu}}-\Delta_{\rm
np}}\right)\right] W^{\rm ab}_{p}.
\eeq
Again, we use the exact expression derived by Horowitz~\cite{Horowitz2002} 
for $W^{\rm ab}_{p}$.

\end{document}